\documentclass[11pt,a4paper]{article}
\pdfoutput=1
\usepackage{jheppub}
\usepackage{slashed}

\makeatletter
\def\@fpheader{\relax}
\makeatother

\usepackage[czech,english]{babel}
\usepackage{graphicx}
\usepackage{amsmath,amsfonts,amssymb}
\DeclareMathOperator{\MyProd}{\scalebox{1.4}{$\mathrm{I\kern-0.2ex I}$}}

\preprint{LCTP-18-27}

\title{Partition Functions of $\mathcal{N}=1$ Gauge Theories on $S^2 \times \mathbb{R}^2_\varepsilon$ and Duality}

\author[a]{Taro Kimura}

\emailAdd{taro.kimura@keio.jp}

\affiliation[a]{Department of Physics, Keio University, Kanagawa 223-8521, Japan}

\author[b, c]{, Jun Nian}

\emailAdd{nian@umich.edu}

\affiliation[b]{Leinweber Center for Theoretical Physics, University of Michigan, Ann Arbor, MI 48109, U.S.A.}

\affiliation[c]{Riemann Center for Geometry and Physics, Leibniz University Hannover, D-30167, Germany}

\author[d]{and Peng Zhao}

\emailAdd{zhaopeng@itp.ac.cn}

\affiliation[d]{Institute of Theoretical Physics, Chinese Academy of Sciences, Beijing 100190, China}

\abstract{We compute the partition functions of $\mathcal{N} = 1$ gauge theories on $S^2 \times \mathbb{R}^2_\varepsilon$ using supersymmetric localization. The path integral reduces to a sum over vortices at the poles of $S^2$ and at the origin of $\mathbb{R}^2_\varepsilon$. The exact partition functions allow us to test Seiberg duality beyond the supersymmetric index. We propose the $\mathcal{N} = 1$ partition functions on the $\Omega$-background, and show that the Nekrasov partition functions can be recovered from these building blocks.}


\arxivnumber{}

\newcommand{\bea}{\begin{eqnarray}}
\newcommand{\eea}{\end{eqnarray}}

\newcommand{\be}{\begin{equation}}
\newcommand{\ee}{\end{equation}}

\begin{document}

\maketitle

%%%%%%%%%%%%%%%%%%%%%%%%%%%%%%%%%%%%%%%%%%%
%%%%%%%%%%%%%%%%%%%%%%%%%%%%%%%%%%%%%%%%%%%
\section{Introduction}\label{sec:introduction}
%%%%%%%%%%%%%%%%%%%%%%%%%%%%%%%%%%%%%%%%%%%
%%%%%%%%%%%%%%%%%%%%%%%%%%%%%%%%%%%%%%%%%%%

The localization technique provides us with an extremely powerful tool for studying physical quantities non-perturbatively \cite{PestunReview}. For theories with extended supersymmetry, it has been applied successfully for the computation of partition functions and expectation values of defect operators, as well as precision tests of dualities. In particular, the low-energy physics of 4d $\mathcal{N}=2$ gauge theories can be studied by computing the Nekrasov partition function \cite{Nekrasov}. Besides flat space, one can also construct supersymmetry and perform localization on curved spacetimes. Following the pioneering work \cite{Pestun}, a plethora of exact results has been obtained in various theories in diverse dimensions.

In minimally supersymmetric theories, which exhibit far richer dynamics, progress has been hindered by the dearth of exact results. Even for 4d $\mathcal{N}=1$ theories, it has been a long-standing problem to localize the theories on the simplest compact manifold $S^4$, aside from indirect attempts using holography \cite{Elvang-2} and analytic continuation in dimensions \cite{Minahan}.  The technical difficulties are explained in \cite{Terashima, LiuZayas}, using the formalism developed in \cite{FestucciaSeiberg, Klare:2012gn, Dumitrescu:2012ha}.  
The compact manifolds on which one can place $\mathcal{N}=1$ theories have been classified in \cite{Closset:2013vra} and exact calculations were performed in \cite{Closset-1, Assel:2014paa, Closset:2014uda, NishiokaYaakov, Benini:2015noa, Honda:2015yha, GRW-2, Benini:2016hjo, Closset:2017bse}. 
However, most available results are generalizations of the Witten index and related to the well-studied superconformal indices \cite{Romelsberger, MaldacenaIndex}. One notable exception is the partition function on $T^2 \times \mathbb{R}^2_\varepsilon$, which has been computed in \cite{Kimura}. By decoupling the Kaluza-Klein modes on $T^2$,
the first author and the collaborators provided a partition function test of the 4d/2d correspondence proposed in \cite{Dorey:1998yh, Dorey:1999zk} (see also \cite{HananyTong-1, Shifman:2004dr, HananyTong-2, DoreyLee-1}).

Inspired by \cite{Shadchin, Kimura}, in this paper we consider $\mathcal{N}=1$ gauge theories on $S^2 \times \mathbb{R}^2_\varepsilon$. From the 4d point of view, the Killing spinor equation is no longer covariant but depends on the directions. This treatment has been used before in the literature \cite{Kawano-1, LY}. We highlight that treating different directions separately is a crucial step, which circumvents previous difficulties and allows us to localize $\mathcal{N}=1$ gauge theories in this case. Following the standard procedure, we construct supersymmetric actions, derive the BPS equations and find the classical solutions. We find vortices and anti-vortices located at the north and the south poles of $S^2$ as well as at the origin of $\mathbb{R}^2_\varepsilon$. This kind of classical configurations is also similar to previous works on Higgs branch localization for other theories and backgrounds \cite{S2-1, S2-2, Benini-3d, Peelaers, Chen, PanPeelaers}. The locations of the (anti-)vortices play the role of the fixed points of the supersymmetry algebra, which are the points where the only contributions to the partition function come from.

Finally, we localize $\mathcal{N} = 1$ gauge theories in the Higgs branch on the background $S^2 \times \mathbb{R}^2_\varepsilon$, and the partition function takes the general form:
\be
  Z = \sum_{\vec m, \vec n} Z_\text{class} Z^\text{vec}_\text{1-loop} Z^\text{chiral}_\text{1-loop}\, ,
\ee
where $\vec m$ and $\vec n$ denote the vortex and the anti-vortex numbers, respectively. $Z_\text{class}$ is the classical contribution. The 1-loop determinants $Z^\text{vec}_\text{1-loop}$ and $Z^\text{chiral}_\text{1-loop}$ can be obtained via the index theorem and the results are expressed as infinite products, which can be suitably regularized.

As an application, we use the partition function to test Seiberg duality between two theories. Seiberg duality has already been subject to numerous tests, most rigorously using superconformal indices \cite{Romelsberger:2007ec, DolanOsborn, Spiridonov-2}. This paper provides another non-index partition function test of Seiberg duality. Moreover, we expect that our $\mathcal{N}=1$ results can be related to the $\mathcal{N}=2$ Nekrasov partition functions. This is because $S^2 \times \mathbb{R}^2_\varepsilon$ can be decomposed into two patches, each identified with a 4d $\Omega$-background with equivariant parameters 
\be
  \varepsilon_1 = \pm \frac{1}{\ell}, \qquad \varepsilon_2 = \varepsilon \, ,
\ee
where $\ell$ denotes the radius of $S^2$. We expect that the results conjectured in this paper can be confirmed by a direct instanton counting in the future.

This paper is organized as follows. In Section~\ref{sec:SUSY}, we construct the $\mathcal{N}=1$ supersymmetry first on $S^2 \times \mathbb{R}^2$, and then on $S^2 \times \mathbb{R}^2_\varepsilon$ through a change of coordinates. In Section~\ref{sec:action}, we construct the actions that are invariant under the supersymmetry transformations. In addition, we derive the BPS equations from the action and find the classical solutions. In Section~\ref{sec:localization}, the partition functions are computed via the Higgs branch localization. In Section~\ref{sec:SeibergDuality}, we test Seiberg duality for 4d $\mathcal{N}=1$ gauge theories with unitary gauge groups. In Section~\ref{sec:NekPartFct} we discuss some possible relations between our results and the $\mathcal{N}=2$ Nekrasov partition functions on the $\Omega$-background. Some possible directions for future research are discussed in Section~\ref{sec:discussion}. Our convention is summarized in Appendix~\ref{app:Convention}. To construct the 4d Killing spinors, we need the 2d Killing spinors on $S^2$, which are discussed in detail in Appendix~\ref{app:S2KS}. The 4d $\mathcal{N}=1$ supersymmetry constructed in Section~\ref{sec:SUSY} can also be written in terms of 2d fields. Assuming that these 2d fields are independent of the coordinates on $\mathbb{R}^2_\varepsilon$, we can also rewrite the 4d $\mathcal{N}=1$ supersymmetry as a 2d $\mathcal{N}=(2,2)$ supersymmetry discussed in Appendix~\ref{app:2dSUSY}. Some identities relevant to deriving the BPS equations are listed in Appendix~\ref{app:BPS}, where we also discuss the classical solutions to the BPS equations. In Appendix~\ref{app:IndexThm}, we apply the index theorem to compute the 1-loop determinants for the partition function. Some special functions used in the paper are collected in Appendix~\ref{app:SpFct}.

%%%%%%%%%%%%%%%%%%%%%%%%%%%%%%%%%%%%%%%%%%%
%%%%%%%%%%%%%%%%%%%%%%%%%%%%%%%%%%%%%%%%%%%
\section{4D $\mathcal{N}=1$ Supersymmetry}\label{sec:SUSY}
%%%%%%%%%%%%%%%%%%%%%%%%%%%%%%%%%%%%%%%%%%%
%%%%%%%%%%%%%%%%%%%%%%%%%%%%%%%%%%%%%%%%%%%

\subsection{Background and Killing Spinors}\label{sec:bgd}

The metric on $S^2 \times \mathbb{R}^2_\varepsilon$ with the 2d $\Omega$-background is given in the coordinates $(\theta,\, \varphi,\, w,\, \bar{w})$ by
\be\label{eq:metric-1}
  ds^2 = \ell^2 (d\theta^2 + \textrm{sin}^2 \theta\, d\varphi^2) + |dw - i w \ell \varepsilon\, d\varphi|^2\, ,
\ee
where $\ell$ denotes the radius of the two-sphere $S^2$, and we assume that $\varepsilon$ is a real parameter. We choose the real vielbeins $e^i$ ($i = 1,\, \cdots,\, 4$) to be
\be\label{eq:vielbein}
  e^1 = \ell\, d\theta\, ,\quad e^2 = \ell\, \textrm{sin} \theta\, d\varphi\, ,\quad e^3 + i e^4 = dw - i w \ell \varepsilon\, d\varphi\, ,\quad e^3 - i e^4 = d\bar{w} + i \bar{w} \ell \varepsilon\, d\varphi\, .
\ee
Consequently, the nonvanishing components of the spin connection are
\begin{align}
\begin{split}\label{eq:spinconn}
  \omega^{12} & = - \omega^{21} = - \textrm{cos} \theta\, d\varphi\, ,\\
  \omega^{34} & = - \omega^{43} = \ell \varepsilon d\varphi\, .
\end{split}
\end{align}

Let us define a set of new coordinates:
\be\label{eq:ChangeCoord}
  \tilde{\theta} = \theta\, ,\quad \tilde{\varphi} = \varphi\, ,\quad z = w\, e^{- i \ell \varepsilon \varphi}\, ,\quad \bar{z} = \bar{w}\, e^{i \ell \varepsilon \varphi}\, .
\ee
Although the $S^2$ part of the coordinates remains the same, to distinguish the new coordinates from the old ones we use $(\tilde{\theta},\, \tilde{\varphi})$ instead of $(\theta,\, \varphi)$. In the new coordinates $(\tilde{\theta},\, \tilde{\varphi},\, z,\, \bar{z})$ we can rewrite the metric \eqref{eq:metric-1} into the following form:
\be\label{eq:metric-2}
  ds^2 = \ell^2 (d\tilde{\theta}^2 + \textrm{sin}^2 \tilde{\theta}\, d\tilde{\varphi}^2) + |d z|^2\, ,
\ee
which is the same as the one for $S^2 \times \mathbb{R}^2$. In fact, a similar coordinate transformation can be applied to $T^2 \times \mathbb{R}^2_\varepsilon$ discussed in \cite{Kimura} and bring it into $T^2 \times \mathbb{R}^2$.\footnote{We would like to thank Cyril Closset for very helpful discussions on this point.}

For the new metric \eqref{eq:metric-2}, we can choose a new set of real vielbeins $\tilde{e}^i$ as follows:
\be\label{eq:newvielbein}
  \tilde{e}^1 = \ell\, d\tilde{\theta}\, ,\quad \tilde{e}^2 = \ell\, \textrm{sin} \tilde{\theta}\, d\tilde{\varphi}\, ,\quad \tilde{e}^3 + i \tilde{e}^4 = dz\, ,\quad \tilde{e}^3 - i \tilde{e}^4 = d\bar{z}\, .
\ee
Within this new frame, the only nonvanishing components of the spin connection are
\be
  \widetilde{\omega}^{12} = - \widetilde{\omega}^{21} = - \textrm{cos} \tilde{\theta}\, d\tilde{\varphi}\, .\label{eq:newspinconn-1}
\ee

To study the supersymmetry defined on the background $S^2 \times \mathbb{R}^2_\varepsilon$, we first consider the supersymmetry on $S^2 \times \mathbb{R}^2$, which is done in Subsection~\ref{sec:SUSYvec} and \ref{sec:SUSYchiral} for the vector multiplet and the gauged (anti-)chiral multiplet respectively, and then we change the coordinates to obtain the supersymmetry on $S^2 \times \mathbb{R}^2_\varepsilon$ in Subsection~\ref{sec:SUSYS2R2}.

Now let us study the supersymmetry on $S^2 \times \mathbb{R}^2$, whose coordinates are denoted by the indices $\{M,\, N,\, \cdots \}$. In addition, we use the indices $\{\mu,\, \nu,\, \cdots \}$ and $\{a,\, b,\, \cdots \}$ to denote the coordinates on $S^2$ and $\mathbb{R}^2$ respectively. We consider the following Killing spinor equations on $S^2 \times \mathbb{R}^2$:
\be\label{eq:4dKSEUpsilon}
  \nabla_\mu \Upsilon = \frac{1}{2 \ell} \Gamma_\mu \Gamma_5 \Upsilon\, ,\quad \nabla_a \Upsilon = 0\, ,
\ee
where the covariant derivatives are defined as
\be
  \nabla_M \Upsilon \equiv \left(\partial_M + \frac{1}{4} \omega_M^{PQ} \Gamma_{PQ}\right) \Upsilon\, .
\ee
We would like to emphasize that the non-covariant expression of Eq.~\eqref{eq:4dKSEUpsilon} is crucial in defining an $\mathcal{N}=1$ supersymmetry, which can be used to localize the theory, while other choices of Killing spinor equations will not work.

The 4d Killing spinor $\Upsilon$ can be decomposed as two 4d Killing spinors:
\be
  \Upsilon = \Sigma + \widetilde{\Sigma}\, ,
\ee
which can be further decomposed into 2d Killing spinors:
\be\label{eq:KillingDecomp}
  \Sigma = \epsilon \otimes \zeta_+\, ,\quad \widetilde{\Sigma} = \tilde{\epsilon} \otimes \zeta_-\, ,
\ee
where $\zeta_\pm$ are eigenvectors of $\sigma_3$, which in practice are chosen to be
\be
  \zeta_+ = \left(\begin{array}{c}
  1\\
  0
  \end{array}\right)\, ,\quad
  \zeta_- = \left(\begin{array}{c}
  0\\
  1
  \end{array}\right)\, ,
\ee
while $\epsilon$ and $\tilde{\epsilon}$ are the 2d Killing spinors on $S^2$ satisfying
\be\label{eq:2dKSE}
  \nabla_\mu \epsilon = \frac{1}{2 \ell} \sigma_\mu \sigma_3 \epsilon\, ,\quad \nabla_\mu \tilde{\epsilon} = - \frac{1}{2 \ell} \sigma_\mu \sigma_3 \tilde{\epsilon}\, .
\ee
These equations are the same as a version of the 2d Killing spinor equations on $S^2$ discussed in \cite{S2-1, S2-2, S2-4}. They can be solved exactly, and the main results of the 2d Killing spinors on $S^2$ are also summarized in Appendix~\ref{app:S2KS}. In this paper, we assume the Killing spinors $\epsilon$ and $\tilde{\epsilon}$ to be commuting spinors, and so are the 4d counter-parts $\Sigma$ and $\widetilde{\Sigma}$.

\subsection{Vector Multiplet on $S^2 \times \mathbb{R}^2$}\label{sec:SUSYvec}

The $\mathcal{N}=1$ vector multiplet on Euclidean $\mathbb{R}^4$ consists of a gauge boson $A_M$, a spinor $\Xi$ and a pseudo-scalar auxiliary field $D$, which are all in the adjoint representation of the gauge group. In the Euclidean signature, all the fields in the vector multiplet are complexified. The vector multiplet satisfies the following off-shell SUSY transformations on $\mathbb{R}^4$ \cite{FreedmanBook}:
\begin{align}
\begin{split}\label{eq:vecR4SUSYtrafo}
  \delta A_M & = - \frac{1}{2} \bar{\Upsilon}\, \Gamma_M \Xi\, ,\\
  \delta \Xi & = \frac{1}{4} \Gamma^{MN} F_{MN} \Upsilon + \frac{i}{2} \Gamma_5 D \Upsilon\, ,\\
  \delta D & = \frac{i}{2} \bar{\Upsilon}\, \Gamma_5 \Gamma^M D_M \Xi\, ,
\end{split}
\end{align}
where
\be
  D_M \Xi \equiv \nabla_M \Xi + [A_M,\, \Xi]\, .
\ee
$\Upsilon$ is the 4d SUSY transformation parameter satisfying the Killing spinor equation \eqref{eq:4dKSEUpsilon}, and its Majorana conjugate is
\be
  \bar{\Upsilon} \equiv \Upsilon^T C_4
\ee
with the 4d charge conjugation matrix $C_4$. As discussed in e.g. Ref.~\cite{PvN}, in the Euclidean spacetime a spinor and its complex conjugate or Hermitian conjugate are independent, and the spinors do not satisfy a reality condition. For a Majorana spinor in the Minkowski signature, one has to perform a careful Wick rotation to obtain a consistent theory in the Euclidean signature.

As we discussed before, we are interested in $\mathcal{N}=1$ theories on $S^2 \times \mathbb{R}^2_\varepsilon$. By changing coordinates one can map the background $S^2 \times \mathbb{R}^2_\varepsilon$ into $S^2 \times \mathbb{R}^2$. Hence, we can first construct $\mathcal{N}=1$ theories on $S^2 \times \mathbb{R}^2$, and then obtain the results in the original background $S^2 \times \mathbb{R}^2_\varepsilon$ by a reverse change of coordinates.

Next, we consider the theory on $S^2 \times \mathbb{R}^2$. We decompose the fields into the components along $S^2$ and the ones along $\mathbb{R}^2$, and then rewrite the transformations \eqref{eq:vecR4SUSYtrafo} in terms of these components. Effectively, we will obtain a 2d gauge theory. This procedure shares the same spirit of the papers \cite{Kawano-1, LY}, which is different from the standard dimensional reduction. The difference is that for the dimensional reduction one assumes that the new fields have no dependence of some spacetime directions, while in \cite{Kawano-1, LY} they rewrite the original theory in terms of the lower-dimensional new fields keeping the dependence of the reduced dimensions. This procedure is compatible with our choice of the Killing spinors discussed in the previous subsection.

The original SUSY transformations \eqref{eq:vecR4SUSYtrafo} do not form a closed algebra on $S^2 \times \mathbb{R}^2$. We have to modify the transformations \eqref{eq:vecR4SUSYtrafo} on $\mathbb{R}^4$ properly to obtain the following transformations on $S^2 \times \mathbb{R}^2$:
\begin{align}
\begin{split}\label{eq:vecS2R2SUSYtrafo}
  \delta A_M & = - \frac{1}{2} \bar{\Upsilon}\, \Gamma_M \Xi\, ,\\
  \delta \Xi & = \frac{1}{4} \Gamma^{MN} F_{MN} \Upsilon + \frac{i}{2} \Gamma_5 D \Upsilon + \frac{1}{2} \Gamma^{M a} A_a D_M \Upsilon\, ,\\
  \delta D & = \frac{i}{2} \bar{\Upsilon}\, \Gamma_5 \Gamma^M D_M \Xi + \frac{i}{2} (D_M \Upsilon)^T C_4 \Gamma_5 \Gamma^M \Xi\, .
\end{split}
\end{align}
Assuming that $\Upsilon$ is a commuting spinor satisfying the 4d Killing spinor equation \eqref{eq:4dKSEUpsilon}
%and all the fields are independent of the coordinates $(z,\, \bar{z})$ on $\mathbb{R}^2$
, we obtain the following relations on $S^2 \times \mathbb{R}^2$:
\be
  \big\{\delta_{\Sigma_1},\, \delta_{\Sigma_2}\big\} = 0\, ,\quad \big\{\delta_{\widetilde{\Sigma}_1},\, \delta_{\widetilde{\Sigma}_2}\big\} = 0\, ,
\ee
and
\begin{align}
\begin{split}\label{eq:vecR4SUSYalg}
  \big\{\delta_\Sigma,\, \delta_{\widetilde{\Sigma}}\big\} A_\mu & = \xi^M \partial_M A_\mu - D_\mu (\xi^M A_M) +(\partial_\mu \xi^\nu) A_\nu\, ,\\
  \big\{\delta_\Sigma,\, \delta_{\widetilde{\Sigma}}\big\} A_a & = \xi^M \partial_M A_a + [\xi^M A_M,\, A_a] + \Theta_a\,^b A_b\, ,\\
  \big\{\delta_\Sigma,\, \delta_{\widetilde{\Sigma}}\big\} \Xi & = \xi^M \partial_M \Xi + [\xi^M A_M,\, \Xi] - \frac{1}{4} \Theta_{\mu\nu} \Gamma^{\mu\nu} \Xi + \frac{1}{4} \Theta_{ab} \Gamma^{ab} \Xi\, ,\\
  \big\{\delta_\Sigma,\, \delta_{\widetilde{\Sigma}}\big\} D & = \xi^M \partial_M D + [\xi^M A_M,\, D]\, ,
\end{split}
\end{align}
where
\be
  \delta = \delta_\Sigma + \delta_{\widetilde{\Sigma}}
\ee
with $\delta_\Sigma$ and $\delta_{\widetilde{\Sigma}}$ denoting the transformations generated by the supercharges $\Sigma$ and $\widetilde{\Sigma}$ defined in Eq.~\eqref{eq:KillingDecomp} respectively, and consequently
\be
  \delta^2 = \big\{\delta_\Sigma,\, \delta_{\widetilde{\Sigma}}\big\}\, ,
\ee
while
\be
  \xi^M \equiv \frac{1}{4} \bar{\Upsilon} \Gamma^M \Upsilon\, ,\quad \Theta_{MN} \equiv \frac{1}{4 \ell} \bar{\Upsilon} \Gamma_{MN} \Gamma_5 \Upsilon
\ee
are the parameters of the translation and the Lorentz rotation respectively. From these relations we see that $A_a$ behaves like a scalar field on $S^2$ but like a vector field on $\mathbb{R}^2$, which is opposite for $A_\mu$, while $\Xi$ and $D$ behave as a spinor and a scalar respectively both on $S^2$ and on $\mathbb{R}^2$.

With our choice of the gamma matrices, we can also express $\delta^2 \Xi$ into the following form:
\be
  \delta^2 \Xi = \xi^M \partial_M \Xi + [\xi^M A_M,\, \Xi] - \frac{1}{4} \Theta_{\mu\nu}^L \Gamma^{\mu\nu} P_R \Xi + \frac{1}{4} \Theta_{\mu\nu}^R \Gamma^{\mu\nu} P_L \Xi + \frac{1}{4} \Theta_{ab}^L \Gamma^{ab} P_R \Xi - \frac{1}{4} \Theta_{ab}^R \Gamma^{ab} P_L \Xi\, ,
\ee
where
\be
  \Theta_{MN}^L \equiv \frac{1}{4 \ell} \bar{\Upsilon} \Gamma_{MN} P_L \Upsilon\, ,\quad \Theta_{MN}^R \equiv \frac{1}{4 \ell} \bar{\Upsilon} \Gamma_{MN} P_R \Upsilon\, .
\ee
Therefore,
\begin{align}
\begin{split}
  \delta^2 P_L \Xi & = \xi^M \partial_M (P_L \Xi) + [\xi^M A_M,\, P_L \Xi] + \frac{1}{4} \Theta_{\mu\nu}^R \Gamma^{\mu\nu} P_L \Xi - \frac{1}{4} \Theta_{ab}^R \Gamma^{ab} P_L \Xi\, ,\\
  \delta^2 P_R \Xi & = \xi^M \partial_M (P_R \Xi) + [\xi^M A_M,\, P_R \Xi] - \frac{1}{4} \Theta_{\mu\nu}^L \Gamma^{\mu\nu} P_R \Xi + \frac{1}{4} \Theta_{ab}^L \Gamma^{ab} P_R \Xi\, ,
\end{split}
\end{align}
where $P_L$ and $P_R$ denote the 4d projection operators, which are given by
\be\label{eq:4dProj}
  P_L \equiv \frac{1}{2} (\mathbb{I} + \Gamma_5)\, ,\quad P_R \equiv \frac{1}{2} (\mathbb{I} - \Gamma_5)\, .
\ee

In the above, we have constructed consistent SUSY transformations on $S^2 \times \mathbb{R}^2$ for the $\mathcal{N}=1$ vector multiplet. The transformations and the algebra can also be expressed in terms of the 2d fields explicitly, which in general also depend on the coordinates along $\mathbb{R}^2$. However, if we assume that these 2d fields are independent of the coordinates along $\mathbb{R}^2$, we obtain a 2d $\mathcal{N}=(2,2)$ supersymmetry similar to the dimensional reduction procedure. We mention in the following some relevant results, and more details can be found in Appendix~\ref{app:2dSUSYvec}.

We use the index $\mu$ to denote the coordinates along the $S^2$-direction, and for the $\mathbb{R}^2$-direction we define $z = x_3 + i x_4$ and $\bar{z} = x_3 - i x_4$. Then we can make the following decompositions of the fields:
\begin{align}
\begin{split}\label{eq:AandXiDecomp}
  A_M & \rightarrow A_\mu\, \textrm{with } \mu \in \{1,\, 2\},\, A_z = \frac{1}{2} (A_3 - i A_4),\, A_{\bar{z}} = \frac{1}{2} (A_3 + i A_4)\, ,\\
  \Xi & = \lambda \otimes \zeta_+ + \widetilde{\lambda} \otimes \zeta_-\, .
\end{split}
\end{align}
Consequently,
\be
  P_L \Xi = P_L \lambda \otimes \zeta_+ + P_R \widetilde{\lambda} \otimes \zeta_-\, ,\quad P_R \Xi = P_R \lambda \otimes \zeta_+ + P_L \widetilde{\lambda} \otimes \zeta_-\, ,
\ee
where $P_L$ and $P_R$ on the right-hand side are the 2d projection operators defined as
\be\label{eq:2dProj}
  P_L \equiv \frac{1}{2} (\mathbb{I} + \sigma_3)\, ,\quad P_R \equiv \frac{1}{2} (\mathbb{I} - \sigma_3)\, .
\ee
We do not use different notations to distinguish the 4d and the 2d projection operators, which can be easily read off from the context.

Together with the decomposition of the 4d Killing spinor \eqref{eq:KillingDecomp}, we can also express various parameters appearing in the 4d algebra in terms of the 2d Killing spinors:
\begin{align}
\begin{split}
  \Theta_{12} & = \frac{i}{2 \ell} \tilde{\epsilon}^T C_2 \epsilon\, ,\quad \Theta_{34} = \frac{i}{2 \ell} \tilde{\epsilon}^T C_2 \sigma_3 \epsilon\, ,\\
  \Theta_{12}^L & = \Theta_{34}^L = \frac{i}{2 \ell} \tilde{\epsilon}^T C_2 P_L \epsilon\, ,\\
  \Theta_{12}^R & = - \Theta_{34}^R = - \frac{i}{2 \ell} \tilde{\epsilon}^T C_2 P_R \epsilon\, .
\end{split}
\end{align}
The Lorentz rotations from the square of the SUSY transformations become more transparent in terms of the 2d fields and the 2d parameters, so we express the supersymmetry algebra relations \eqref{eq:vecR4SUSYalg} on $S^2 \times \mathbb{R}^2$ using the new fields and parameters as follows:
\begin{align}
\begin{split}
  \big\{\delta_\Sigma,\, \delta_{\widetilde{\Sigma}}\big\} A_u & = \xi^M \partial_M A_u - D_u \left(\xi^M A_M \right) + \alpha A_u\, ,\\
  \big\{\delta_\Sigma,\, \delta_{\widetilde{\Sigma}}\big\} A_{\bar{u}} & = \xi^M \partial_M A_{\bar{u}} - D_{\bar{u}} \left(\xi^M A_M \right) - \alpha A_{\bar{u}}\, ,\\
  \big\{\delta_\Sigma,\, \delta_{\widetilde{\Sigma}}\big\} A_z & = \xi^M \partial_M A_z + [\xi^M A_M,\, A_z] + \rho A_z\, ,\\
  \big\{\delta_\Sigma,\, \delta_{\widetilde{\Sigma}}\big\} A_{\bar{z}} & = \xi^M \partial_M A_{\bar{z}} + [\xi^M A_M,\, A_{\bar{z}}] - \rho A_{\bar{z}}\, ,\\
  \big\{\delta_\Sigma,\, \delta_{\widetilde{\Sigma}}\big\} P_L \lambda & = \xi^M \partial_M P_L \lambda + [\xi^M A_M,\, P_L \lambda] + \frac{1}{2} \alpha P_L \lambda + \frac{1}{2} \rho P_L \lambda\, ,\\
  \big\{\delta_\Sigma,\, \delta_{\widetilde{\Sigma}}\big\} P_R \widetilde{\lambda} & = \xi^M \partial_M P_R \widetilde{\lambda} + [\xi^M A_M,\, P_R \widetilde{\lambda}] - \frac{1}{2} \alpha P_R \widetilde{\lambda} - \frac{1}{2} \rho P_R \widetilde{\lambda}\, ,\\
  \big\{\delta_\Sigma,\, \delta_{\widetilde{\Sigma}}\big\} P_R \lambda & = \xi^M \partial_M P_R \lambda + [\xi^M A_M,\, P_R \lambda] - \frac{1}{2} \alpha P_R \lambda + \frac{1}{2} \rho P_R \lambda\, ,\\
  \big\{\delta_\Sigma,\, \delta_{\widetilde{\Sigma}}\big\} P_L \widetilde{\lambda} & = \xi^M \partial_M P_L \widetilde{\lambda} + [\xi^M A_M,\, P_L \widetilde{\lambda}] + \frac{1}{2} \alpha P_L \widetilde{\lambda} - \frac{1}{2} \rho P_L \widetilde{\lambda}\, ,\\
  \big\{\delta_\Sigma,\, \delta_{\widetilde{\Sigma}}\big\} D & = \xi^M \partial_M D + [\xi^M A_M,\, D]\, ,
\end{split}
\end{align}
where
\begin{align}
\begin{split}
  A_u \equiv \frac{1}{2} (A_1 - i A_2)\, , & \quad A_{\bar{u}} \equiv \frac{1}{2} (A_1 + i A_2)\, ,\\
  D_u \equiv \frac{1}{2} (D_1 - i D_2)\, , & \quad D_{\bar{u}} \equiv \frac{1}{2} (D_1 + i D_2)\, ,
\end{split}
\end{align}
and
\be\label{eq:alpharho}
  \alpha \equiv - i \Theta_{12} = \frac{1}{2 \ell} (\tilde{\epsilon}^T C_2 \epsilon)\, ,\quad \rho \equiv i \Theta_{34} = - \frac{1}{2 \ell} (\tilde{\epsilon}^T C_2 \sigma_3 \epsilon)
\ee
play the role of the rotation parameters on $S^2$ and on $\mathbb{R}^2$ respectively.

From the relations above, we can clearly see the spins of various fields on $S^2$ and $\mathbb{R}^2$. To summarize, for the vector multiplet on $S^2 \times \mathbb{R}^2$ we find that the SUSY transformations satisfy
\be\label{eq:SUSYalg}
  \delta^2 = \xi^M \tilde{\partial}_M + \xi^M A_M + \alpha J_s^u + \rho J_s^z\, ,
\ee
where $J_s^u$ and $J_s^z$ denote the spins of the field on $S^2$ and $\mathbb{R}^2$ respectively. The square of the 
supersymmetry generates a Lorentz transformation, a gauge transformation, a vector R-rotation by $\rho$ and an axial R-rotation by $\alpha$. As we will see later, it takes this universal form for all fields.

\subsection{Gauged (Anti-)Chiral Multiplet on $S^2 \times \mathbb{R}^2$}\label{sec:SUSYchiral}

The gauged chiral multiplet on Euclidean $\mathbb{R}^4$ consists of a complex scalar $\Phi$, a left-projected spinor $P_L \Psi$ and a complex auxiliary field $F$, while the gauged anti-chiral multiplet includes a complex scalar $\bar{\Phi}$, a right-projected spinor $P_R \Psi$ and a complex auxiliary field $\bar{F}$. In the Euclidean signature, all the fields in the gauged (anti-)chiral multiplet are complexified. They have the following SUSY transformations on $\mathbb{R}^4$ \cite{FreedmanBook}:
\begin{align}
\begin{split}\label{eq:chiralR4SUSYtrafo}
  \delta \Phi & = \frac{1}{\sqrt{2}} \bar{\Upsilon} P_L \Psi\, ,\\
  \delta P_L \Psi & = \frac{1}{\sqrt{2}} P_L (\Gamma^M D_M \Phi + F) \Upsilon\, ,\\
  \delta F & = \frac{1}{\sqrt{2}} \bar{\Upsilon} P_R \Gamma^M D_M \Psi - \bar{\Upsilon} P_R \Xi \Phi\, ,
\end{split}
\end{align}
and
\begin{align}
\begin{split}\label{eq:antichiralR4SUSYtrafo}
  \delta \bar{\Phi} & = \frac{1}{\sqrt{2}} \bar{\Upsilon} P_R \Psi\, ,\\
  \delta P_R \Psi & = \frac{1}{\sqrt{2}} P_R (\Gamma^M D_M \bar{\Phi} + \bar{F}) \Upsilon\, ,\\
  \delta \bar{F} & = \frac{1}{\sqrt{2}} \bar{\Upsilon} P_L \Gamma^M D_M \Psi - \bar{\Upsilon} P_L \Xi \bar{\Phi}\, .
\end{split}
\end{align}

The modified transformations on $S^2 \times \mathbb{R}^2$ are
\begin{align}
\begin{split}\label{eq:chiralS2R2SUSYtrafo}
  \delta \Phi & = \frac{1}{\sqrt{2}} \bar{\Upsilon} P_L \Psi\, ,\\
  \delta P_L \Psi & = \frac{1}{\sqrt{2}} P_L (\Gamma^M D_M \Phi + F) \Upsilon + \frac{q}{\sqrt{2}} (P_L \Gamma^M D_M \Upsilon) \Phi\, ,\\
  \delta F & = \frac{1}{\sqrt{2}} \bar{\Upsilon} P_R \Gamma^M D_M \Psi - \bar{\Upsilon} P_R \Xi \Phi + \frac{q}{\sqrt{2}} (D_M \Upsilon)^T C_4 P_R \Gamma^M \Psi\, ,
\end{split}
\end{align}
and for the gauged anti-chiral multiplet $(\bar{\Phi},\, P_R \Psi,\, \bar{F})$:
\begin{align}
\begin{split}\label{eq:antichiralS2R2SUSYtrafo}
  \delta \bar{\Phi} & = \frac{1}{\sqrt{2}} \bar{\Upsilon} P_R \Psi\, ,\\
  \delta P_R \Psi & = \frac{1}{\sqrt{2}} P_R (\Gamma^M D_M \bar{\Phi} + \bar{F}) \Upsilon + \frac{q}{\sqrt{2}} (P_R \Gamma^M D_M \Upsilon) \bar{\Phi}\, ,\\
  \delta \bar{F} & = \frac{1}{\sqrt{2}} \bar{\Upsilon} P_L \Gamma^M D_M \Psi - \bar{\Upsilon} P_L \Xi \bar{\Phi} + \frac{q}{\sqrt{2}} (D_M \Upsilon)^T C_4 P_L \Gamma^M \Psi\, .
\end{split}
\end{align}
Assuming that $\Upsilon$ is a commuting spinor satisfying the 4d Killing spinor equation \eqref{eq:4dKSEUpsilon},
% and all the fields are independent of the coordinates $(z,\, \bar{z})$ on $\mathbb{R}^2$
we find that these modified transformations on $S^2 \times \mathbb{R}^2$ satisfy the following relations:
\be
  \big\{\delta_{\Sigma_1},\, \delta_{\Sigma_2}\big\} = 0\, ,\quad \big\{\delta_{\widetilde{\Sigma}_1},\, \delta_{\widetilde{\Sigma}_2}\big\} = 0\, ,
\ee
and
\begin{align}
\begin{split}\label{eq:chiralR4SUSYalg}
  \big\{\delta_\Sigma,\, \delta_{\widetilde{\Sigma}}\big\} \Phi & = \xi^M \partial_M \Phi + [\xi^M A_M,\, \Phi]\, ,\\
  \big\{\delta_\Sigma,\, \delta_{\widetilde{\Sigma}}\big\} P_L \Psi & = \xi^M \partial_M (P_L \Psi) + [\xi^M A_M,\, P_L \Psi] + \frac{1}{4} \Theta_{\mu\nu}^R \Gamma^{\mu\nu} P_L \Psi - \frac{1}{4} \Theta_{ab}^R \Gamma^{ab} P_L \Psi\, ,\\
  \big\{\delta_\Sigma,\, \delta_{\widetilde{\Sigma}}\big\} F & = \xi^M \partial_M F + [\xi^M A_M,\, F]\, ,\\
  \big\{\delta_\Sigma,\, \delta_{\widetilde{\Sigma}}\big\} \bar{\Phi} & = \xi^M \partial_M \bar{\Phi} + [\xi^M A_M,\, \bar{\Phi}]\, ,\\
  \big\{\delta_\Sigma,\, \delta_{\widetilde{\Sigma}}\big\} P_R \Psi & = \xi^M \partial_M (P_R \Psi) + [\xi^M A_M,\, P_R \Psi] - \frac{1}{4} \Theta_{\mu\nu}^L \Gamma^{\mu\nu} P_R \Psi + \frac{1}{4} \Theta_{ab}^L \Gamma^{ab} P_R \Psi\, ,\\
  \big\{\delta_\Sigma,\, \delta_{\widetilde{\Sigma}}\big\} \bar{F} & = \xi^M \partial_M \bar{F} + [\xi^M A_M,\, \bar{F}]\, .
\end{split}
\end{align}
From these relations we see that $P_L \Psi$ and $P_R \Psi$ behave as spinors both on $S^2$ and on $\mathbb{R}^2$, while $\Phi$, $\bar{\Phi}$, $F$ and $\bar{F}$ transform as scalars both on $S^2$ and on $\mathbb{R}^2$.

In the above, we have constructed consistent SUSY transformations on $S^2 \times \mathbb{R}^2$ for the (anti-)chiral multiplet. Like the vector multiplet, the transformations and the algebra for the (anti-)chiral multiplet can also be expressed in terms of the 2d fields explicitly, which in general also depend on the coordinates along $\mathbb{R}^2$. If we assume that these 2d fields are independent of the coordinates along $\mathbb{R}^2$, we obtain a 2d $\mathcal{N}=(2,2)$ supersymmetry similar to the dimensional reduction procedure. Besides some relevant results mentioned in the following, more details can be found in Appendix~\ref{app:2dSUSYchiral}.

Like for the vector multiplet, we decompose the 4d fields into the 2d fields:
\begin{align}
\begin{split}\label{eq:PsiLRDecomp}
  \Phi \to \phi\, ,\quad \bar{\Phi} \to \bar{\phi}\, , & \quad F \to F\, ,\quad \bar{F} \to \bar{F}\, ,\\
  P_L \Psi = P_L \psi \otimes \zeta_+ + P_R \widetilde{\psi} \otimes \zeta_-\, , & \quad P_R \Psi = P_R \psi \otimes \zeta_+ + P_L \widetilde{\psi} \otimes \zeta_-\, .
\end{split}
\end{align}
Based on our choice of the gamma matrices and the previously defined parameters, we can express the SUSY algebra relations \eqref{eq:chiralR4SUSYalg} on $S^2 \times \mathbb{R}^2$ in terms of the 2d fields and parameters as follows:
\begin{align}
\begin{split}
  \big\{\delta_\Sigma,\, \delta_{\widetilde{\Sigma}}\big\} \phi & = \xi^M \partial_M \phi + [\xi^M A_M,\, \phi]\, ,\\
  \big\{\delta_\Sigma,\, \delta_{\widetilde{\Sigma}}\big\} P_L \psi & = \xi^M \partial_M P_L \psi + [\xi^M A_M,\, P_L \psi] + \frac{1}{2} \alpha P_L \psi + \frac{1}{2} \rho P_L \psi\, ,\\
  \big\{\delta_\Sigma,\, \delta_{\widetilde{\Sigma}}\big\} P_R \widetilde{\psi} & = \xi^M \partial_M P_R \widetilde{\psi} + [\xi^M A_M,\, P_R \widetilde{\psi}] - \frac{1}{2} \alpha P_R \widetilde{\psi} - \frac{1}{2} \rho P_R \widetilde{\psi}\, ,\\
  \big\{\delta_\Sigma,\, \delta_{\widetilde{\Sigma}}\big\} F & = \xi^M \partial_M F + [\xi^M A_M,\, F]\, ,\\
  \big\{\delta_\Sigma,\, \delta_{\widetilde{\Sigma}}\big\} \bar{\phi} & = \xi^M \partial_M \bar{\phi} + [\xi^M A_M,\, \bar{\phi}]\, ,\\
  \big\{\delta_\Sigma,\, \delta_{\widetilde{\Sigma}}\big\} P_R \psi & = \xi^M \partial_M P_R \psi + [\xi^M A_M,\, P_R \psi] - \frac{1}{2} \alpha P_R \psi + \frac{1}{2} \rho P_R \psi\, ,\\
  \big\{\delta_\Sigma,\, \delta_{\widetilde{\Sigma}}\big\} P_L \widetilde{\psi} & = \xi^M \partial_M P_L \widetilde{\psi} + [\xi^M A_M,\, P_L \widetilde{\psi}] + \frac{1}{2} \alpha P_L \widetilde{\psi} - \frac{1}{2} \rho P_L \widetilde{\psi}\, ,\\
  \big\{\delta_\Sigma,\, \delta_{\widetilde{\Sigma}}\big\} \bar{F} & = \xi^M \partial_M \bar{F} + [\xi^M A_M,\, \bar{F}]\, .
\end{split}
\end{align}
From the relations above, we summarize that the SUSY transformations for the (anti-)chiral multiplets also satisfy \eqref{eq:SUSYalg}.

\subsection{Supersymmetry on $S^2 \times \mathbb{R}^2_\varepsilon$}\label{sec:SUSYS2R2}

In the previous subsections, we have constructed the SUSY transformations for the vector multiplet and the gauged (anti-)chiral multiplet on $S^2 \times \mathbb{R}^2$. To obtain the SUSY transformations on $S^2 \times \mathbb{R}^2_\varepsilon$, we have to apply an inverse change of coordinates due to Eq.~\eqref{eq:ChangeCoord}.

First, we choose the Killing spinors $\epsilon$ and $\tilde{\epsilon}$ on $S^2$ part of $S^2 \times \mathbb{R}^2$ to be
\be\label{eq:ExplicitKS}
  \epsilon = (\mathbb{I} + i \sigma_3) e^{-i \tilde{\varphi} / 2} \left(\begin{array}{c}
    \textrm{sin} \frac{\tilde{\theta}}{2} \\
    -i\, \textrm{cos} \frac{\tilde{\theta}}{2}
  \end{array}\right)\, ,\quad
  \tilde{\epsilon} = (\mathbb{I} + i \sigma_3) e^{i \tilde{\varphi} / 2} \left(\begin{array}{c}
    \textrm{cos} \frac{\tilde{\theta}}{2} \\
    - i\, \textrm{sin} \frac{\tilde{\theta}}{2}
  \end{array}\right)\, ,
\ee
which are specific solutions to the Killing spinor equations \eqref{eq:2dKSE} on $S^2$ \cite{S2-1, S2-2}. We make this choice, because $\epsilon$ and $\tilde{\epsilon}$ will have definite chirality at the poles of $S^2$.

With these Killing spinors, we obtain the explicit expression of the Killing vector $\xi^M$:
\be
  \xi^1 = 0\, ,\quad \xi^2 = - \textrm{sin}\, \tilde{\theta}\, ,\quad \xi^3 = i\, \textrm{cos} \tilde{\varphi}\, \textrm{sin} \tilde{\theta}\, ,\quad \xi^4 = i\, \textrm{sin} \tilde{\varphi}\, \textrm{sin} \tilde{\theta}\, .
\ee
Hence, using the inverse vielbeins obtained from the vielbeins \eqref{eq:newvielbein}, we can express the Killing vector in the coordinates $(\tilde{\theta},\, \tilde{\varphi},\, z,\, \bar{z})$ as follows:
\be\label{eq:KillingVecS2R2}
  \xi^M \tilde{\partial}_M = -\frac{1}{\ell} \partial_{\tilde{\varphi}} + i\, \textrm{cos} \tilde{\varphi}\, \textrm{sin} \tilde{\theta} (\partial_z + \partial_{\bar{z}}) - \textrm{sin} \tilde{\varphi}\, \textrm{sin} \tilde{\theta} (\partial_z - \partial_{\bar{z}})\, .
\ee

After changing the coordinates $(\tilde{\theta},\, \tilde{\varphi},\, z,\, \bar{z})$ to $(\theta,\, \varphi,\, w,\, \bar{w})$ as discussed in Subsection~\ref{sec:bgd}, the background $S^2 \times \mathbb{R}^2$ becomes $S^2 \times \mathbb{R}^2_\varepsilon$. The SUSY algebra on $S^2 \times \mathbb{R}^2_\varepsilon$ formally remains the same as \eqref{eq:SUSYalg}, but there are a few changes.

First, the Killing spinors \eqref{eq:ExplicitKS} on $S^2 \times \mathbb{R}^2$ are no longer solutions to the Killing spinor equations on $S^2 \times \mathbb{R}^2_\varepsilon$, unless we turn on a background gauge field to cancel the $\omega^{34}$ and the $\omega^{43}$ components of the spin connection \eqref{eq:spinconn}.
\be\label{eq:CovDerWithV}
  D_M \Sigma = \left(\partial_M + \frac{1}{4} \omega_M^{PQ} \Gamma_{PQ} - i V_M \right) \Sigma\, ,\quad D_M \widetilde{\Sigma} = \left(\partial_M + \frac{1}{4} \omega_M^{PQ} \Gamma_{PQ} + i V_M \right) \widetilde{\Sigma}\, ,
\ee
where we have introduced a background gauge field
\be
  V_M dx^M = \frac{1}{2} \omega_M^{34} dx^M = \frac{1}{2} \ell \varepsilon d\varphi\, .\label{eq:bgdGaugeV}
\ee
In Eqs.~\eqref{eq:CovDerWithV}, the opposite signs in front of $V_M$ is due to the fact that $\Gamma_{34} = \mathbb{I} \otimes i \sigma_3$ acting on $\Sigma = \epsilon \otimes \zeta_+$ and $\widetilde{\Sigma} = \tilde{\epsilon} \otimes \zeta_-$ has opposite eigenvalues. We will interpret $V_M$ as the background gauge field for the R-symmetry, and correspondingly the Killing spinors $\Sigma$ and $\widetilde{\Sigma}$ have opposite R-charges. For the physical fields in the theory, their R-charges turn out to be a combination of the standard R-charges with their spins on $\mathbb{R}^2_\varepsilon$. In terms of the 2d field components \eqref{eq:KillingDecomp}\eqref{eq:AandXiDecomp}\eqref{eq:PsiLRDecomp}, we found a consistent R-charge assignment given in Table~\ref{Rcharge}. Introducing the background field $V_M$ can be viewed as a partial topological twist on $\mathbb{R}^2_\varepsilon$.

\begin{table}[h!]
\begin{center}
\begin{tabular}{c|c|c|c}
  $P_L \epsilon$ & $P_R \epsilon$ & $P_L \tilde{\epsilon}$ & $P_R \tilde{\epsilon}$ \\
  \hline
  $-1$ & $-1$ & $+1$ & $+1$
\end{tabular}

\vspace{2mm}
\begin{tabular}{c|c|c|c|c|c}
  $A_\mu$ & $A_w$ & $A_{\bar{w}}$ & $\lambda$ & $\widetilde{\lambda}$ & $D$ \\
  \hline
  $0$ & $-2$ & $+2$ & $-1$ & $+1$ & $0$
\end{tabular}

\vspace{2mm}
\begin{tabular}{c|c|c|c||c|c|c|c}
  $\phi$ & $P_L \psi$ & $P_R \widetilde{\psi}$ & $F$ & $\bar{\phi}$ & $P_R \psi$ & $P_L \widetilde{\psi}$ & $\bar{F}$ \\
  \hline
  $q$ & $q-1$ & $q+1$ & $q$ & $-q$ & $-q-1$ & $-q+1$ & $-q$
\end{tabular}
\end{center}

\caption{R-charge Assignments for the 2-Component Fields}\label{Rcharge}
\end{table}

Second, in the coordinates $(\theta,\, \varphi,\, w,\, \bar{w})$, instead of \eqref{eq:KillingVecS2R2} now $\xi^M \partial_M$ is given by
\begin{align}
  \xi^M \partial_M & = -\frac{1}{\ell} \partial_\varphi - i \varepsilon (w \partial_w - \bar{w} \partial_{\bar{w}}) + i\, \textrm{cos} \varphi\, \textrm{sin} \theta (\partial_w + \partial_{\bar{w}}) - \textrm{sin} \varphi\, \textrm{sin} \theta (\partial_w - \partial_{\bar{w}})\, .\label{eq:KillingVecS2R2epsilon}
\end{align}
Note that $\partial_\varphi$ generates the $U(1)$ rotation on $S^2$, and $(w \partial_w - \bar{w} \partial_{\bar{w}})$ generates the $U(1)$ rotation on $\mathbb{R}^2_\varepsilon$. The Killing vector has two fixed points located at the north pole ($\theta = 0$) and at the south pole ($\theta = \pi$) on $S^2$, and both of them are located at the origin of $\mathbb{R}^2_\varepsilon$ at the same time. We would like to highlight that the expression \eqref{eq:KillingVecS2R2epsilon} is one of the main results of this paper, which is crucial to perform the localization for $\mathcal{N}=1$ gauge theories.

Moreover, besides the background gauge field $V_M$ corresponding to the R-symmetry, we can introduce another background gauge field $\widetilde{A}_M^I$ corresponding to the flavor symmetry:
\be\label{eq:U(1) flavor}
  \widetilde{A}_M^I dx^M = - \ell\, m_I d \varphi\, ,
\ee
where the index $I \in \{ 1,\, \cdots,\, \textrm{rank} (G_\mathcal{F})\}$. For a fixed $I$, $\widetilde{A}_M^I$ is in general a complex background gauge field for a $U(1)$ subgroup in the Cartan of the flavor symmetry $G_\mathcal{F}$. The complex background gauge field has appeared in the literature before, in particular, the 3d case has been studied in Ref.~\cite{SqS3-4} in great detail. We denote the complex parameters $m_I$ as the twisted masses.

Up to now we have used anti-Hermitian generators for the Lie algebra. In order to be consistent with the literature, we will use the Hermitian generators in the following of this paper, which can be obtained by simply replacing the gauge field $A_M$ with $i A_M$ and similar for other background gauge fields. Hence, we can define a parameter $\Lambda$ to denote the gauge transformation:
\be\label{eq:defineLambda}
  \Lambda \equiv i \xi^M A_M\, ,
\ee
which will appear in the SUSY algebra relation \eqref{eq:SUSYalg} from now on.

The full covariant derivative acting on the fields should include the connections of the background gauge fields $V_M$ and $\widetilde{A}_M^I$. For instance, the full covariant derivative acting on the spinor in the chiral multiplet is
\be
  D_M \Psi^I = \left(\nabla_M + i A_M + i R V_M + i \mathcal{F}_I \widetilde{A}_M^I \right) \Psi^I\, .
\ee
Correspondingly, the full SUSY algebra on $S^2 \times \mathbb{R}^2_\varepsilon$ becomes
\be\label{eq:FullSUSYalgTemp}
  \delta^2 = \xi^M \partial_M + \Lambda + \alpha J_s^u + \rho J_s^w + i R\, \xi^M V_M + i \mathcal{F}_I\, \xi^M \widetilde{A}_M^I\, ,
\ee
where $R$ and $\mathcal{F}_I$ denote the R-charge and the $I$-th flavor charge respectively, while the explicit expression of $\xi^M \partial_M$ is given by Eq.~\eqref{eq:KillingVecS2R2epsilon}. We can use the explicit expressions of the Killing spinors \eqref{eq:ExplicitKS} to compute the parameters:
\be
  \alpha = -\frac{i}{\ell}\, \textrm{cos} \theta\, ,\quad \rho = -\frac{i}{\ell}\, .
\ee

A careful analysis shows that the R-charge assignments listed in Table~\ref{Rcharge} can be separated into the $q$-dependent part $R_1$ and the $q$-independent part $R_2$. On the one hand, the $q$-independent R-charges $R_2$ are proportional to the spins in $\mathbb{R}^2_\varepsilon$, more precisely,
\be
  R_2 = -2 J_s^w\, .
\ee
On the other hand, due to the fact that the two background gauge fields $V_M$ and $\widetilde{A}_M^I$ are proportional to each other, the $q$-dependent R-charges $R_1$ can be absorbed into the twisted masses, which means that for $\mathcal{F}_I \neq 0$ we can define new twisted masses $\widetilde{m}_I$ as
\be\label{eq:RshiftMass}
  \widetilde{m}_I \equiv m_I - \frac{\varepsilon R_1}{2 \mathcal{F}_I}\, .
\ee
In the following, we choose $\mathcal{F}_I = 1$ for flavors in the fundamental representation, while $\mathcal{F}_I = 0$ for the adjoint representation. Hence, in the presence of fundamental chiral multiplets the spins on $\mathbb{R}^2_\varepsilon$, i.e. $J_s^w$, do not need to show up in the SUSY algebra. In principle, the R-charges can be affected by the superpotential. In this paper, we do not consider superpotential for simplicity, and we will set $q=0$ in the following. Using the explicit expressions of $\alpha$, $\rho$, $V_M$ and $\widetilde{A}_M^I$, we can now express the SUSY algebra \eqref{eq:FullSUSYalgTemp} as
\be\label{eq:FullSUSYalg}
  \delta^2 = \xi^M \partial_M + \Lambda - \frac{i}{\ell}\, \textrm{cos} \theta\, J_s^u - \frac{i}{2} (\varepsilon - \frac{1}{\ell}) R_2 + i \mathcal{F}_I \widetilde{m}_I\, .
\ee

So far we have constructed a 4d $\mathcal{N}=1$ background explicitly. It would be nice to obtain it in a more systematic way, similar to the 4d $\mathcal{N}=1$ backgrounds from supergravity models \cite{FestucciaSeiberg, Klare:2012gn, Dumitrescu:2012ha} or the 4d $\mathcal{N}=2$ $\Omega$-background introduced in \cite{Nekrasov, NekrasovMarshakov, NekrasovOkounkov}. However, as discussed before it is crucial for our purpose to have a non-covariant expression \eqref{eq:4dKSEUpsilon} of the Killing spinor equation, which does not fit into the framework of the $\mathcal{N}=1$ backgrounds from supergravity models. Hence, we provide an explicit example beyond the class from supergravity. Whether this example belongs to a larger class and can be obtained more systematically, requires further studies in the future.

Finally in this section, let us make some comments on the global anomalies of the theory, similar to the analysis in \cite{Closset-1}. The $U(1)$ flavor symmetries and the $U(1)$ R-symmetry can potentially have some cubic and mixed anomalies with the following coefficients:
\be
  \textrm{Tr} (Q_R^3) = R^3\, ,\quad \textrm{Tr} (Q_F^3) = \mathcal{F}^3\, ,\quad \textrm{Tr} (Q_R) = R\, ,\quad \textrm{Tr} (Q_F) = \mathcal{F}\, ,
\ee
as well as $\textrm{Tr} (Q_R^2 Q_F)$ and $\textrm{Tr} (Q_F^2 Q_R)$. The violation of the classical conservation laws are given by these anomaly coefficients and the topological densities:
\be
  \mathcal{P}^{(F)} = \epsilon^{MNPQ} \widetilde{F}_{MN} \widetilde{F}_{PQ}\, ,\quad \mathcal{P}^{(R)} = \epsilon^{MNPQ} F^R_{MN} F^R_{PQ}\, ,\quad \mathcal{P}^{(g)} = \epsilon^{MNPQ} R_{MNRS} R_{PQ}\,^{RS}\, ,
\ee
where $\widetilde{F}_{MN}$ and $F^R_{MN}$ are the field strengths of the $U(1)$ flavor symmetry and the $U(1)$ R-symmetry respectively. Using the background gauge fields $V_M$ \eqref{eq:bgdGaugeV} and $\widetilde{A}_M^I$ \eqref{eq:U(1) flavor}, one can find by direct computations:
\be
  \widetilde{F}_{MN} = 0 = F^R_{MN}\, .
\ee
Hence, the topological densities $\mathcal{P}^{(F)}$ and $\mathcal{P}^{(R)}$ vanish. Similarly, one can show by direct computation that the topological density $\mathcal{P}^{(g)}$ also vanishes on the curved space $S^2 \times \mathbb{R}^2$. Hence, all the $U(1)$ flavor symmetry and $U(1)$ R-symmetry currents are conserved. Therefore, we do not need to worry about the global anomalies for the gauge theories constructed on the background discussed in this paper.

%%%%%%%%%%%%%%%%%%%%%%%%%%%%%%%%%%%%%%%%%%%
%%%%%%%%%%%%%%%%%%%%%%%%%%%%%%%%%%%%%%%%%%%
\section{Supersymmetric Action}\label{sec:action}
%%%%%%%%%%%%%%%%%%%%%%%%%%%%%%%%%%%%%%%%%%%
%%%%%%%%%%%%%%%%%%%%%%%%%%%%%%%%%%%%%%%%%%%

\subsection{$\delta$-Exact Action and BPS Equations}\label{sec:ExactAction}

To localize the $\mathcal{N}=1$ supersymmetric theories on $S^2 \times \mathbb{R}^2_\varepsilon$, we should introduce a $\delta$-exact term to deform the original theory, which is obtained by generalizing the standard $\mathcal{N}=1$ gauge theory on $\mathbb{R}^4$ to $S^2 \times \mathbb{R}^2$ and is also $\delta$-exact. Hence, the original theory together with the deformation can be written into a $\delta$-exact form:
\be
  \mathscr{L}_{\textrm{exact}} = \delta \mathcal{V}\, ,
\ee
which consists of several parts:
\be
  \mathcal{V} = \mathcal{V}_{\textrm{gauge}} + \mathcal{V}_{\textrm{chiral}} + \mathcal{V}_H\, .
\ee
Following the standard approach, we can choose
\begin{align}
\begin{split}\label{eq:VH}
  \mathcal{V}_{\textrm{gauge}} & = \frac{1}{2 g_{YM}^2} \textrm{Tr} \left[(\delta \Xi)^\dagger \Xi \right]\, ,\\
  \mathcal{V}_{\textrm{chiral}} & = \frac{1}{2} \Big[ (\delta P_L \Psi)^\dagger P_L \Psi + (\delta P_R \Psi)^\dagger P_R \Psi \Big]\, ,\\
  \mathcal{V}_H & = \frac{i}{2} \textrm{Tr} \left[ \Big(\Sigma^\dagger \Gamma_5 \Xi + \Xi^\dagger \Gamma_5 \widetilde{\Sigma} \Big) H(\Phi,\, \bar{\Phi}) \right]\, ,
\end{split}
\end{align}
where $H(\Phi,\, \bar{\Phi})$ is a real function of $\Phi$ and $\bar{\Phi}$, and the dagger ($\phantom{|}^\dagger$) denotes the Hermitian conjugate. We emphasize that although similar to \cite{S2-1, Benini-3d, Kimura} this particular form of $\mathcal{V}_H$ with the insertion of the $\Gamma_5$-matrix as well as the separation of the supercharges $\Sigma$ and $\widetilde{\Sigma}$ is carefully chosen and has not appeared in the literature before. Moreover, we choose $H(\Phi,\, \bar{\Phi})$ to be
\be
  H(\Phi,\, \bar{\Phi}) = (\bar{\Phi}^I \Phi^I - \eta)\, ,
\ee
with the index $I$ denoting the flavor, and $\eta$ is the Fayet-Iliopoulos (FI) parameter, which controls the size of the (anti-)vortices that will be discussed later in this section. The standard $\mathcal{N}=1$ gauge theory on $\mathbb{R}^4$ can be recovered from the flat-space limit of $(\delta \mathcal{V}_{\textrm{gauge}} + \delta \mathcal{V}_{\textrm{chiral}})$.

Based on the constructions above, the bosonic part of the $\delta$-exact Lagrangian is given by
\be
  \mathscr{L}_{\textrm{exact}}^b = (\delta \mathcal{V}_{\textrm{gauge}})^b + (\delta \mathcal{V}_{\textrm{chiral}})^b + (\delta \mathcal{V}_H)^b\, ,
\ee
where
\begin{align}
\begin{split}
  (\delta \mathcal{V}_{\textrm{gauge}})^b & = \frac{1}{2 g_{YM}^2} \textrm{Tr} \left[ (\delta \Xi)^\dagger (\delta \Xi) \right]\, ,\\
  (\delta \mathcal{V}_{\textrm{chiral}})^b & = \frac{1}{2} \Big[(\delta P_L \Psi)^\dagger (\delta P_L \Psi) + (\delta P_R \Psi)^\dagger (\delta P_R \Psi) \Big]\, ,\\
  (\delta \mathcal{V}_H)^b & = \frac{i}{2} \textrm{Tr} \left[ \Big(\Sigma^\dagger \Gamma_5 (\delta \Xi) + (\delta \Xi)^\dagger \Gamma_5 \widetilde{\Sigma} \Big) (\bar{\Phi}^I \Phi^I - \eta) \right]\, .
\end{split}
\end{align}
Using the choice of the Killing spinors \eqref{eq:ExplicitKS}, one can work out the explicit expressions of these terms. Meanwhile, one needs some identities of the Killing spinors \eqref{eq:ExplicitKS}, which are summarized in Appendix~\ref{app:BPS}.

For simplicity, we first focus on an Abelian gauge theory, and the non-Abelian generalization will be discussed in Subsection~\ref{sec:nonAbelian}. Because the fermionic part $\mathscr{L}_{\textrm{exact}}^f$ is irrelevant for our later discussions, we only list the explicit expressions of the bosonic part $\mathscr{L}_{\textrm{exact}}^b$ in the following:
\begin{align}
  \mathscr{L}_{\textrm{exact}}^b & = \frac{1}{4 g_{YM}^2} F^{MN} F_{MN} - \frac{1}{2 g_{YM}^2} D^2 + \frac{1}{2 g_{YM}^2 \ell^2} (A_3^2 + A_4^2) + \frac{1}{g_{YM}^2 \ell}\, \textrm{sin}\, \theta\, (F_{23} A_4 - F_{24} A_3) \nonumber\\
  {} & \quad - \frac{1}{8 g_{YM}^2} \, \textrm{cos}\, \theta\, \epsilon^{MN}\,_{PQ}\, F_{MN} F^{PQ} - (F_{12} - \textrm{cos}\, \theta\, F_{34} + D) (| \Phi^I |^2 - \eta) \nonumber\\
  {} & \quad + (D^M \Phi^I)^\dagger (D_M \Phi^I) + \left(F^I - \frac{q}{\ell}\, \textrm{cos}\, \theta\, \Phi^I\right)^\dagger \left(F^I - \frac{q}{\ell}\, \textrm{cos}\, \theta\, \Phi^I\right) + \frac{q^2\, \textrm{sin}^2 \theta}{\ell^2} (\Phi^I)^\dagger \Phi^I\, ,\label{eq:LexactWithD}
\end{align}
where to obtain a positive semi-definite action we have assumed that $D$ is an anti-hermitian field, and we have imposed the following reality conditions:
\be
  \bar{\Phi}^I = (\Phi^I)^\dagger\, ,\quad \bar{F}^I = (F^I)^\dagger\, .
\ee

By integrating out the auxiliary fields $D$ and $F$, we see that $\mathscr{L}_{\textrm{exact}}^b$ can be written as a sum of several squared terms:
\begin{align}
  \mathscr{L}_{\textrm{exact}}^b & = \frac{1}{2 g_{YM}^2} \Big(F_{12} - \textrm{cos}\, \theta\, F_{34} - g_{YM}^2 (| \Phi^I |^2 - \eta) \Big)^2 + \frac{\textrm{sin}^2 \theta}{2 g_{YM}^2} (F_{34})^2 \nonumber\\
  {} & \quad + \frac{1}{2 g_{YM}^2} \left(F_{13} + \textrm{cos}\, \theta\, F_{24} \right)^2 +  \frac{1}{2 g_{YM}^2} \left(\textrm{sin}\, \theta\, F_{24} - \frac{1}{\ell} A_3 \right)^2 \nonumber\\
  {} & \quad + \frac{1}{2 g_{YM}^2} \left(F_{14} - \textrm{cos}\, \theta\, F_{23} \right)^2 +  \frac{1}{2 g_{YM}^2} \left(\textrm{sin}\, \theta\, F_{23} + \frac{1}{\ell} A_4 \right)^2 \nonumber\\
  {} & \quad + (D^{\bar{u}} \Phi^I)^\dagger (D_{\bar{u}} \Phi^I) + (D^{\bar{w}} \Phi^I)^\dagger (D_{\bar{w}} \Phi^I) + \frac{q^2\, \textrm{sin}^2 \theta}{\ell^2} (\Phi^I)^\dagger \Phi^I\, ,\label{eq:LexactWithoutD}
\end{align}
where
\be
  D_{\bar{w}} \Phi^I \equiv \frac{1}{2} (D_3 + i D_4) \Phi^I\, .
\ee
In fact, the term $(D^{\bar{w}} \Phi^I)^\dagger (D_{\bar{w}} \Phi^I)$ in $\mathscr{L}_{\textrm{exact}}^b$ can be replaced by $(D^w \Phi^I)^\dagger (D_w \Phi^I)$, and they differ by a total derivative term. Therefore, for $q = 0$ we obtain the BPS equations by setting each squared term to be zero:
\be
  F_{12} - \textrm{cos}\, \theta\, F_{34} - g_{YM}^2 (| \Phi^I |^2 - \eta) = 0\, ,\quad \textrm{sin}\, \theta\, F_{34} = 0\, ,\label{eq:AbelianBPS-1}
\ee
\be
  F_{13} + \textrm{cos}\, \theta\, F_{24} = 0\, ,\quad \textrm{sin}\, \theta\, F_{24} - \frac{1}{\ell} A_3 = 0\, ,\label{eq:AbelianBPS-2-1}
\ee
\be
\quad F_{14} - \textrm{cos}\, \theta\, F_{23} = 0\, ,\quad \textrm{sin}\, \theta\, F_{23} + \frac{1}{\ell} A_4 = 0\, ,\label{eq:AbelianBPS-2-2}
\ee
\be
  D_{\bar{u}} \Phi^I = 0\, ,\quad \left(D_w \Phi^I = 0 \textrm{ or } D_{\bar{w}} \Phi^I = 0\right)\, ,\label{eq:AbelianBPS-3}
\ee
where for the last equation we have two choices, which correspond to anti-vortex and vortex solutions respectively, as we will see in the next subsection. Note that in the flat-space limit ($\ell \to \infty$) and near the poles, the system of equations reduce to the instanton-vortex equations first found by \cite{HananyTong-2} and extensively studied in \cite{Nitta}. The classical solutions to these BPS equations will be discussed in the next subsection.

\subsection{Classical Solutions}\label{sec:ClassicalSol}

In this subsection, we discuss the classical solutions to the BPS equations \eqref{eq:AbelianBPS-1} $\sim$ \eqref{eq:AbelianBPS-3}. First, the second equation of \eqref{eq:AbelianBPS-1} leads to
\be
  F_{34} = 0\quad \textrm{for } 0 < \theta < \pi\, .
\ee
Assuming that $A_1$ and $A_2$ are independent of $(w,\, \bar{w})$, Eq.~\eqref{eq:AbelianBPS-2-1} and Eq.~\eqref{eq:AbelianBPS-2-2} imply
\be
  \frac{1}{\ell}\, \partial_\theta A_3 = - \frac{1}{\ell} \textrm{cot}\, \theta\, A_3\, ,\quad \frac{1}{\ell}\, \partial_\theta A_4 = - \frac{1}{\ell} \textrm{cot}\, \theta\, A_4\, ,
\ee
which can be solved by
\be
  A_3 = \frac{f_1 (\varphi)\, g_1 (w,\, \bar{w})}{\textrm{sin}\, \theta}\, ,\quad A_4 = \frac{f_2 (\varphi)\, g_2 (w,\, \bar{w})}{\textrm{sin}\, \theta}\, ,
\ee
with arbitrary functions $f_{1, 2} (\varphi)$ and $g_{1, 2} (w,\, \bar{w})$. The consistency of $F_{23}$ and $F_{24}$ requires
\be
  f_1 (\varphi) = f'_2 (\varphi)\, ,\quad f'_1 (\varphi) = - f_2 (\varphi)\, ,\quad g_1 (w,\, \bar{w}) = g_2 (w,\, \bar{w})\, ,
\ee
which can be solved by
\be
  f_1 (\varphi) = C_1\, \textrm{cos}\, \varphi - C_2\, \textrm{sin}\, \varphi\, ,\quad f_2 (\varphi) = C_2\, \textrm{cos}\, \varphi + C_1\, \textrm{sin}\, \varphi\, ,
\ee
where $C_1$ and $C_2$ are constants. However, if we require $A_3$ and $A_4$ to be regular for $\theta \in [0, \, \pi]$, there are only the trivial solution
\be\label{eq:SolA3A4}
  A_w = A_{\bar{w}} = 0 \quad \Longleftrightarrow \quad A_3 = A_4 = 0\quad \textrm{for } 0 \leq \theta \leq \pi\, .
\ee
Consequently, the regular solutions also imply that
\be\label{eq:BPS-F34}
  F_{34} = 0\quad \textrm{for } 0 \leq \theta \leq \pi\, .
\ee
For the non-Abelian case, there can also be non-trivial solutions for $A_3$ and $A_4$, but if we restrict to continuous and regular configurations, Eq.~\eqref{eq:BPS-F34} still holds. We cannot, at present, rule out the existence of singular solutions such as a point-like instanton to our BPS equation. If such solutions exist, they should also contribute
to the partition function.

With trivial solutions for $A_3$ and $A_4$, the BPS equations reduce to the vortex equation
\be
  F_{12} - g_{YM}^2 (| \Phi^I |^2 - \eta) = 0\, .\label{eq:AbelianBPS-4}
\ee
This one is the most important BPS equation, we will analyze it carefully in the following.

The vortex equation admits the so-called Higgs branch solutions
\be
  F_{12} = 0\, ,\quad \Phi^I \textrm{ are constants with } | \Phi^I | = \sqrt{\eta}\, .
\ee
In this background, the equation $D_{\bar{w}} \Phi^I = 0$ is automatically satisfied. Then in the gauge $A_\theta = 0$, we can solve the equation $D_{\bar{u}} \Phi^I = 0$, which can be written more explicitly as
\be
  \left[\frac{1}{\ell} \partial_\theta + \frac{i}{\ell\, \textrm{sin}\, \theta} (\partial_\varphi + i A_\varphi + i \widetilde{A}_\varphi^I) - \frac{\varepsilon}{\textrm{sin}\, \theta} w \partial_w + \frac{\varepsilon}{\textrm{sin}\, \theta} \bar{w} \partial_{\bar{w}} \right] \Phi^I = 0\, .
\ee
For constant $\Phi^I$, the solution to this equation is
\be
  A_\varphi = - \widetilde{A}_\varphi^I = \ell\, \widetilde{m}_I\, .
\ee
The Higgs branch solutions described above are valid for $0 \leq \theta \leq \pi$, including the poles of $S^2$.

With the solution \eqref{eq:SolA3A4}, the equation $D_{\bar{w}} \Phi^I = 0$ implies that $\Phi^I$ are holomorphic functions of $w$. For the $k$-vortex configuration in the $I$-th flavor on $\mathbb{R}^2_\varepsilon$, $\Phi^I$ can be chosen to be
\be\label{eq:VortexSol}
  \Phi^I = f(\theta,\, \varphi)\, w^k\, ,\quad \Phi^J = 0\, \textrm{ for } J \neq I\, ,
\ee
where $f(\theta,\, \varphi)$ is a factor depending on the coordinates on $S^2$. Similarly, the equation $D_{w} \Phi^I = 0$ implies that $\Phi^I$ are anti-holomorphic functions of $\bar{w}$. For the $k'$-anti-vortex configuration in the $I$-th flavor on $\mathbb{R}^2_\varepsilon$, $\Phi^I$ can be chosen to be
\be\label{eq:AntiVortexSol}
  \Phi^I = f(\theta,\, \varphi)\, \bar{w}^{k'}\, ,\quad \Phi^J = 0\, \textrm{ for } J \neq I\, .
\ee

Besides the Higgs branch solutions, Eq.~\eqref{eq:AbelianBPS-4} also has infinite towers of vortex solutions located at the poles of $S^2$, which is a common feature for Higgs branch localization on the spheres \cite{S2-1, S2-2, Benini-3d, Peelaers, Chen, PanPeelaers}. The new phenomenon in our case is that from the $\mathbb{R}^2_\varepsilon$ point of view all these vortices are also located at the origin of $\mathbb{R}^2_\varepsilon$ at the same time.

We first consider the vortex solution throughout the whole $S^2$, and list its asymptotic behaviors at different points in the following. In principle, we can also consider the anti-vortex solution throughout $S^2$, but for the partition function this configuration effectively corresponds to flipping the signs of the equivariant parameters.
\begin{itemize}
\item $\theta \approx 0$ (near the core of the vortex at the north pole):
\be
  \Phi^I \simeq (\theta\, e^{i \varphi})^m w^k\, ,\quad A_\varphi \simeq \ell \widetilde{m}_I - k \ell \varepsilon \quad (k \geq 0)\, ;
\ee

\item $\theta \approx \pi$ (near the core of the anti-vortex at the south pole):
\be
  \Phi^I \simeq (\hat{\theta}\, e^{i \varphi})^n w^{k'}\, ,\quad A_\varphi \simeq \ell \widetilde{m}_I - k' \ell \varepsilon \quad (k' \geq 0)\, ;
\ee

\item $\theta \approx \frac{\pi}{2}$ (on the northern hemisphere far from the core of the vortex):
\be
  \Phi^I \simeq \sqrt{\eta}\, e^{i m \varphi} w^k\, ,\quad A_\varphi \simeq \ell \widetilde{m}_I - m - k \ell \varepsilon \quad (k \geq 0)\, ;
\ee

\item $\theta \approx \frac{\pi}{2}$ (on the southern hemisphere far from the core of the anti-vortex):
\be
  \Phi^I \simeq \sqrt{\eta}\, e^{i n \varphi} w^{k'}\, ,\quad A_\varphi \simeq \ell \widetilde{m}_I - n - k' \ell \varepsilon \quad (k' \geq 0)\, ,
\ee
\end{itemize}
where $\hat{\theta} \equiv \pi - \theta$, and the continuity of $\Phi^{I}$ at the equator will impose
\be
  k = k'\, .
\ee
The integers $m$, $n$, $k = k'$ denote the numbers of vortices at the north pole of $S^2$, vortices at the south pole of $S^2$, vortices at the origin of $\mathbb{R}^2_\varepsilon$ respectively.

In principle there can also be a configuration with a vortex solution \eqref{eq:VortexSol} at the north pole and an anti-vortex solution \eqref{eq:AntiVortexSol} at the south pole, or vice versa. More details of these solutions can be found in Appendix~\ref{app:BPS}. Gluing the solutions from the north and the south patches of $S^2$ together, the continuity of $\Phi^I$ at the equator will exclude all the non-trivial configurations.

From these asymptotic solutions we can compute the fluxes through the northern and the southern hemispheres respectively:
\begin{align}
\begin{split}
  \frac{1}{2 \pi} F_N & = \frac{1}{2 \pi} \int_{\textrm{north}} d\theta d\varphi F_{\theta\varphi} = A_\varphi \Big|_{\theta=0}^{\pi/2} = - m\, ,\\
  \frac{1}{2 \pi} F_S & = \frac{1}{2 \pi} \int_{\textrm{south}} d\theta d\varphi F_{\theta\varphi} =  A_\varphi \Big|_{\theta=\pi/2}^{\pi} = n\, .
\end{split}
\end{align}
Therefore, the total flux through $S^2$ is
\be
  \frac{1}{2 \pi} F_{S^2} = \frac{1}{2 \pi} (F_N + F_S) = n - m\, .
\ee

\subsection{$\delta$-Closed Action}\label{sec:ClosedAction}

Besides the $\delta$-exact action discussed in Subsection~\ref{sec:ExactAction}, we can also introduce some $\delta$-closed terms.

As discussed in Appendix~\ref{app:2dSUSY}, under certain assumptions the 4d $\mathcal{N}=1$ gauge theory that we consider in this paper can also be formulated in terms of the 2d fields. Hence, similar to Ref.~\cite{Kimura}, we can construct a $\delta$-closed term from a twisted superpotential in the 2d superspace:
\be\label{eq:4dTwistedSuperpot}
  S_{\widetilde{W}} = \frac{2}{V_{\mathbb{R}^2_\varepsilon}} \int d^2 w\, \int d^2 u\, d\bar{\theta}\, d\theta\, \widetilde{W}\, ,
\ee
where $V_{\mathbb{R}^2_\varepsilon}$ is the equivariant volume of $\mathbb{R}^2_\varepsilon$, and
\be
  \widetilde{W} = i \tau_0 \Sigma\, ,
\ee
\be
  \Sigma = \sigma + \frac{1}{2} \lambda_u \theta_{\bar{u}} + \frac{1}{2} \bar{\lambda}_{\bar{u}} \bar{\theta}_u + i F_{u \bar{u}} \theta_{\bar{u}} \bar{\theta}_u\, .
\ee
Hence, at the classical vortex solutions discussed in the previous subsection $S_{\widetilde{W}}$ takes the value:
\be\label{eq:SWtilde}
  S_{\widetilde{W}} = 2 \int d^2 u\, d\bar{\theta}\, d\theta\, i \tau_0 \Sigma = -i \tau_0 \int d\theta\, d\varphi\, F_{\theta\varphi} = 2 i \pi \tau_0 (m - n)\, ,
\ee
where
\be\label{eq:2d tau0}
  \tau_0 = \frac{\vartheta}{2 \pi} + i r_{2d}\, ,
\ee
with $\vartheta$ and $r_{2d}$ denoting the theta angle and the 2d FI parameter respectively. This 2d coupling can be lifted to a 4d coupling by taking into account the volume factor of $\mathbb{R}^2_\varepsilon$ via Eq.~\eqref{eq:4dTwistedSuperpot}, and consequently $r_{2d}$ can be related to the 4d gauge coupling:
\be\label{eq:4d tau0}
  \tau_0 = \frac{\vartheta}{2 \pi} + \frac{4 \pi i}{g_{YM}^2} \, .
\ee
In the following, we will apply this expression to the $\delta$-closed action \eqref{eq:SWtilde}.

Besides the $\delta$-closed term from the twisted superpotential discussed above, in principle we can also introduce $\delta$-closed terms from the superpotential as follows:
\be
  S_W = \int d^4 x\, d^2 \theta\, W + \int d^4 x\, d^2 \bar{\theta}\, \overline{W}\, .
\ee
Since these terms are also $\delta$-exact, as long as they do not change the charge assignments of the fields, they will not affect the result of the localization. In this paper, for simplicity we will not consider the terms from the superpotential.

\subsection{Non-Abelian Generalization}\label{sec:nonAbelian}

In Subsections~\ref{sec:ExactAction} and \ref{sec:ClassicalSol}, we have discussed the Abelian gauge theory and its classical solutions. In order to discuss some non-perturbative effects, we have to generalize the theory to the non-Abelian case.

Consider a theory with gauge group $G$ of rank $r$ on $S^2 \times \mathbb{R}^2_\varepsilon$. It has $N_F$ chiral multiplets, whose representations under the gauge group are $\bigotimes_{I=1}^{N_F} \mathfrak{R}^I$.

Now, we should solve the non-Abelian version of the previously discussed Abelian BPS equations \eqref{eq:AbelianBPS-1} $\sim$ \eqref{eq:AbelianBPS-3}. The non-Abelian BPS equations are as follows:
\begin{align}
\begin{split}
  F_{12}^\alpha - \textrm{cos}\, \theta\, F_{34}^\alpha - g_{YM}^2 \big((\Phi^I)^\dagger T^\alpha_{\mathfrak{R}^I} \Phi^I & - \eta \big) = 0\, , \quad \textrm{sin}\, \theta\, F_{34}^\alpha = 0\, ,\\
  F_{13}^\alpha + \textrm{cos}\, \theta\, F_{24}^\alpha = 0\, , & \quad \textrm{sin}\, \theta\, F_{24}^\alpha - \frac{1}{\ell} A_3^\alpha = 0\, ,\\
  \quad F_{14}^\alpha - \textrm{cos}\, \theta\, F_{23}^\alpha = 0\, , & \quad \textrm{sin}\, \theta\, F_{23}^\alpha + \frac{1}{\ell} A_4^\alpha = 0\, ,\\
  D_{\bar{u}} \Phi^I = 0\, ,\quad (D_w \Phi^I & = 0 \textrm{ or } D_{\bar{w}} \Phi^I = 0)\, ,
\end{split}
\end{align}
where $\alpha \in \{1,\, \cdots,\, \textrm{dim} (G)\}$, and $T^\alpha_{\mathfrak{R}^I}$ denotes the generator of the Lie group $G$ in the representation $\mathfrak{R}^I$, while $\eta$ is the FI parameter which is nonvanishing only for the $U(1)$ part of the gauge group. Although most of the equations above formally look the same as their Abelian counter-parts, the Lie group structure is implied for the non-Abelian generalization.

Before discussing the non-Abelian solutions, let us make a short note on the indices. We use $(I_1,\, I_2,\, \cdots,\, I_r)$, $(\vec{\mu}_1,\, \vec{\mu}_2,\, \cdots,\, \vec{\mu}_r)$ and $(k_1,\, k_2,\, \cdots,\, k_r)$ to denote the flavor indices, the weight vectors and the (anti-)vortex numbers at the origin of $\mathbb{R}^2_\varepsilon$ corresponding to different Cartan generators of the gauge group, which are labelled by the index $i \in \{1,\, \cdots,\, r \}$. For a fixed $i$, $\vec{\mu}_i \in \mathfrak{R}^{I_i}$ denotes the $r$-dimensional weight vector of the representation $\mathfrak{R}^{I_i}$.

The classical solutions to the non-Abelian BPS equations can be obtained similar to the Abelian solutions discussed in Subsection~\ref{sec:ClassicalSol}. We list them in the following. First, due to the regularity of the classical gauge fields we still have for the non-Abelian case:
\be
  A_3^\alpha = A_4^\alpha = 0\, .
\ee
Consequently, the remaining non-trivial BPS equations are
\be\label{eq:nontrivialBPS-1}
  2 i F_{u \bar{u}}^\alpha + g_{YM}^2 \left((\Phi^I)^\dagger T^\alpha_{\mathfrak{R}^I} \Phi^I - \eta \right) = 0\, ,
\ee
\be\label{eq:nontrivialBPS-2}
  D_{\bar{u}} \Phi^I = 0\, ,\quad \left(D_w \Phi^I = 0 \textrm{ or } D_{\bar{w}} \Phi^I = 0\right)\, .
\ee
Formally, these equations can be solved by \cite{Kimura}:
\be\label{eq:FormalSol}
  A_{\bar{u}} = -\frac{i}{2} \vec{H} \cdot \partial_{\bar{u}} \vec{\omega}\, ,\quad \Phi^J_\rho = \sqrt{\eta}\, \textrm{exp} \left(-\frac{1}{2} \vec{\rho}\cdot \vec{\omega} \right)\, h^J_\rho\, ,
\ee
where $J$ and $\rho$ denote a general flavor and a general gauge group representation respectively, and $\vec{H}$ are the Cartan generators, while $\vec{\omega} = (\omega_1,\, \omega_2,\, \cdots,\, \omega_r)$ are the profile functions, whose explicit form will not be important. The factor $h^J_\rho$ is a polynomial given by
\be\label{eq:factor-h}
  h^J_\rho = \Bigg\{
  \begin{array}{ll}
    u^{m_i}\, w^{k_i}\, , & \textrm{ for vortex with } \vec{\rho} = \vec{\mu}_i,\, J = I_i\, ;\\
    u^{n_i}\, \bar{w}^{k'_i}\, , & \textrm{ for anti-vortex with } \vec{\rho} = \vec{\mu}_i,\, J = I_i\, ;\\
    0\, , & \textrm{ otherwise}\, .
  \end{array}
\ee
Like the Abelian case, in the non-Abelian case we also have the Higgs branch solutions and the vortex solutions for the other components of the gauge field $A_M^\alpha$ and the scalar field $\Phi^J$. The Higgs branch solutions are characterized by
\be
  F_{12}^\alpha = 0\, ,\quad (\Phi^I)^\dagger T^\alpha_{\mathfrak{R}^I} \Phi^I - \eta = 0\, .
\ee
The infinite towers of the vortex solutions are located at the north pole ($\theta = 0$) and the south pole ($\theta = \pi$) on $S^2$, while at the origin $(w = 0)$ of $\mathbb{R}^2$ at the same time.

Similar to the discussions in Subsection~\ref{sec:ClassicalSol}, we consider a non-Abelian vortex solution to Eq.~\eqref{eq:factor-h} throughout the whole $S^2$. Its asymptotic behaviors at different points are listed as follows:
\begin{itemize}
\item $\theta \approx 0$ (near the core of the vortex at the north pole):
\be
  \Phi^{I_i} \simeq (\theta\, e^{i \varphi})^{m_i} w^{k_i}\, ,\quad A_\varphi^i \simeq \ell \widetilde{m}_{I_i} - k_i \ell \varepsilon \quad (k_i \geq 0)\, ;
\ee

\item $\theta \approx \pi$ (near the core of the anti-vortex at the south pole):
\be
  \Phi^{I_i} \simeq (\hat{\theta}\, e^{i \varphi})^{n_i} w^{k'_i}\, ,\quad A_\varphi^i \simeq \ell \widetilde{m}_{I_i} - k'_i \ell \varepsilon \quad (k'_i \geq 0)\, ;
\ee

\item $\theta \approx \frac{\pi}{2}$ (on the northern hemisphere far from the core of the vortex):
\be
  \Phi^{I_i} \simeq \sqrt{\eta}\, e^{i m_i \varphi} w^{k_i}\, ,\quad A_\varphi^i \simeq \ell \widetilde{m}_{I_i} - m_i - k_i \ell \varepsilon \quad (k_i \geq 0)\, ;
\ee

\item $\theta \approx \frac{\pi}{2}$ (on the southern hemisphere far from the core of the anti-vortex):
\be
  \Phi^{I_i} \simeq \sqrt{\eta}\, e^{i n_i \varphi} w^{k'_i}\, ,\quad A_\varphi^i \simeq \ell \widetilde{m}_{I_i} - n_i - k'_i \ell \varepsilon \quad (k'_i \geq 0)\, ,
\ee
\end{itemize}
where $\hat{\theta} \equiv \pi - \theta$, and the continuity of $\Phi^{I_i}$ at the equator will impose
\be
  k_i = k'_i\, .
\ee
The integers $m_i$, $n_i$, $k_i = k'_i$ denote the numbers of the vortices at the north pole of $S^2$, the vortices at the south pole of $S^2$, the vortices at the origin of $\mathbb{R}^2_\varepsilon$ for the $i$-th Cartan generator respectively.

In principle there can also a configuration with a vortex solution at the north pole and an anti-vortex solution at the south pole, or vice versa. Gluing the solutions from the north and the south patches of $S^2$ together, the continuity of $\Phi^{I_i}$ at the equator will exclude all the non-trivial solutions.

%%%%%%%%%%%%%%%%%%%%%%%%%%%%%%%%%%%%%%%%%%%
%%%%%%%%%%%%%%%%%%%%%%%%%%%%%%%%%%%%%%%%%%%
\section{Supersymmetric Localization}\label{sec:localization}
%%%%%%%%%%%%%%%%%%%%%%%%%%%%%%%%%%%%%%%%%%%
%%%%%%%%%%%%%%%%%%%%%%%%%%%%%%%%%%%%%%%%%%%

\subsection{1-Loop Determinants}\label{sec:1-loop}

The partition function is given by the classical solutions and the 1-loop determinants of fluctuations around them. They can be conveniently computed using the index theorem by summing over contributions at the fixed points. We list the results in the following, and more details can be found in Appendix~\ref{app:IndexThm}.

We compute the index of the Dolbeault operator for the chiral multiplet. The Abelian case is discussed in Appendix~\ref{app:IndexThm}, and now we generalize the results there to the non-Abelian case. For the $m_i$-vortices in the $I_i$-th flavor background the index at the north pole of $S^2$ ($\theta = 0$) becomes
\be\label{eq:indexChiralNP}
  \textrm{ind}_N^{\textrm{chiral}} = - \sum_{J=1}^{N_F} \sum_{\vec{\rho} \in \mathfrak{R}^J} \sum_{p \in \mathbb{Z}} e^{- i p \varepsilon} \sum_{q \geq 0} e^{-i q / \ell}\, e^{- i (\varepsilon - 1 / \ell) R_2 / 2}\, e^{i \mathcal{F}_J \widetilde{m}_J}\, e^{i \vec{\rho} \cdot \vec{\sigma}^N}\, ,
\ee
where $\vec{\sigma}^N$ are constants defined by
\be\label{eq:FixConstNP}
  \vec{\mu}_i \cdot \vec{\sigma}^N \equiv - \mathcal{F}_{I_i} \widetilde{m}_{I_i} + \frac{1}{\ell} m_i + k_i \varepsilon\, ,
\ee
and $\vec{\sigma}^N$ can be understood as a special value of the Coulomb branch parameter \cite{Kimura}, whose meaning will be clear in the ``Coulomb branch'' localization.\footnote{We put ``Coulomb branch'' in quotation mark, because 4d $\mathcal{N}=1$ gauge theories do not have a standard Coulomb branch like in $\mathcal{N}=2$ theories.}

For the $n_i$-anti-vortices in the $I_i$-th flavor background the index at the south pole of $S^2$ ($\theta = \pi$) becomes
\be\label{eq:indexChiralSP}
  \textrm{ind}_S^{\textrm{chiral}} = \sum_{J=1}^{N_F} \sum_{\vec{\rho} \in \mathfrak{R}^J} \sum_{p \in \mathbb{Z}} e^{i p \varepsilon} \sum_{q \geq 1} e^{i q / \ell}\, e^{- i (\varepsilon - 1 / \ell) R_2 / 2}\, e^{i \mathcal{F}_J \widetilde{m}_J}\, e^{i \vec{\rho} \cdot \vec{\sigma}^S}\, ,
\ee
where $\vec{\sigma}^S$ are constants defined by
\be\label{eq:FixConstSP}
  \vec{\mu}_i \cdot \vec{\sigma}^S \equiv - \mathcal{F}_{I_i} \widetilde{m}_{I_i} + \frac{1}{\ell} n_i + k_i \varepsilon\, .
\ee
In both Eq.~\eqref{eq:FixConstNP} and Eq.~\eqref{eq:FixConstSP}, $\vec{\mu}_i$ denotes the weight vectors in the representation $\mathfrak{R}^{I_i}$ of the gauge group, while $k_i$ can be absorbed into a redefinition of $p$, hence does not appear in the final result.

Combining the indices \eqref{eq:indexChiralNP} and \eqref{eq:indexChiralSP}, we obtain the 1-loop determinant for the generic $J$-th chiral multiplet ($J \neq I_i$) around the vortex solution characterized by the (anti-)vortex numbers $(m_i,\, n_i)$ with the $I_i$-th flavor and the weight vector $\vec{\mu}_i$:
\be\label{eq:nonAbelChiralPartFctProduct}
  \left(Z_{I_i,\, \vec{\mu}_i}^{(m_i, n_i)} \right)_{\textrm{1-loop}}^{\textrm{$J$-th chiral}} = \prod_{\vec{\rho} \in \mathfrak{R}^J} \prod_{p \in \mathbb{Z}} \prod_{q \geq 0} \frac{p \varepsilon + \frac{1}{\ell} (q + 1) - \frac{1}{2} (\varepsilon - \frac{1}{\ell}) R_2 + \mathcal{F}_J \widetilde{m}_J + \vec{\rho} \cdot \vec{\sigma}^S}{- p \varepsilon - \frac{1}{\ell} q - \frac{1}{2} (\varepsilon - \frac{1}{\ell}) R_2 + \mathcal{F}_J \widetilde{m}_J + \vec{\rho} \cdot \vec{\sigma}^N}\, .
\ee
In the absence of superpotential, we will set the R-charge $R_2 = 0$ for simplicity in the following, which can be easily recovered for general R-charge assignments.

Similar to the chiral multiplet, we can compute the indices of the de Rham operator for the vector multiplet. The indices at the north and the south poles are
\be\label{eq:indexVec}
  \textrm{ind}_N^{\textrm{vec}} = \frac{1}{2} \sum_{\vec{\alpha} \in \Delta_G} \sum_{p \in \mathbb{Z}} e^{- i p \varepsilon}\, e^{i \vec{\alpha} \cdot \vec{\sigma}^N}\, ,\quad \textrm{ind}_S^{\textrm{vec}} = \frac{1}{2} \sum_{\vec{\alpha} \in \Delta_G} \sum_{p \in \mathbb{Z}} e^{i p \varepsilon}\, e^{i \vec{\alpha} \cdot \vec{\sigma}^S}\, .
\ee
In Eq.~\eqref{eq:indexVec}, $\vec{\alpha}$ takes value in the set $\Delta_G$ of the root vectors of the gauge group, which generalizes the Abelian case discussed in the previous section to the non-Abelian case. For the Abelian case, $i \vec{\alpha} \cdot \vec{\sigma}^N$ and $i \vec{\alpha} \cdot \vec{\sigma}^S$ both vanish, and the sum over the roots can be removed.

Combining the indices in \eqref{eq:indexVec}, we obtain the 1-loop determinant for the vector multiplet around the vortex solution characterized by the (anti-)vortex numbers $(m_i,\, n_i)$ in the $I_i$-th flavor with the representation $\vec{\mu}_i$:
\be\label{eq:nonAbelVecPartFctProduct}
  \left(Z_{I_i}^{(m_i, n_i)} \right)_{\textrm{1-loop}}^{\textrm{vec}} = \prod_{\vec{\alpha} \in \Delta_G} \prod_{p \in \mathbb{Z}} \Big[-i p \varepsilon + i \vec{\alpha} \cdot \vec{\sigma}^N \Big]^{\frac{1}{2}} \Big[i p \varepsilon + i \vec{\alpha} \cdot \vec{\sigma}^S \Big]^{\frac{1}{2}}\, .
\ee
Similar to Refs.~\cite{DrukkerOkuda, Benini-3d}, the expression \eqref{eq:nonAbelVecPartFctProduct} can be regularized as
\be\label{eq:nonAbelVecPartFctProductReg}
  \left(Z_{I_i}^{(m_i, n_i)} \right)_{\textrm{1-loop}}^{\textrm{vec}} = \prod_{\vec{\alpha} \in \Delta_G} x^{- \frac{|\alpha (m - n)|}{2}} \left(1 - e^{\frac{2 \pi i}{\varepsilon} \alpha (\widetilde{m})} x^{|\alpha (m - n)| - \alpha (m + n)} \right)\, ,
\ee
where the parameter $x$ is defined as
\be\label{eq:define_x}
  x \equiv e^{- \frac{i \pi}{\ell \varepsilon}}\, .
\ee

In Eq.~\eqref{eq:nonAbelChiralPartFctProduct}, we considered a generic flavor $J \neq I_i$. As discussed in Ref.~\cite{Kimura}, for $J = I_i$ there is a subtlety about the contribution from the unphysical modes to the 1-loop determinant, which we would like to clarify now.

As we discussed, the non-trivial BPS equations \eqref{eq:nontrivialBPS-1} \eqref{eq:nontrivialBPS-2} have the formal solutions \eqref{eq:FormalSol}.
The fluctuations around these solutions can be obtained by linearizing the solutions \eqref{eq:FormalSol}:
\be
  \delta A_{\bar{u}} = e^{-\frac{1}{2} \vec{H}\cdot \vec{\omega}} (i \partial_{\bar{u}} \delta \Omega)\, e^{\frac{1}{2} \vec{H}\cdot \vec{\omega}}\, ,\quad \delta \Phi_\rho^J = \sqrt{\eta}\, e^{-\frac{1}{2} \vec{\rho} \cdot \vec{\omega}} \left[\delta h^J + \delta\Omega\cdot h^J \right]_\rho\, ,
\ee
where $\Omega(u, \bar{u})$ is a $\mathfrak{g}$-valued function. The non-trivial BPS equation \eqref{eq:nontrivialBPS-2} includes both the holomorphic case and the anti-holomorphic case. To simplify our discussions, let us focus on the holomorphic case in the following, and the anti-holomorphic case is similar.

These fluctuations are invariant under the following transformations:
\be\label{eq:nonAbelGaugeTrafo}
  \delta h^J \to \delta h^J + v(u, w, \bar{w})\, h^J\, ,\quad \delta \Omega \to \delta \Omega - v(u, w, \bar{w})\, ,
\ee
where $v(u, w, \bar{w})$ is a $\mathfrak{g}$-valued regular function, which is holomorphic in $u$ and has the decomposition:
\be\label{eq:vDecomp}
  v(u, w, \bar{w}) = \vec{P}(u, w, \bar{w}) \cdot \vec{H} + \sum_{\vec{\alpha} \in \Delta_G} Q_{\vec{\alpha}} (u, w, \bar{w})\, E_{\vec{\alpha}}\, .
\ee
The transformations \eqref{eq:nonAbelGaugeTrafo} introduce some unphysical modes, which should be removed from the final result of the partition function.

Due to the decomposition given by Eq.~\eqref{eq:vDecomp}, we distinguish the unphysical modes caused by the Cartan part $\sim H_i$ and the off-diagonal part $\sim E_{\vec{\alpha}}$. Because the unphysical modes caused by the Cartan part are flavor-dependent, removing them will contribute to the partition function of the chiral multiplet. On the contrary, since the off-diagonal part of the unphysical modes is independent of the flavor, when removing these modes, the possible additional contribution can be assigned to the vector multiplet partition function.

For the vortex solution in the Cartan part, due to $h_{\vec{\mu}_i}^{I_i} \sim u^{m_i}$, $v(u)$ can be used to remove all the powers $\geq m_i$ in the expansion of $\delta h_{\vec{\mu}_i}^{I_i}$, i.e.,
\be
  \textrm{for $\vec{\rho} = \vec{\mu}_i,\, J = I_i$:}\quad \delta h_{\vec{\mu}_i}^{I_i} = \sum_{q=0}^{m_i - 1} c_{\vec{\mu}_i,\, q}^{I_i} (w, \bar{w})\, u^q\, .
\ee
Together with the modes around the anti-vortex solution in the Cartan part, the product in the 1-loop determinant \eqref{eq:nonAbelChiralPartFctProduct} for $\vec{\rho} = \vec{\mu}_i,\, J = I_i$ will be restricted to:
\be
  \prod_{p \in \mathbb{Z}} \frac{\prod_{q = 0}^{- n_i - 1} \left(p \varepsilon + \frac{1}{\ell} q + \mathcal{F}_{I_i} \widetilde{m}_{I_i} + \vec{\mu}_i \cdot \vec{\sigma}^S \right)}{ \prod_{q = 0}^{m_i - 1} \left(- p \varepsilon - \frac{1}{\ell} q + \mathcal{F}_{I_i} \widetilde{m}_{I_i} + \vec{\mu}_i \cdot \vec{\sigma}^N \right)} = \prod_{p \in \mathbb{Z}} \frac{\prod_{q = 0}^{- n_i - 1} \left(p \varepsilon + \frac{1}{\ell} q + \frac{1}{\ell} n_i \right)}{ \prod_{q = 0}^{m_i - 1} \left(- p \varepsilon - \frac{1}{\ell} q + \frac{1}{\ell} m_i \right)} \, ,
\ee
where we have used Eqs.~\eqref{eq:FixConstNP} and \eqref{eq:FixConstSP}, and the minus sign of $- n_i$ is due to the regularity requirement of the configurations at the south pole ($u = \infty$). However, for $n_i \geq 0$ the modes around the anti-vortex cannot be removed by the redundancy \eqref{eq:nonAbelGaugeTrafo}, which is consistent with the fact that the range of the infinite product in the numerator is empty for $n_i > 0$. Also, the range of the infinite product in the denominator is empty for $m_i = 0$, hence the expression above should not enter the perturbative part of the partition function.

Finally, the 1-loop determinant of the physical $N_F$ pairs of chiral and anti-chiral multiplets around the vortex solution characterized by the (anti-)vortex numbers $(m_i,\, n_i) \in \mathbb{Z}_{\geq 0}^2$ with the $I_i$-th flavor and the weight vector $\vec{\mu}_i$ is given by
\begin{align}
  \left(Z_{I_i,\, \vec{\mu}_i}^{(m_i, n_i)} \right)_{\textrm{1-loop}}^{\textrm{chiral}} & = \left[\sideset{}{'}\prod_{J=1}^{N_F} \sideset{}{'}\prod_{\vec{\rho} \in \mathfrak{R}^J} \prod_{p \in \mathbb{Z}} \prod_{q \geq 0} \frac{p \varepsilon + \frac{1}{\ell} (q + 1) + \mathcal{F}_J \widetilde{m}_J + \vec{\rho} \cdot \vec{\sigma}^S}{- p \varepsilon - \frac{1}{\ell} q + \mathcal{F}_J \widetilde{m}_J + \vec{\rho} \cdot \vec{\sigma}^N} \right] \nonumber\\
  {} & \quad \cdot \left[\prod_{i=1}^r \prod_{p \in \mathbb{Z}} \frac{\prod_{q \geq 0} \left(p \varepsilon + \frac{1}{\ell} (q + 1) + \frac{1}{\ell} n_i \right)}{\prod_{q = 0}^{m_i - 1} \left(- p \varepsilon - \frac{1}{\ell} q + \frac{1}{\ell} m_i \right)} \right]\, ,\label{eq:nonAbelChiralPartFctProductNF}
\end{align}
where in the first line the prime on the products denotes the condition $(J,\, \vec{\rho}) \neq (I_i,\, \vec{\mu}_i)$. Similar to Refs.~\cite{DrukkerOkuda, Benini-3d}, the expression \eqref{eq:nonAbelChiralPartFctProductNF} can be regularized as
\begin{align}
  {} & \left(Z_{I_i,\, \vec{\mu}_i}^{(m_i, n_i)} \right)_{\textrm{1-loop}}^{\textrm{chiral}} \nonumber\\
  = & \left[\sideset{}{'}\prod_{J=1}^{N_F} \sideset{}{'}\prod_{\vec{\rho} \in \mathfrak{R}^J} \left(x^{1 + \rho (n) + \rho (m) + 2 \ell \mathcal{F}_J \widetilde{m}_J}\, e^{\frac{2 \pi i}{\varepsilon} \rho (\widetilde{m})} \right)^{\frac{\rho(n) - \rho(m)}{2}} \frac{\left(x^{2 + 2 \rho(n) + 2 \ell \mathcal{F}_J \widetilde{m}_J} e^{\frac{2 \pi i}{\varepsilon} \rho(\widetilde{m})};\, x^2 \right)_\infty}{\left(x^{-2 \rho(m) - 2 \ell \mathcal{F}_J \widetilde{m}_J} e^{- \frac{2 \pi i}{\varepsilon} \rho (\widetilde{m})};\, x^2 \right)_\infty} \right] \nonumber\\
  {} & \cdot \left[\prod_{i=1}^r \frac{x^{\frac{1}{2} (n^2 + n + \frac{1}{6})} (x^{2+2n};\, x^2)_\infty}{\prod_{q = 0}^{m_i - 1} \textrm{sinh} \left[\frac{i \pi}{\ell \varepsilon} (q - m_i) \right]} \right]\, ,\label{eq:nonAbelChiralPartFctProductNFReg}
\end{align}
where we used the $q$-Pochhammer symbol $(a;\, q)_n$ defined as
\be
  (a;\, q)_n \equiv \prod_{k = 0}^{n-1} (1 - a\, q^k)\, ,
\ee
and the Hurwitz zeta function regularization:
\be
  \sum_{q \geq 0} (q + x) = \zeta_H (-1, x) = -\frac{1}{2} (x^2 - x + \frac{1}{6})\, .
\ee

There is another possible contribution to the 1-loop determinant of the vector multiplet, which originates from the off-diagonal part of $v(u, w, \bar{w})$ in Eq.~\eqref{eq:vDecomp}. Consider a basis for the off-diagonal part of $v(u, w, \bar{w})$ given by $u^q g(w, \bar{w}) E_{\vec{\alpha}}$ ($q \geq 0$). The corresponding unphysical modes are
\be
  \delta h^J = u^q g(w, \bar{w}) E_{\vec{\alpha}}\cdot h^J\, ,
\ee
%\be
%  [H_i, E_{\vec{\alpha}} \cdot h^J] = \vec{\alpha} \cdot h^J = \vec{\mu}_i \cdot h^i - \vec{\mu}_j \cdot h^j = (0, \cdots , u^{m_i} w^k, \cdots, - u^{m_j} w^k, \cdots, 0)
%\ee
%(always consider $\vec{\alpha}$ as $\vec{\mu}_i - \vec{\mu}_j$ )
where $g(w, \bar{w})$ is a regular function. The anti-holomorphic part has a similar expansion.

We should remove these unphysical modes in the partition function. Taking into account the expression of $h^J$ given by Eq.~\eqref{eq:factor-h}, one can show that to remove the contribution from the vortex solution is equivalent to multiplying the 1-loop determinant with
\be\label{eq:unphysicalNP}
  \prod_{\vec{\alpha}\in \Delta_G} \prod_{p \in \mathbb{Z}} \prod_{q = 0}^\infty \left(- p \varepsilon - \frac{1}{\ell} (q + \alpha (m)) + \vec{\alpha} \cdot \vec{\sigma}^N\right)^{\frac{1}{2}}\, .
\ee
To remove the contribution from the anti-vortex solution, we should also multiply the 1-loop determinant with
\be\label{eq:unphysicalSP}
  \prod_{\vec{\alpha}\in \Delta_G} \prod_{p \in \mathbb{Z}} \prod_{q = 0}^\infty \left(\frac{1}{p \varepsilon + \frac{1}{\ell} (q - \alpha(n)) + \vec{\alpha}\cdot \vec{\sigma}^S} \right)^{\frac{1}{2}}\, .
\ee
Up to a constant these two factors cancel each other. Therefore, in our case the unphysical modes from the off-diagonal part do not contribute to the 1-loop determinant, and the 1-loop determinant of the physical vector multiplet is just given by Eqs.~\eqref{eq:nonAbelVecPartFctProduct} and \eqref{eq:nonAbelVecPartFctProductReg}.

\subsection{Vortex Partition Functions}\label{sec:vortexPartFct}

Based on the results from the previous subsection, now we can write down the full partition functions for 4d $\mathcal{N} = 1$ gauge theories. For simplicity, in this subsection we consider two special examples. The generalization to more complicated theories is straightforward.

For the Abelian case, we consider a $U(1)$ gauge group with one vector multiplet and $N_F$ pairs of chiral and anti-chiral multiplets. The full partition function of this 4d $\mathcal{N}=1$ gauge theory with $(m, n) \in \mathbb{Z}_{\geq 0}^2$ is
\be
  Z_I = \sum_{m, n} Z_{\textrm{class}}\, Z^{\textrm{vec}}_{\textrm{1-loop}}\, Z^{\textrm{chiral}}_{\textrm{1-loop}}
\ee
with
\begin{align}
\begin{split}
  Z_{\textrm{class}} & = e^{- 2 \pi i (m - n) \tau_0}\, ,\\
  Z^{\textrm{vec}}_{\textrm{1-loop}} & = 1\, ,\\
  Z^{\textrm{chiral}}_{\textrm{1-loop}} & = \left[\prod_{\substack{J=1\\ J \neq I}}^{N_F} \left(x^{1 + n + m - 2 \ell \widetilde{m}_{IJ}} \right)^{\frac{n - m}{2}} \frac{\left(x^{2 + 2 m - 2 \ell \widetilde{m}_{IJ}};\, x^2 \right)_\infty}{\left(x^{-2 m + 2 \ell \widetilde{m}_{IJ}};\, x^2 \right)_\infty} \right] \cdot \left[\frac{1}{\prod_{q = 0}^{m - 1} \textrm{sinh} \left[\frac{i \pi}{\ell \varepsilon} (q-m) \right]} \right]\, ,
\end{split}
\end{align}
where $\widetilde{m}_{IJ} \equiv \widetilde{m}_I - \widetilde{m}_J$. The full partition function can be decomposed into the perturbative part and the vortex part:
\be
  Z_I = \left(Z_I \right)_{\textrm{pert}} \cdot \left(Z_I \right)_{\textrm{vortex}}\, ,
\ee
where
\be
  \left(Z_I \right)_{\textrm{pert}} = Z_I \Big|_{m = n = 0}\, .
\ee

For the non-Abelian case, we consider a $U(N)$ gauge group with one vector multiplet and $N_F$ pairs of chiral and anti-chiral multiplets in the fundamental representation. The full partition function of this 4d $\mathcal{N}=1$ gauge theory with $(m_i, n_i) \in \mathbb{Z}_{\geq 0}^2$ is
\be\label{eq:nonAbelvortexPartFct-1}
  Z_{I_i,\, \vec{\mu}_i} = \sum_{m_i, n_i} Z_{\textrm{class}}\, Z^{\textrm{vec}}_{\textrm{1-loop}}\, Z^{\textrm{chiral}}_{\textrm{1-loop}}
\ee
with
\begin{align}
\begin{split}\label{eq:nonAbelvortexPartFct-2}
  Z_{\textrm{class}} & = \prod_{i=1}^N e^{- 2 \pi i (m_i - n_i) \tau_0}\, ,\\
  Z^{\textrm{vec}}_{\textrm{1-loop}} & = \prod_{\substack{i, j = 1\\ i \neq j}}^N x^{- \frac{|(m_i - n_i) - (m_j - n_j)|}{2}} \left(1 - e^{\frac{2 \pi i}{\varepsilon} (\widetilde{m}_{I_i} - \widetilde{m}_{I_j})} x^{|(m_i - n_i) - (m_j - n_j)| - (m_i + n_i) + (m_j + n_j)} \right)\, ,\\
  Z^{\textrm{chiral}}_{\textrm{1-loop}} & = \left[\prod_{i=1}^N \prod_{\substack{J=1\\ J\neq I_i}}^{N_F} \left(x^{1 + n_i + m_i + 2 \ell (\widetilde{m}_J - \widetilde{m}_{I_i})} \right)^{\frac{\rho(n) - \rho(m)}{2}} \frac{\left(x^{2 + 2 m_i + 2 \ell (\widetilde{m}_J - \widetilde{m}_{I_i})};\, x^2 \right)_\infty}{\left(x^{-2 m_i - 2 \ell (\widetilde{m}_J - \widetilde{m}_{I_i})};\, x^2 \right)_\infty} \right] \\
  {} & \quad\cdot \left[\prod_{i=1}^N \frac{1}{\prod_{q = 0}^{m_i - 1} \textrm{sinh} \left[\frac{i \pi}{\ell \varepsilon} (q-m_i) \right]} \right]\, ,
\end{split}
\end{align}
where the fundamental weight vectors $\vec{\mu}_i$ ($i = 1,\, \cdots,\, r$) are chosen to be the standard basis vectors of $\mathbb{R}^r$, while for the root vector:
\be
  \vec{\alpha} = \vec{\mu}_i - \vec{\mu}_j \quad \textrm{with } 1 \leq i \neq j \leq r\, .
\ee
Similar to the Abelian case, for the non-Abelian case the full partition function can also be decomposed into the perturbative part and the vortex part:
\be
  Z_{I_i,\, \vec{\mu}_i} = \left(Z_{I_i,\, \vec{\mu}_i} \right)_{\textrm{pert}} \cdot \left(Z_{I_i,\, \vec{\mu}_i} \right)_{\textrm{vortex}}\, ,
\ee
where
\be
 \left(Z_{I_i,\, \vec{\mu}_i} \right)_{\textrm{pert}} = Z_{I_i,\, \vec{\mu}_i} \Big|_{m_i = n_i = 0}\, .
\ee

We would like to emphasize that the results of the partition functions discussed in this section are discussed in the Higgs branch. In principle, the localization can also be performed in the ``Coulomb branch'', and the result can be expressed as an integral over the Coulomb branch parameter $\sigma$, which should be equal to the one from the Higgs branch discussed in this paper. We leave this study for future research.

\subsection{Some Discussions}

As pointed out in \cite{GomisKomargodski-1}, the partition functions of $\mathcal{N} = 1$ theories on curved manifolds suffer from ambiguities, and the results depend on the renormalization scheme. However, for certain $\mathcal{N}=1$ theories such as class $\mathcal{S}_k$ theories \cite{Pomoni-1, Pomoni-2} and some mass-deformed $\mathcal{N}=1^*$ theories, the partition functions can still be unambiguously defined.

In this paper we have chosen a scheme implicitly during our calculations. Nevertheless, we can still compute within this scheme and extract some universal contributions independent of the scheme. What has been computed in this paper can be viewed as building blocks that can be used to construct arbitrary $\mathcal{N} = 1$ theories within the chosen scheme. The same treatment has been done in a previous work \cite{Minahan}.

%%%%%%%%%%%%%%%%%%%%%%%%%%%%%%%%%%%%%%%%%%%
%%%%%%%%%%%%%%%%%%%%%%%%%%%%%%%%%%%%%%%%%%%
\section{Seiberg Duality}\label{sec:SeibergDuality}
%%%%%%%%%%%%%%%%%%%%%%%%%%%%%%%%%%%%%%%%%%%
%%%%%%%%%%%%%%%%%%%%%%%%%%%%%%%%%%%%%%%%%%%

In this section, we use the partition functions obtained in the previous section to check an infrared duality between two theories. This is a duality first studied by Seiberg between a 4d $\mathcal{N}=1$ $SU(N)$ gauge theory with $N_F\, (> N)$ flavors and an $SU(N_F - N)$ gauge theory with $N_F$ flavors plus gauge singlets \cite{Seiberg}. Later, many similar dualities have been discovered in various dimensions by changing gauge groups and matter contents. For simplicity, we will study the $U(N)$ version of the original duality, which was first introduced in \cite{Kutasov} using brane constructions and later studied in \cite{Shifman:2007kd} in more details. The $U(N)$ Seiberg duality can be related to the $SU(N)$ Seiberg duality by gauging the $U(1)$ baryonic symmetry. Various Seiberg-like dualities have been extensively tested using the supersymmetric index \cite{Romelsberger:2007ec, DolanOsborn, Spiridonov-2}. Here we will apply our partition functions to test the $U(N)$ duality beyond the supersymmetric index.

Technically, our approach is a 4d uplift of the test of the 2d Seiberg-like duality for $U(N)$ gauge theories in \cite{S2-1}, in which the generalization to 2d $SU(N)$ gauge theories has also been discussed. In principle, we can also generalize the discussions in this section to 4d $SU(N)$ gauge theories. For simplicity, we will skip this step and save it for the future research.

The partition function of a $U(N)$ gauge theory with $N_F$ pairs of fundamental chiral and anti-chiral multiplets is given by Eq.~\eqref{eq:nonAbelvortexPartFct-1} and Eq.~\eqref{eq:nonAbelvortexPartFct-2}. We observe that the full partition function can also be factorized into the perturbative part, the vortex part and the anti-vortex part as follows:
\be
  Z_{I_i,\, \vec{\mu}_i} = Z_{\textrm{pert}} \cdot Z_{\textrm{v}} \cdot Z_{\textrm{av}}\, .
\ee
To test Seiberg duality, we should prove that each part on the right-hand side of the equation above is invariant under the duality transformations:
\be\label{eq:dualityTrafo}
   - \widetilde{m}_{T}^D = \Bigg\{
   \begin{array}{ll}
     \widetilde{m}_{T} + \frac{1}{4 \ell} & \textrm{ for fundamental in $SU(N_F)$}\, ,\\
     \widetilde{m}_{T} - \frac{1}{4 \ell} & \textrm{ for anti-fundamental in $SU(N_F)$}\, ,
   \end{array}
\ee
\be
  \vec{L} \leftrightarrow \vec{L}^D\, ,\nonumber
\ee
where we adopt a slightly different notation as follows. Because $N_F > N$, we can pick up $N$ out of $N_F$ flavors, which are previously labelled by the index $I_i$ $(i = 1,\, \cdots,\, N)$. From now on, let us denote a flavor $J$ within the $N$ selected ones by $J \in \vec{L}$, otherwise $J \notin \vec{L}$. We also use $\vec{L}^D$ to denote the complement set of $\vec{L}$. The shifts $\sim \frac{1}{\ell}$ in the dual masses $\widetilde{m}^D_T$ are due to \eqref{eq:RshiftMass} and the R-charge assignment of the duality, while $\widetilde{m}_T$ are unshifted masses in the original theory. If we require that the expressions in \eqref{eq:RshiftMass} and \eqref{eq:dualityTrafo} are equal for a special case $\frac{1}{\ell} = \hbar = - \varepsilon$ considered later in the paper, then we obtain $N_F / N = 2$, which strictly stays in the conformal window $3 N / 2 < N_F < 3 N$. Hence, the duality map \eqref{eq:dualityTrafo} is consistent.

Some attention should be paid for the gauge couplings. Since Seiberg duality is an electro-magnetic duality, there should be
\be\label{eq:g and gD}
  g_{YM} \propto \frac{1}{g_{YM}^D}\, .
\ee
The gauge couplings and the theta angles only appear in the complex coupling $\tau_0$ \eqref{eq:4d tau0} in the classical part of partition function. In principle, $\tau_0$ should be renormalized, and consequently the partition function depends on the energy scale. Since Seiberg duality is an IR duality, as a test of Seiberg duality we should compare the partition functions of two dual theories at the IR fixed point. We choose at the IR fixed point that
\be
  \textrm{at IR fixed point:}\quad g_{YM}^D = g_{YM}\, ,\quad \vartheta^D = \vartheta\, ,
\ee
which are not true for the dual theories in the UV. The precise relation between the couplings in the UV should be obtained using the RG equation, and is consistent with \eqref{eq:g and gD}.

Let us first consider the perturbative part of the partition function, $Z_\textrm{pert}$. In the new notation, it can be written as
\be\label{eq:vortexPartFctZpert}
  Z_{\textrm{pert}} = \left[\prod_{\substack{I, J \in \vec{L}\\ I \neq J}} \prod_{p \in \mathbb{Z}} (p \varepsilon + \widetilde{m}_I - \widetilde{m}_J) \right]\cdot \left[\prod_{T \in \vec{L}} \prod_{\substack{S=1\\ S \neq T}}^{N_F} \prod_{p \in \mathbb{Z}} \prod_{q \geq 0} \frac{p \varepsilon + (q+1) \frac{1}{\ell} + \widetilde{m}_S - \widetilde{m}_T}{p \varepsilon + q \frac{1}{\ell} - \widetilde{m}_S + \widetilde{m}_T} \right]\, .
\ee
The second product of the expression above can be further factorized into two products due to
\be
  \prod_{T \in \vec{L}} \prod_{\substack{S=1\\ S \neq T}}^{N_F} = \prod_{\substack{S, T \in \vec{L}\\ S \neq T}} \prod_{\substack{T \in \vec{L}\\ S \notin \vec{L}}}\, .
\ee
The product with $S, T \in \vec{L}$ ($S \neq T$) exactly cancels the first product in Eq.~\eqref{eq:vortexPartFctZpert}. Hence, $Z_{\textrm{pert}}$ is equal to the product with $T \in \vec{L}$, $S \notin \vec{L}$, i.e.,
\be\label{eq:SeibergDualityZpert}
  Z_{\textrm{pert}} = \prod_{\substack{T \in \vec{L}\\ S \notin \vec{L}}} \prod_{p \in \mathbb{Z}} \prod_{q \geq 0} \frac{p \varepsilon + (q+1) \frac{1}{\ell} + \widetilde{m}_S - \widetilde{m}_T}{p \varepsilon + q \frac{1}{\ell} - \widetilde{m}_S + \widetilde{m}_T}\, .
\ee
This expression can be regularized using the $\textrm{sinh}$ function and the $q$-Pochhammer symbol as before. After the transformations \eqref{eq:dualityTrafo}, it becomes the combination of the W-boson and the fundamental chiral multiplets in the dual theory. Moreover, the dual theory should also receive the contribution from the gauge singlets
\begin{align}
  Z^{\textrm{singlet}} & = \prod_{S, T=1}^{N_F} \prod_{p \in \mathbb{Z}} \prod_{q \geq 0} \frac{p \varepsilon + (q + 1) \frac{1}{\ell} + \widetilde{m}_S^D - \widetilde{m}_T^D}{p \varepsilon + q \frac{1}{\ell} - \widetilde{m}_S^D + \widetilde{m}_T^D} \nonumber\\
  {} & = \prod_{S, T=1}^{N_F} \prod_{p \in \mathbb{Z}} \prod_{q \geq 0} \frac{p \varepsilon + (q + 1) \frac{1}{\ell} - (\widetilde{m}_S + \frac{1}{4 \ell}) + (\widetilde{m}_T - \frac{1}{4 \ell})}{p \varepsilon + q \frac{1}{\ell} + (\widetilde{m}_S + \frac{1}{4 \ell}) - (\widetilde{m}_T - \frac{1}{4 \ell})} \, ,
\end{align}
which contributes a trivial factor to the partition function of the dual theory. Hence, the perturbative partition function of the dual theory is exactly equal to \eqref{eq:SeibergDualityZpert}.

Now let us consider the vortex and the anti-vortex parts of the partition function, which are given by
\begin{align}
  Z_{\textrm{v}} & = \sum_{\vec{m}} e^{2 \pi i \tau_0 |\vec{m}|} \left[\prod_{\vec{\alpha} \in \Delta_G} \prod_{p \in \mathbb{Z}} \left(\frac{p \varepsilon + \vec{\alpha} \cdot \vec{\sigma}^N}{p \varepsilon + \vec{\alpha} \cdot \vec{\sigma}^N \Big|_{m_i = 0}} \right)^{\frac{1}{2}} \right] \cdot \left[\prod_{i=1}^N \prod_{p \in \mathbb{Z}} \prod_{q = 0}^{m_i - 1} \frac{1}{p \varepsilon + \frac{1}{\ell} (q - m_i)} \right] \nonumber\\
  {} & \quad \cdot \left[\prod_{i=1}^N \prod_{\substack{J=1\\ J \neq I_i}}^{N_F} \prod_{p \in \mathbb{Z}} \prod_{q \geq 0} \frac{p \varepsilon + \frac{1}{\ell} q - \widetilde{m}_J + \widetilde{m}_{I_i}}{p \varepsilon + \frac{1}{\ell} q - \widetilde{m}_J + \widetilde{m}_{I_i} - \frac{1}{\ell} m_i}\right]\, ,\\
  Z_{\textrm{av}} & = \sum_{\vec{n}} e^{- 2 \pi i \tau_0 |\vec{n}|} \left[\prod_{\vec{\alpha} \in \Delta_G} \prod_{p \in \mathbb{Z}} \left(\frac{p \varepsilon + \vec{\alpha} \cdot \vec{\sigma}^S}{p \varepsilon + \vec{\alpha} \cdot \vec{\sigma}^S \Big|_{n_i = 0}} \right)^{\frac{1}{2}} \right] \cdot \left[\prod_{i=1}^N \prod_{p \in \mathbb{Z}} \prod_{q = 0}^{\infty} \frac{p \varepsilon + \frac{1}{\ell} (q+1) + \frac{1}{\ell} n_i}{p \varepsilon + \frac{1}{\ell} (q + 1)}\right] \nonumber\\
  {} & \quad \cdot \left[\prod_{i=1}^N \prod_{\substack{J=1\\ J \neq I_i}}^{N_F} \prod_{p \in \mathbb{Z}} \prod_{q \geq 0} \frac{p \varepsilon + \frac{1}{\ell} (q+1) + \widetilde{m}_J - \widetilde{m}_{I_i} + \frac{1}{\ell} n_i}{p \varepsilon + \frac{1}{\ell} (q + 1) + \widetilde{m}_J - \widetilde{m}_{I_i}}\right] \, .
\end{align}
After a few steps, $Z_\textrm{v}$ with $m_i > 0$ can be simplified as
\be
  Z_\textrm{v} = \sum_{m = 0}^\infty e^{2 \pi i \tau_0 m} Z_\textrm{v}^{m}
\ee
with $m \equiv |\vec{m}|$ and
\be
  Z_\textrm{v}^m = \sum_{\substack{\vec{m} \in \mathbb{Z}_{\geq 0}^N\\ |\vec{m}| = m}} \frac{\prod_{\substack{i, j = 1 \\ i \neq j}}^N \prod_{p \in \mathbb{Z}} \left(\frac{p \varepsilon + \widetilde{m}_{I_i} - \frac{1}{\ell} m_i - \widetilde{m}_{I_j} + \frac{1}{\ell} m_j}{p \varepsilon + \widetilde{m}_{I_i} - \widetilde{m}_{I_j}} \right)^{\frac{1}{2}}}{\prod_{i = 1}^N \prod_{J = 1}^{N_F} \prod_{p \in \mathbb{Z}} \left(\frac{1}{\ell}\right)^{m_i} \left(p \ell \varepsilon - \ell \widetilde{m}_J + \ell \widetilde{m}_{I_i} - m_i \right)_{m_i}}\, ,\label{eq:Zvm-temp}
\ee
where we used the Pochhammer symbol, which is defined as
\be
  (x)_m \equiv \prod_{k=0}^{m-1} (x + k)\, . 
\ee
Similarly, $Z_{\textrm{av}}$ with $n_i > 0$ can be rewritten as
\be
  Z_{\textrm{av}} = \sum_{n = 0}^\infty e^{- 2 \pi i \tau_0 n} Z_{\textrm{av}}^{n}
\ee
with $n \equiv |\vec{n}|$ and
\be
  Z_{\textrm{av}}^n = \sum_{\substack{\vec{n} \in \mathbb{Z}_{\geq 0}^N\\ |\vec{n}| = n}} \frac{\prod_{\substack{i, j = 1 \\ i \neq j}}^N \prod_{p \in \mathbb{Z}} \left(\frac{p \varepsilon + \widetilde{m}_{I_i} - \frac{1}{\ell} n_i - \widetilde{m}_{I_j} + \frac{1}{\ell} n_j}{p \varepsilon + \widetilde{m}_{I_i} - \widetilde{m}_{I_j}} \right)^{\frac{1}{2}}}{\prod_{i = 1}^N \prod_{J = 1}^{N_F} \prod_{p \in \mathbb{Z}} \left(- \frac{1}{\ell}\right)^{n_i} \left(p \ell \varepsilon - \ell \widetilde{m}_J + \ell \widetilde{m}_{I_i} - n_i \right)_{n_i}}\, .
\ee
In the following let us focus on the vortex part of the partition function, i.e. $Z_{\textrm{v}}$. The anti-vortex part $Z_{\textrm{av}}$ can be treated similarly.

We observe that the expression \eqref{eq:Zvm-temp} is very similar to the vortex partition function on $S^2$ studied in Refs.~\cite{S2-1, S2-2}, and the only difference is that in our case there is an extra product over $p \in \mathbb{Z}$. In principle, we can regularize this infinite product and test the 4d Seiberg duality directly, however, it is technically easier to keep the unregularized infinite product, so that the problem of testing the 4d Seiberg duality becomes a problem of testing the 2d Seiberg-like duality. Hence, we can apply the same trick as in Ref.~\cite{S2-1} to test Seiberg duality in the following.

Applying the following identity of the Pochhammer symbol
\be
  (a - n)_m (- a - m)_n = \left(1 + \frac{m - n}{a} \right)^{-1} (a + 1)_m (- a + 1)_n\, ,
\ee
in the new notation we can rewrite the numerator of the summand in Eq.~\eqref{eq:Zvm-temp} as
%\be
%  \prod_{\substack{i, j = 1 \\ i \neq j}}^N \prod_{p \in \mathbb{Z}} \left(\frac{(-1)^{m_i + m_j} (- p \varepsilon - M_{I_i}^{I_j} - m_j)_{m_j} (p \varepsilon + M_{I_i}^{I_j} - m_i)_{m_i}}{(p \varepsilon + M_{I_i}^{I_j} - m_i)_{m_j} (- p \varepsilon - M_{I_i}^{I_j} - m_j)_{m_i}} \right)^{\frac{1}{2}}
%\ee
\be\label{eq:ZvmNum}
  (-1)^{(N - 1) m} \prod_{\substack{S, T \in \vec{L} \\ S \neq T}} \prod_{p \in \mathbb{Z}} \frac{ (p \ell \varepsilon + \ell \widetilde{m}_{ST} - m_S)_{m_S}}{(p \ell \varepsilon + \ell \widetilde{m}_{ST} - m_S)_{m_T}}\, ,
\ee
and the denominator as
\be\label{eq:ZvmDenom}
  \prod_{S \in \vec{L}} \left[\prod_{T \in \vec{L}} \prod_{p \in \mathbb{Z}} (p \ell \varepsilon + \ell \widetilde{m}_{ST} - m_S)_{m_S} \right] \cdot \left[\prod_{T \notin \vec{L}} \prod_{p \in \mathbb{Z}} (p \ell \varepsilon + \ell \widetilde{m}_{ST} - m_S)_{m_S} \right]\, .
\ee
Using Eqs.~\eqref{eq:ZvmNum} and \eqref{eq:ZvmDenom}, we can express Eq.~\eqref{eq:Zvm-temp} as
\begin{align}
  Z_\textrm{v}^{m} & = \sum_{\substack{\vec{m} \in \mathbb{Z}_{\geq 0}^N\\ |\vec{m}| = m}} (-1)^{(N - 1) m} \prod_{S \in \vec{L}} \frac{1}{\left[\prod_{T \in \vec{L}} \prod_{p \in \mathbb{Z}} (p \ell \varepsilon + \ell \widetilde{m}_{ST} - m_S)_{m_T} \right]} \nonumber\\
  {} & \qquad\qquad\qquad\qquad\qquad \cdot \frac{1}{\left[\prod_{T \notin \vec{L}} \prod_{p \in \mathbb{Z}} (p \ell \varepsilon + \ell \widetilde{m}_{ST} - m_S)_{m_S} \right]}\, ,
\end{align}
which can be regularized as
\begin{align}
  Z_\textrm{v}^{m} & = \sum_{\substack{\vec{m} \in \mathbb{Z}_{\geq 0}^N\\ |\vec{m}| = m}} (-1)^{(N - 1) m}\, b^{A N_F m} \prod_{S \in \vec{L}} \frac{1}{\left[\prod_{T \in \vec{L}} \textrm{sin} (\ell b \widetilde{m}_{ST} - b m_S)_{(b,\, m_T)} \right]} \nonumber\\
  {} & \qquad\qquad\qquad\qquad\qquad\qquad \cdot \frac{1}{\left[\prod_{T \notin \vec{L}} \textrm{sin} (\ell b \widetilde{m}_{ST} - b m_S)_{(b,\, m_S)} \right]}\, ,\label{eq:Zvm}
\end{align}
where $b \equiv \frac{\pi}{\ell \varepsilon}$, and $A$ is the regularized value of $\sum_{p \in \mathbb{Z}} 1$. We have also introduced the modified sine-Pochhammer symbol:
\be
  \textrm{sin} (x)_{(a,\, m)} \equiv \prod_{k=0}^{m-1} \textrm{sin} (x + a\, k)\, .
\ee
Because $\frac{1}{\textrm{sin} (x)}$ has a simple pole at $x = 0$, we can rewrite the expression \eqref{eq:Zvm} into a contour integral:
\begin{align}
  Z_\textrm{v}^m & = \frac{(-1)^{(N - 1) m}\, b^{A N_F m} }{m! \left[\textrm{sin} (b) \right]^m} \int_C \left[\prod_{j=1}^m \frac{d \varphi_j}{2 \pi i} \right] \left[\prod_{i < j}^m \frac{\left(\textrm{sin} (\ell b \varphi_i - \ell b \varphi_j) \right)^2}{\textrm{sin} (\ell b \varphi_i - \ell b \varphi_j + b) \cdot \textrm{sin} (\ell b \varphi_i - \ell b \varphi_j - b)} \right] \nonumber\\
  {} & \quad\quad \cdot \left[\prod_{j=1}^m \frac{1}{\prod_{S \in \vec{L}} \textrm{sin} (\ell b \varphi_j - \ell b \widetilde{m}_S)\cdot \prod_{T \notin \vec{L}} \textrm{sin} (\ell b \widetilde{m}_T - \ell b \varphi_j - b)} \right]\, .\label{eq:ZvmContourIntegral}
\end{align}
where the contour $C$ is chosen in the following way. First, we assume that the parameter $\frac{1}{\ell}$, the masses $\widetilde{m}_{S \in \vec{L}}$ and $\widetilde{m}_{T \notin \vec{L}}$ have small positive imaginary parts, and the imaginary part of $\frac{1}{\ell}$ is larger than the ones for the masses. Consequently, we find in the upper half-plane the following poles:
\be\label{eq:Poles}
  \widetilde{m}_S\, ,\, \widetilde{m}_S + \frac{1}{\ell}\, ,\, \cdots\, ,\, \widetilde{m}_S + (m_S - 1) \frac{1}{\ell}\quad (S \in \vec{L})\, .
\ee
Second, since $\frac{1}{\textrm{sin} (x)}$ has poles at $x = l \pi$ with $l \in \mathbb{Z}$, 
\be
  \widetilde{m}_S + l \pi\, ,\, \widetilde{m}_S + \frac{1}{\ell} + l \pi\, ,\, \cdots\, ,\, \widetilde{m}_S + (m_S - 1) \frac{1}{\ell} + l \pi \quad (S \in \vec{L})
\ee
are also poles in the upper 
 plane. We take the contour $C$ along the real axis and closed in the upper plane, but then we deform it in such a way that it encloses only the poles with $l = 0$, i.e. the poles given by Eq.~\eqref{eq:Poles}.

We observe that the contour integral \eqref{eq:ZvmContourIntegral} can be viewed as a trigonometric version of the contour integral discussed in Ref.~\cite{S2-1}. Similar to that case, the contour integral \eqref{eq:ZvmContourIntegral} has no poles at infinity when $N_F > 1$. We define the integration variables for the dual theory as
\be
  \varphi_j^D = - \varphi_j - \frac{1}{\ell}\, ,
\ee
and the other parameters in the dual theory are still related to the ones in the original theory through the duality transformations \eqref{eq:dualityTrafo}. In terms of the dual integration variables and parameters, the contour integral \eqref{eq:ZvmContourIntegral} can be rewritten as
\begin{align}
  Z_\textrm{v}^m & = \frac{(-1)^{(N - 1) m}\, b^{A N_F m} }{m! \left[\textrm{sin} (b) \right]^m} \int_{C^D} \left[\prod_{j=1}^m \frac{d \varphi_j^D}{2 \pi i} \right] \left[\prod_{i < j}^m \frac{\left(\textrm{sin} (\ell b \varphi_i^D - \ell b \varphi_j^D) \right)^2}{\textrm{sin} (\ell b \varphi_i^D - \ell b \varphi_j^D + b) \cdot \textrm{sin} (\ell b \varphi_i^D - \ell b \varphi_j^D - b)} \right] \nonumber\\
  {} & \quad\quad \cdot \left[\prod_{j=1}^m \frac{1}{\prod_{T \notin \vec{L}} \textrm{sin} (\ell b \varphi_j^D - \ell b \widetilde{m}_T^D)\cdot \prod_{S \in \vec{L}} \textrm{sin} (\ell b \widetilde{m}_S^D - \ell b \varphi_j^D - b)} \right]\, ,\label{eq:ZvmDualContourIntegral}
\end{align}
where the contour $C^D$ is now chosen to be along the real axis and closed in the upper half-plane, and then deformed in such a way that it picks up only the poles at
\be\label{eq:DualPoles}
  \widetilde{m}_T^D\, ,\, \widetilde{m}_T^D + \frac{1}{\ell}\, ,\, \cdots\, ,\, \widetilde{m}_T^D + (m_T - 1) \frac{1}{\ell}\quad (T \in \vec{L}^D)\, .
\ee
Now the expression \eqref{eq:ZvmDualContourIntegral} describes the vortex part of the partition function of an $\mathcal{N}=1$ gauge theory on $S^2 \times \mathbb{R}^2_\varepsilon$ with $N_F$ flavors and a gauge group $U(N_F - N)$.

Combining the perturbative part, the vortex part and the anti-vortex part of the partition function on $S^2 \times \mathbb{R}^2_\varepsilon$, we conclude that at the IR fixed point
\be
  Z_{U(N)} (g_{YM},\, \vartheta,\, \widetilde{m}_i) = Z_{U(N_F - N)} (g_{YM}^D,\, \vartheta^D\, , \widetilde{m}_i^D)\, .
\ee

The original 4d Seiberg duality has special unitary gauge groups instead of the unitary gauge groups. In principle, we can apply the approaches in Refs.~\cite{Kutasov, S2-1} to obtain the results for the special unitary gauge groups from our results. It is more convenient to use the integral expression of the partition function obtained in the ``Coulomb branch''. Also, using the $\mathcal{N}=1$ partition functions on $S^2 \times \mathbb{R}^2_\varepsilon$ we can also consider other dualities. We would like to leave these for future work.

%%%%%%%%%%%%%%%%%%%%%%%%%%%%%%%%%%%%%%%%%%%
%%%%%%%%%%%%%%%%%%%%%%%%%%%%%%%%%%%%%%%%%%%
\section{Comparison with Nekrasov Partition Functions}\label{sec:NekPartFct}
%%%%%%%%%%%%%%%%%%%%%%%%%%%%%%%%%%%%%%%%%%%
%%%%%%%%%%%%%%%%%%%%%%%%%%%%%%%%%%%%%%%%%%%

Because $S^2 \times \mathbb{R}^2_\varepsilon$ approaches the 4d $\Omega$-background near the poles of $S^2$, we expect that our results can be interpreted as a product of two partition functions on the $\Omega$-background, each from one patch of $S^2 \times \mathbb{R}^2_\varepsilon$. This picture is consistent with previous works in the literature \cite{Pestun, HamaHosomichi, S2S2}. Because $\mathcal{N}=2$ theories can be formulated in terms of $\mathcal{N}=1$ multiplets, from these elementary building blocks we should be able to construct the $\mathcal{N}=2$ partition functions on the $\Omega$-background, also known as the Nekrasov partition function \cite{Nekrasov, NekrasovMarshakov, NekrasovOkounkov}.

There is a technical subtlety. Due to the SUSY algebra given by Eqs.~\eqref{eq:FullSUSYalg} and \eqref{eq:KillingVecS2R2epsilon}, all the eigenmodes $e^{i p \psi}$ to the operator $\partial_\psi$ on $\mathbb{R}^2_\varepsilon$ should contribute to the 1-loop determinant, which corresponds to the product over $p \in \mathbb{Z}$ in Eqs.~\eqref{eq:nonAbelChiralPartFctProduct} and \eqref{eq:nonAbelChiralPartFctProductNF}. However, to compare with the Nekrasov partition functions, we should focus on the vortex sector without anti-vortices on $\mathbb{R}^2_\varepsilon$. Hence, the products should now be taken over $p \geq 0$ instead of $p \in \mathbb{Z}$ due to the expansion in the basis of holomorphic functions $\{1,\, w,\, w^2,\, \cdots \}$.

After taking care of this subtlety, we can propose the $\mathcal{N}=1$ partition functions on the $\Omega$-background, and then compare them with the $\mathcal{N}=2$ Nekrasov partition functions. To do so, we need to identify 
the parameters $(1 / \ell,\, \varepsilon)$ from $S^2 \times \mathbb{R}^2_\varepsilon$ with the equivariant parameters
$(\varepsilon_1,\, \varepsilon_2)$ of the $\Omega$-background, i.e.,
\be
  \varepsilon_1 \equiv \pm \frac{1}{\ell}\, ,\quad \varepsilon_2 \equiv \varepsilon\, ,
\ee
and also the twisted masses with the Coulomb branch parameters:
\be
  a_i \equiv - \widetilde{m}_{I_i}\, .
\ee
Note that in a general $\Omega$-background, $(\varepsilon_1,\, \varepsilon_2)$ can be complex. Here we identify $(\varepsilon_1,\, \varepsilon_2)$ with the real parameters in our theory, and conjecture that the results can be analytically continued to complex parameters.

With these identifications, let us first consider the $\mathcal{N}=2$ vector multiplet, which can be decomposed into an $\mathcal{N}=1$ vector multiplet and an adjoint chiral multiplet. From Eq.~\eqref{eq:nonAbelChiralPartFctProduct} and the expressions \eqref{eq:nonAbelVecPartFctProduct} \eqref{eq:unphysicalNP} with $\prod_{p \in \mathbb{Z}}$ replaced by $\prod_{p \geq 0}$, we first propose the following expressions for the partition functions of the $\mathcal{N}=1$ adjoint chiral and the vector multiplets on the $\Omega$-background:
\be
  Z_{\textrm{$\Omega$}}^{\textrm{adj}} = \prod_{i \neq j} \prod_{p \geq 0} \prod_{q \geq 0} \Big[(q+1) \varepsilon_1 + p \varepsilon_2 + \sigma_i - \sigma_j \Big]^{\frac{1}{2}}\, ,
\ee
\be
  Z_{\textrm{$\Omega$}}^{\textrm{vec}} = \left(\prod_{i \neq j} \prod_{p \geq 0} (p \varepsilon_2 + \sigma_i - \sigma_j)^{\frac{1}{2}} \right) \cdot \left(\prod_{i \neq j} \prod_{p \geq 0} \prod_{q \geq 0} (q \varepsilon_1 + p \varepsilon_2 + \sigma_i - \sigma_j)^{\frac{1}{2}} \right)\, .
\ee
Therefore,
\be\label{eq:NekPartFctVec-temp}
  Z_{\textrm{$\Omega$}}^{\textrm{adj}} \cdot Z_{\textrm{$\Omega$}}^{\textrm{vec}} = \prod_{i \neq j} \prod_{p \geq 0} \prod_{q \geq 0} \Big[q \varepsilon_1 + p \varepsilon_2 + \sigma_i - \sigma_j \Big]\, ,
\ee
where $\sigma_i$ ($i = 1,\, \cdots,\, r$) include the contributions from vortices on $\mathbb{R}^2_\varepsilon$:
\be
  \sigma_i = - \widetilde{m}_{I_i} + k_i\, \varepsilon\, .
\ee
Moreover, we make a partition of the vortex number $k_i$:
\be\label{eq:k-partition}
  k_i^1 \geq k_i^2 \geq \cdots \geq k_i^{l (k_i)} \geq k_i^{l (k_i) + 1} = k_i^{l (k_i) + 2} = \cdots = 0\, ,
\ee
which form $r$ Young diagrams for $i = 1,\, \cdots,\, r$, and $l (k_i)$ denotes the length of the $i$-th Young diagram. We treat $k_i^q$ ($q \geq 1$) with different $q$'s as solutions in different sectors, and
\be
  k_i = \sum_{q=1}^\infty k_i^q\, ,\quad k_i^0 = 0\, .
\ee
Hence, in the $q$-sector we have
\be\label{eq:sigma-q-sector}
  \sigma_i = a_i + k_i^q \, \varepsilon_2\, .
\ee
% Because $k_i$ labels the vortex in $\mathbb{R}^2_\varepsilon$, and $q$ labels the winding modes in $S^2$, we need to use $q$ as internal index for $k_i$.
Consequently, Eq.~\eqref{eq:NekPartFctVec-temp} becomes
\be\label{eq:NekPartFctVec}
  Z_{\textrm{$\Omega$}}^{\textrm{adj}} \cdot Z_{\textrm{$\Omega$}}^{\textrm{vec}} = \prod_{\substack{i, j = 1\\ (i, p) \neq (j, q)}}^r \prod_{p \geq 0} \prod_{q \geq 0} \Big[q \varepsilon_1 + p \varepsilon_2 + a_i - a_j + k_i^{q+1}\, \varepsilon_2 - k_j^{p+1}\, \varepsilon_2 \Big]\, .
\ee
When $- \varepsilon_1 = \hbar = \varepsilon_2$, by setting the vortex numbers to zero and applying a regularization similar to Ref.~\cite{Nekrasov} using the Barnes double zeta function $\zeta_2 (s; x | \varepsilon_1, \varepsilon_2)$ defined in \eqref{eq:defineBarnesDoubleZeta}, we can also extract the perturbative part of the partition function from Eq.~\eqref{eq:NekPartFctVec}:
% See Eq.(3.27) in Nekrasov's instanton counting paper.
\be
  \textrm{exp} \left(\sum_{i, j = 1}^r \gamma_\hbar (a_i - a_j) \right)\, ,
\ee
% whether the expression above is in the numerator or denominator, depends on the signs of $\varepsilon_{1, 2}$.
where $\gamma_\hbar (x) \equiv \frac{d}{ds} \big|_{s=0} \zeta_2 (s; x | -\hbar, \hbar)$. This is the same as the perturbative partition function for the $\mathcal{N}=2$ vector multiplet \cite{Nekrasov, NekrasovMarshakov, NekrasovOkounkov}. \footnote{We notice that there is a little discrepancy among literatures. Our results are the same as the ones obtained in Ref.~\cite{NekrasovOkounkov}, which differs from the ones used in Ref.~\cite{AGT} by some shifts of $\varepsilon_{1, 2}$.} After subtracting the perturbative contribution from the expression above, what remains is
\begin{align}
  {} & \prod_{\substack{i, j = 1\\ (i, p) \neq (j, q)}}^r \prod_{p \geq 0} \prod_{q \geq 0} \frac{a_i - a_j + \hbar (k_i^{q+1} - k_j^{p+1} + p - q)}{a_i - a_j + \hbar (p - q)} \nonumber\\
  = \, & \prod_{\substack{i, j = 1\\ (i, p) \neq (j, q)}}^r \prod_{p \geq 1} \prod_{q \geq 1} \frac{a_i - a_j + \hbar (k_i^q - k_j^p + p - q)}{a_i - a_j + \hbar (p - q)}\, ,
\end{align}
which is exactly the same as the instanton partition function for the $\mathcal{N}=2$ vector multiplet \cite{Nekrasov, NekrasovMarshakov, NekrasovOkounkov}.

Let us move on to discuss the hypermultiplet, which can be decomposed into a pair of chiral and anti-chiral multiplets. To be precise, the partition functions of the chiral multiplet obtained in Section~\ref{sec:localization} consists of the contributions from both the chiral and the anti-chiral multiplets. Hence, we expect that the partition function obtained before for the (anti-)chiral multiplet should coincide with the one for the hypermultiplet.

From Eq.~\eqref{eq:nonAbelChiralPartFctProduct} we read off the partition function on the $\Omega$-background for the fundamental (anti-)chiral multiplet:
\be\label{eq:matchNekPartFctHypFull}
  Z_{\textrm{$\Omega$}}^{\textrm{fun (anti-)chiral}} = \prod_{i=1}^r \prod_{p \geq 0} \prod_{q \geq 0} \frac{1}{- q \varepsilon_1 - p \varepsilon_2 + \widetilde{m}_J + \sigma_i}\, ,
\ee
where now the constants $\sigma_i$ are given by
\be
  \sigma_i = - \widetilde{m}_{I_i} + \frac{1}{\ell} m_i
\ee
with the partition of the vortex number $m_i$:
\be\label{eq:m-partition}
  m_i^1 \geq m_i^2 \geq \cdots \geq m_i^{l (m_i)} \geq m_i^{l (m_i) + 1} = m_i^{l (m_i) + 2} = \cdots = 0\, ,
\ee
and
\be
  m_i = \sum_{p=1}^\infty m_i^p\, ,\quad m_i^0 = 0\, .
\ee
Hence, in the $p$-sector we have
\be\label{eq:sigma-p-sector}
  \sigma_i = a_i - m_i^p\, \varepsilon_1\, .
\ee
% Because $m_i$ labels the vortex in $S^2$, and $p$ labels the winding modes in $\mathbb{R}^2$, we need to use $p$ as internal index for $m_i$.
We perform a regularization for the infinite product \eqref{eq:matchNekPartFctHypFull} as follows:
\be
  Z_{\textrm{$\Omega$}}^{\textrm{fun (anti-)chiral}} = \prod_{i=1}^r \prod_{p \geq 1} \prod_{q \geq 0} \frac{1}{\hbar (q - p + 1) + \widetilde{m}_J + a_i + m_i^p \hbar}\, , \label{eq:regular}
\ee
where we have applied $- \varepsilon_1 = \hbar = \varepsilon_2$, and used the expression \eqref{eq:sigma-p-sector} for the constants $\sigma_i$ in the $p$-sector of the partition \eqref{eq:m-partition}. Moreover, by setting the vortex numbers to zero and applying a regularization similar to Ref.~\cite{Nekrasov}, we can also extract the perturbative part of the partition function from Eq.~\eqref{eq:matchNekPartFctHypFull}:
\be
  \textrm{exp} \left(- \sum_{i=1}^r \gamma_\hbar (\widetilde{m}_J + a_i) \right)\, ,
\ee
which is consistent with the result in Ref.~\cite{NekrasovOkounkov}, although there is a sign typo there. After subtracting the perturbative contribution from the expression \eqref{eq:regular}, what remains is
\begin{align}
  {} & \prod_{i=1}^r \prod_{p \geq 1} \prod_{q \geq 0} \frac{\hbar (q - p + 1) + \widetilde{m}_J + a_i}{\hbar (q - p + 1) + \widetilde{m}_J + a_i + m_i^p \hbar} \nonumber\\
  = \, & \prod_{i=1}^r \prod_{p \geq 1} \frac{\Gamma \left(\frac{\widetilde{m}_J + a_i}{\hbar} + m_i^p + 1 - p\right)}{\Gamma \left(\frac{\widetilde{m}_J + a_i}{\hbar} + 1 - p\right)}\, ,\label{eq:InstantonForHyper}
\end{align}
where we have performed a partial regularization using
\be
  \prod_{q \geq 0} \frac{1}{q + x} \simeq \frac{\Gamma (x)}{\sqrt{2 \pi}}\, .
\ee
The expression \eqref{eq:InstantonForHyper} is exactly the same as the instanton partition function for the hypermultiplet in the fundamental representation \cite{NekrasovOkounkov}.

To summarize, we have made some conjectures for the $\mathcal{N}=1$ partition functions on the $\Omega$-background based on the results from previous sections on $S^2 \times \mathbb{R}^2_\varepsilon$, and then checked their relation with the $\mathcal{N}=2$ Nekrasov partition functions. These conjectures should be tested using a direct $\mathcal{N}=1$ instanton counting approach in the future.

%%%%%%%%%%%%%%%%%%%%%%%%%%%%%%%%%%%%%%%%%%%
%%%%%%%%%%%%%%%%%%%%%%%%%%%%%%%%%%%%%%%%%%%
\section{Discussions}\label{sec:discussion}
%%%%%%%%%%%%%%%%%%%%%%%%%%%%%%%%%%%%%%%%%%%
%%%%%%%%%%%%%%%%%%%%%%%%%%%%%%%%%%%%%%%%%%%

In this paper we have discussed 4d $\mathcal{N}=1$ gauge theories on $S^2 \times \mathbb{R}^2_\varepsilon$, and computed their partition functions using supersymmetric localization. We provided a non-index partition function test for Seiberg duality. Moreover, under some assumptions we have compared our conjectured $\mathcal{N}=1$ partition functions on the $\Omega$-background with the $\mathcal{N}=2$ Nekrasov partition functions.

We have shown that under reasonable assumptions, the BPS equations admit vortex solutions at the poles of $S^2$ and at the origin of $\mathbb{R}^2_\varepsilon$. We have computed the 1-loop determinants of fluctuations around these configurations. However, the instanton-vortex equations suggest the possibility that point-like instanton solutions may exist. It will be important to further study whether the equations admit extra solutions other than the ones we found, and their contributions to the partition function.

There are some natural generalizations and applications of our work. 
\begin{enumerate}
\item In this paper, we performed the Higgs branch localization and the final result is written as a sum over vortex configurations. We should be able to perform the ``Coulomb branch'' localization to obtain an integral expression of the partition function.

\item The generalization to other backgrounds (e.g. $D^2 \times \mathbb{R}^2_\varepsilon$, $S^2 \times S^2$) and other matter contents, for instance the anti-fundamental chiral multiplet and the (anti-)symmetric tensor field, should be straightforward. With these new ingredients we can rigorously test more conjectured dualities.

\item We hope that our work can shed light on the still-mysterious rigid $\mathcal{N}=1$ theories on curved space. For example, it gives us hints to previously intractable cases (e.g. $\mathcal{N}=1$ localization on $S^2 \times S^2$), and also helps us better understand some notoriously hard cases  (e.g. $\mathcal{N}=1$ localization on $S^4$) and conceptual issues like the ambiguity of the sphere partition function \cite{GomisKomargodski-1}. The study can also provide us with more rigorous tests of the AdS/CFT correspondence in the same spirit of \cite{Elvang-1, Elvang-2}.

\item More mathematically, it would be nice to localize $\mathcal{N}=1$ gauge theories on generic toric 4-manifolds, just like for $\mathcal{N}=2$ gauge theories \cite{NekrasovICMP, BonelliP2}.

\item We can compute the partition functions of $\mathcal{N}=1$ superconformal field theories and study their anomalies \cite{GomisKomargodski-2}, which complements previous studies using superconformal indices \cite{ArashLiu-1, ArashLiu-2}.

\item Just as the $\mathcal{N}=2$ prepotential can be computed from the Nekrasov partition function, we expect the free energy of our $\mathcal{N}=1$ partition function to be related to the effective superpotential. This should be compared with other approaches such as \cite{Tanzini, Ferrari}.

\item Some $\mathcal{N}=1$ theories flow to known $\mathcal{N}=2$ theories under renormalization group flows. The superconformal indices of the Minahan-Nemeschansky theories \cite{GRW-1, MS-4} and the Argyres-Douglas theories have been computed in this way \cite{MS-1, MS-2, MS-3}. By following such RG flows, we can also compute the non-index partition functions of non-Lagrangian theories.

\item There have been recent progress in obtaining 4d $\mathcal{N}=1$ gauge theories from 6d $\mathcal{N}= (1,0)$ theory with fluxes \cite{N=1from6d-1, N=1from6d-2}. We believe that these new ideas can provide us with a more direct 4d $\mathcal{N}=1$ instanton counting. Eventually, we should be able to study the 4d $\mathcal{N}=1$ gauge theories directly on the $\Omega$-background based on the recent progress.

\item Our results, in particular the relations with the Nekrasov partition functions, can facilitate recent studies on the conjectured $\mathcal{N} = 1$ AGT relation \cite{Pomoni-1, Pomoni-2}. 
\end{enumerate}
The topics listed above are under investigation and results will be published elsewhere.

\section*{Acknowledgements}

We would like to thank Eoin \'O Colg\'ain, Nick Dorey, Henriette Elvang, Rouven Frassek, Jie Gu, Song He, Hans Jockers, Olaf Lechtenfeld, Jim Liu, Vladimir Mitev, Satoshi Nawata, Vasily Pestun, Elli Pomoni, Silviu Pufu, Savdeep Sethi, Huajia Wang, Brian Willett, Wenbin Yan, Gang Yang, Hong Zhang, Xinyu Zhang and Yang Zhou for many useful discussions. Special thanks to Jin Chen, Cyril Closset and Sungjay Lee for enlightening discussions that help resolve some key issues in the paper. JN's work was supported in part by the U.S. Department of Energy under grant DE-SC0007859 and by a Van Loo Postdoctoral Fellowship. JN also would like to thank the CERN theory group, the Interdisciplinary Center for Theoretical Study at the University of Science and Technology of China, Tsinghua Sanya International Mathematics Forum, the Institute of Theoretical Physics at Chinese Academy of Sciences, the Korea Institute for Advanced Study, the Leinweber Center for Theoretical Physics and the Riemann Center for Geometry and Physics for hospitality during various stages of this work. PZ would like to thank Leibniz University Hannover, Rutgers University, the Simons Center for Geometry and Physics and the Yau Mathematical Sciences Center for warm hospitality. The work of TK was supported in part by Keio Gijuku Academic Development Funds, JSPS Grant-in-Aid for Scientific Research (No.~JP17K18090), MEXT-Supported Program for the Strategic Research Foundation at Private Universities ``Topological Science'' (No.~S1511006), JSPS Grant-in-Aid for Scientific Research on Innovative Areas ``Topological Materials Science'' (No.~JP15H05855), and ``Discrete Geometric Analysis for Materials Design'' (No.~JP17H06462).

\appendix

%%%%%%%%%%%%%%%%%%%%%%%%%%%%%%%%%%%%%%%%%%%
%%%%%%%%%%%%%%%%%%%%%%%%%%%%%%%%%%%%%%%%%%%
\section{Convention}\label{app:Convention}
%%%%%%%%%%%%%%%%%%%%%%%%%%%%%%%%%%%%%%%%%%%
%%%%%%%%%%%%%%%%%%%%%%%%%%%%%%%%%%%%%%%%%%%

We follow closely the notations introduced in the book by Freedman and Van Proeyen \cite{FreedmanBook}. For the 4d Euclidean space $\mathbb{R}^4$, the gamma matrices are chosen to be
\be
  \Gamma_\mu = \sigma_\mu \otimes \mathbb{I}\, \textrm{ with } \mu \in \{1,\, 2\}\, ,\quad \Gamma_3 = \sigma_3 \otimes \sigma_1\, ,\quad \Gamma_4 = \sigma_3 \otimes \sigma_2\, ,
\ee
where $\sigma_i$ ($i = 1,\, 2,\, 3$) are the Pauli matrices, which play the role of the 2d gamma matrices. The 4d and the 2d charge conjugation matrices are related by
\be
  C_4 = C_2 \otimes \sigma_1\, ,
\ee
and they satisfy
\be
  C_4\, \Gamma_M\, C_4^{-1} = - \Gamma_M^T\, ,\quad C_2\, \sigma_\mu\, C_2^{-1} = - \sigma_\mu^T\, ,\quad C_2\, \sigma_3\, C_2^{-1} = - \sigma_3^T\, ,
\ee
with $M \in \{1,\, \cdots,\, 4\}$ and $\mu \in \{1,\, 2\}$. In practice, we can choose $C_2$ to be $i \sigma_2$. With the choice above, the matrix $\Gamma_5$ is given by
\be
  \Gamma_5 = - \Gamma_1\, \Gamma_2\, \Gamma_3\, \Gamma_4\, = \sigma_3 \otimes \sigma_3\, .
\ee
Moreover, for later convenience we obtain explicitly the following expressions :
\be
  \Gamma_{\mu\nu} = \sigma_{\mu\nu} \otimes \mathbb{I}\, ,\quad \Gamma_{\mu 3} = \sigma_\mu \sigma_3 \otimes \sigma_1\, ,\quad \Gamma_{\mu 4} = \sigma_\mu \sigma_3 \otimes \sigma_2\, ,\quad \Gamma_{3 4} = \mathbb{I} \otimes i \sigma_3\, .
\ee

In this paper, we use both the 4d and the 2d commuting Killing spinors. The 4d commuting spinors satisfy the following Fierz identity:
\begin{align}
  (\bar{\lambda} A \chi) (\bar{\psi} B \varphi) & = \frac{1}{4} (\bar{\lambda} A B \varphi) (\bar{\psi} \chi) + \frac{1}{4} (\bar{\lambda} A \Gamma^M B \varphi) (\bar{\psi} \Gamma_M \chi) \nonumber\\
  {} & \quad - \frac{1}{8} (\bar{\lambda} A \Gamma^{MN} B \varphi) (\bar{\psi} \Gamma_{MN} \chi) - \frac{1}{4} (\bar{\lambda} A \Gamma^M \Gamma_5 B \varphi) (\bar{\psi} \Gamma_M \Gamma_5 \chi) \nonumber\\
  {} & \quad + \frac{1}{4} (\bar{\lambda} A \Gamma_5 B \varphi) (\bar{\psi} \Gamma_5 \chi)\, ,\label{eq:4dFierz}
\end{align}
and the 2d commuting spinors satisfy the following Fierz identity:
\be\label{eq:2dFierz}
  (\epsilon \eta) \chi = \frac{1}{2} \eta\, (\epsilon \chi) + \frac{1}{2} \sigma_3\, \eta\, (\epsilon \sigma_3 \chi) + \frac{1}{2} \sigma_\mu \eta\, (\epsilon \sigma^\mu \chi)\, .
\ee
For the 4d and 2d anti-commuting spinors, the corresponding Fierz identities are just the expressions above with a global minus sign on the right-hand side.

In our convention, the transpose of a bilinear of commuting spinors has an extra minus sign, i.e.,
\be
  (\epsilon^T M \eta)^T = - \eta^T M^T \epsilon\, .
\ee

%%%%%%%%%%%%%%%%%%%%%%%%%%%%%%%%%%%%%%%%%%%
%%%%%%%%%%%%%%%%%%%%%%%%%%%%%%%%%%%%%%%%%%%
\section{Killing Spinors on $S^2$}\label{app:S2KS}
%%%%%%%%%%%%%%%%%%%%%%%%%%%%%%%%%%%%%%%%%%%
%%%%%%%%%%%%%%%%%%%%%%%%%%%%%%%%%%%%%%%%%%%

To define the theory on $S^2 \times \mathbb{R}^2$ and still preserve $\mathcal{N} = 1$ supersymmetry, we require that $\Upsilon$ satisfy the following equations:
\be
  \nabla_\mu \Upsilon = \pm \frac{1}{2 \ell} \Gamma_\mu \Gamma_5 \Upsilon\, ,\quad \nabla_z \Upsilon = 0 = \nabla_{\bar{z}} \Upsilon\, ,
\ee
where $\ell$ denotes the radius of $S^2$. In this paper, we make the following choice of the sign:
\be
  \nabla_\mu \Upsilon = \frac{1}{2 \ell} \Gamma_\mu \Gamma_5 \Upsilon\, ,\quad \nabla_z \Upsilon = 0 = \nabla_{\bar{z}} \Upsilon\, .
\ee
Under the decomposition \eqref{eq:KillingDecomp}, the Killing spinor equations above can be written in terms of the 2d Killing spinors:
\begin{align}\label{eq:S2KillingSpinor-1}
\begin{split}
  \nabla_\mu \epsilon = \frac{1}{2 \ell} \sigma_\mu \sigma_3\, \epsilon\, , & \quad \nabla_z \epsilon = 0 = \nabla_{\bar{z}} \epsilon\, ,\\
  \nabla_\mu \tilde{\epsilon} = - \frac{1}{2 \ell} \sigma_\mu \sigma_3\, \tilde{\epsilon}\, , & \quad \nabla_z \tilde{\epsilon} = 0 = \nabla_{\bar{z}} \tilde{\epsilon}\, ,
\end{split}
\end{align}
where $\epsilon$ and $\tilde{\epsilon}$ are Killing spinors on $S^2$.

The Killing spinor equation on $S^2$ has an alternative expression:
\be\label{eq:S2KillingSpinor-2}
  \nabla_\mu \epsilon'_\pm = \pm \frac{i}{2 \ell} \sigma_\mu \epsilon'_\pm\, .
\ee
The general solution $\epsilon'_+$ is given by
\be
  \epsilon'_+ = b_1\, e^{- i \frac{\varphi}{2}}
  \left( \begin{array}{c}
    \textrm{sin} \frac{\theta}{2} \\
    -i\, \textrm{cos} \frac{\theta}{2}
  \end{array} \right) + b_2\, e^{i \frac{\varphi}{2}}
  \left( \begin{array}{c}
    \textrm{cos} \frac{\theta}{2} \\
    i\, \textrm{sin} \frac{\theta}{2}
  \end{array} \right)\, ,
\ee
where $b_1$ and $b_2$ are two complex constants. The other solution $\epsilon'_-$ can be obtained by $\epsilon'_- = \sigma_3 \epsilon'_+$. The conjugate spinors $\epsilon'^c \equiv C_2^{-1} \epsilon'^*$ satisfies
\be
  \nabla_\mu \epsilon'^c = \mp \frac{i}{2 \ell} \sigma_\mu \epsilon'^c\, .
\ee
Moreover, $\sigma_3\, \epsilon'$ satisfies the same equations above as $\epsilon'^c$.

From the Killing spinors $\epsilon'_\pm$ on $S^2$ satisfying \eqref{eq:S2KillingSpinor-2}, one can construct
\be
  \epsilon_\pm = (\mathbb{I} + i \sigma_3) \epsilon'_\pm\, ,
\ee
which satisfy the other Killing spinor equations \eqref{eq:S2KillingSpinor-1} on $S^2$.

%%%%%%%%%%%%%%%%%%%%%%%%%%%%%%%%%%%%%%%%%%%
%%%%%%%%%%%%%%%%%%%%%%%%%%%%%%%%%%%%%%%%%%%
\section{2D $\mathcal{N}=(2,2)$ Supersymmetry}\label{app:2dSUSY}
%%%%%%%%%%%%%%%%%%%%%%%%%%%%%%%%%%%%%%%%%%%
%%%%%%%%%%%%%%%%%%%%%%%%%%%%%%%%%%%%%%%%%%%

\subsection{Vector Multiplet}\label{app:2dSUSYvec}

As discussed in Subsection~\ref{sec:SUSYvec}, under certain assumptions we can rewrite the 4d $\mathcal{N}=1$ supersymmetry in terms of 2d fields with $\mathcal{N} = (2, 2)$ supersymmetry, similar to the one studied in Refs.~\cite{S2-1, S2-2, S2-4}.

The index $\mu$ denotes the coordinates along the $S^2$-direction, while $z = x_3 + i x_4$ and $\bar{z} = x_3 - i x_4$ denote the $\mathbb{R}^2$ directions. The fields in the 4d vector multiplet can be decomposed into the 2d fields in the following way:
\begin{align}
  A_M & \rightarrow A_\mu\, \textrm{with } \mu \in \{1,\, 2\},\, A_z = \frac{1}{2} (A_3 - i A_4),\, A_{\bar{z}} = \frac{1}{2} (A_3 + i A_4)\, ,\\
  \Xi & = \lambda \otimes \zeta_+ + \widetilde{\lambda} \otimes \zeta_-\, .
\end{align}

In principle, these 2d fields also depend on the coordinates $(z,\, \bar{z})$ along the $\mathbb{R}^2$ directions, which implies their origin of the 4d $\mathcal{N}=1$ supersymmetry constructed Section~\ref{sec:SUSY}. However, we can further require that they are independent of the coordinates $(z,\, \bar{z})$, then we will obtain a 2d $\mathcal{N} = (2, 2)$ supersymmetry, similar to the one from dimensional reduction. Under these assumptions, we obtain the supersymmetry transformations for the vector multiplet on $\mathbb{R}^4$ as follows:
\begin{align}
\begin{split}\label{eq:vecR22SUSYtrafo}
  \delta A_\mu & = -\frac{1}{2} \epsilon^T C_2 \sigma_\mu \widetilde{\lambda} - \frac{1}{2} \tilde{\epsilon}^T C_2 \sigma_\mu \lambda\, ,\\
  \delta A_z & = - \frac{1}{2} \epsilon^T C_2 \sigma_3 \lambda\, ,\\
  \delta A_{\bar{z}} & = - \frac{1}{2} \tilde{\epsilon}^T C_2 \sigma_3 \widetilde{\lambda}\, ,\\
  \delta \lambda & = \frac{1}{4} F_{\mu\nu}\, \sigma^{\mu\nu} \epsilon + F_{\mu z} \sigma^\mu \sigma_3 \tilde{\epsilon} + F_{z \bar{z}}\, \epsilon + \frac{i}{2} D \sigma_3\, \epsilon\, ,\\
  \delta \widetilde{\lambda} & = \frac{1}{4} F_{\mu\nu}\, \sigma^{\mu\nu} \tilde{\epsilon} + F_{\mu \bar{z}} \sigma^\mu \sigma_3 \epsilon - F_{z \bar{z}}\, \tilde{\epsilon} - \frac{i}{2} D \sigma_3\, \tilde{\epsilon}\, ,\\
  \delta D & = -\frac{i}{2} \epsilon^T C_2 \sigma_3 \sigma^\mu D_\mu \widetilde{\lambda} + \frac{i}{2} \tilde{\epsilon}^T C_2 \sigma_3 \sigma^\mu D_\mu \lambda - i \epsilon^T C_2 D_{\bar{z}} \lambda + i \tilde{\epsilon}^T C_2 D_z \widetilde{\lambda}\, ,
\end{split}
\end{align}
where the auxiliary field $D$ is assumed to be anti-hermitian. These transformations on $\mathbb{R}^4$ satisfy the following SUSY algebra relations:
\be
  \{\delta_{\epsilon_1},\, \delta_{\epsilon_2}\} = 0 = \{\delta_{\tilde{\epsilon}_1},\, \delta_{\tilde{\epsilon}_2}\}\, ,
\ee
and
\begin{align}
\begin{split}
  \{\delta_\epsilon,\, \delta_{\tilde{\epsilon}}\} A_\mu & = \xi^\nu \partial_\nu A_\mu - D_\mu (\xi^\nu A_\nu)\, ,\\
  \{\delta_\epsilon,\, \delta_{\tilde{\epsilon}}\} A_z & = \xi^\nu \partial_\nu A_z - D_z (\xi^\nu A_\nu)\, ,\\
  \{\delta_\epsilon,\, \delta_{\tilde{\epsilon}}\} A_{\bar{z}} & = \xi^\nu \partial_\nu A_{\bar{z}} - D_{\bar{z}} (\xi^\nu A_\nu)\, ,\\
  \{\delta_\epsilon,\, \delta_{\tilde{\epsilon}}\} \lambda & = \xi^\nu \partial_\nu \lambda + [\xi^\nu A_\nu,\, \lambda]\, ,\\
  \{\delta_\epsilon,\, \delta_{\tilde{\epsilon}}\} \widetilde{\lambda} & = \xi^\nu \partial_\nu \widetilde{\lambda} + [\xi^\nu A_\nu,\, \widetilde{\lambda}]\, ,\\
  \{\delta_\epsilon,\, \delta_{\tilde{\epsilon}}\} D & = \xi^\nu \partial_\nu D + [\xi^\nu A_\nu,\, D]\, ,
\end{split}
\end{align}
where
\begin{align}
  \xi^\mu & \equiv \frac{1}{2} (\tilde{\epsilon}^T C_2 \sigma^\mu \epsilon)\, .
\end{align}
To obtain these commutation relations, we use the fact that $\epsilon$ and $\tilde{\epsilon}$ are two independent solutions to the same 2d Killing spinor equation, and
\be
  \epsilon_1 \propto \epsilon_2\, ,\quad \tilde{\epsilon}_1 \propto \tilde{\epsilon}_2\, ,
\ee
which are all constant spinors for $\mathbb{R}^2$. Moreover, we need the Fierz identity for the 2d commuting spinors \eqref{eq:2dFierz}.

To define a consistent supersymmetry on $S^2 \times \mathbb{R}^2$, we first see that the transformations \eqref{eq:vecR22SUSYtrafo} on $\mathbb{R}^4$ do not form a closed algebra on the space $S^2 \times \mathbb{R}^2$. To obtain a closed algebra, we have to add additional terms to $\delta \lambda$, $\delta \widetilde{\lambda}$ and $\delta D$, which are
\begin{align}
\begin{split}
  \delta' \lambda & = - b A_z \sigma^\mu \sigma_3 D_\mu \tilde{\epsilon}\, ,\\
  \delta' \widetilde{\lambda} & = - \tilde{b} A_{\bar{z}} \sigma^\mu \sigma_3 D_\mu \epsilon\, ,\\
  \delta' D & = \frac{i \tilde{a}}{2} (D_\mu \epsilon)^T C_2 \sigma_3 \sigma^\mu \widetilde{\lambda} - \frac{i a}{2} (D_\mu \tilde{\epsilon})^T C_2 \sigma_3 \sigma^\mu \lambda\, ,
\end{split}
\end{align}
while the other transformations remain the same. Moreover, due to the non-trivial 2d Killing spinor equations there are some additional terms appearing in $\delta F_{\mu \nu}$, $\delta F_{\mu z}$ and $\delta F_{\mu \bar{z}}$. By requiring the closure of the algebra on $S^2 \times \mathbb{R}^2$, one can fix the constants in the additional terms $\delta' \lambda$, $\delta' \widetilde{\lambda}$ and $\delta' D$ to be
\be
  a = \tilde{a} = b = \tilde{b} = -1\, .
\ee
Hence, the new SUSY transformations on $S^2 \times \mathbb{R}^2$ become
\begin{align}
\begin{split}\label{eq:vecS2R2SUSYtrafoNew}
  \delta A_\mu & = -\frac{1}{2} \epsilon^T C_2 \sigma_\mu \widetilde{\lambda} - \frac{1}{2} \tilde{\epsilon}^T C_2 \sigma_\mu \lambda\, ,\\
  \delta A_z & = - \frac{1}{2} \epsilon^T C_2 \sigma_3 \lambda\, ,\\
  \delta A_{\bar{z}} & = - \frac{1}{2} \tilde{\epsilon}^T C_2 \sigma_3 \widetilde{\lambda}\, ,\\
  \delta \lambda & = \frac{1}{4} F_{\mu\nu}\, \sigma^{\mu\nu} \epsilon + F_{\mu z} \sigma^\mu \sigma_3 \tilde{\epsilon} + F_{z \bar{z}}\, \epsilon + \frac{i}{2} D \sigma_3\, \epsilon + A_z \sigma^\mu \sigma_3 D_\mu \tilde{\epsilon}\, ,\\
  \delta \widetilde{\lambda} & = \frac{1}{4} F_{\mu\nu}\, \sigma^{\mu\nu} \tilde{\epsilon} + F_{\mu \bar{z}} \sigma^\mu \sigma_3 \epsilon - F_{z \bar{z}}\, \tilde{\epsilon} - \frac{i}{2} D \sigma_3\, \tilde{\epsilon} + A_{\bar{z}} \sigma^\mu \sigma_3 D_\mu \epsilon\, ,\\
  \delta D & = -\frac{i}{2} \epsilon^T C_2 \sigma_3 \sigma^\mu D_\mu \widetilde{\lambda} + \frac{i}{2} \tilde{\epsilon}^T C_2 \sigma_3 \sigma^\mu D_\mu \lambda - i \epsilon^T C_2 D_{\bar{z}} \lambda + i \tilde{\epsilon}^T C_2 D_z \widetilde{\lambda} \\
  {} & \quad - \frac{i}{2} (D_\mu \epsilon)^T C_2 \sigma_3 \sigma^\mu \widetilde{\lambda} + \frac{i}{2} (D_\mu \tilde{\epsilon})^T C_2 \sigma_3 \sigma^\mu \lambda\, .
\end{split}
\end{align}
The commutation relations of these transformations on $S^2 \times \mathbb{R}^2$ are
\be
  \{\delta_{\epsilon_1},\, \delta_{\epsilon_2}\} = 0 = \{\delta_{\tilde{\epsilon}_1},\, \delta_{\tilde{\epsilon}_2}\}\, ,
\ee
and
\begin{align}
\begin{split}
  \{\delta_\epsilon,\, \delta_{\tilde{\epsilon}}\} A_u & = \xi^\nu \partial_\nu A_u - D_u \left(\xi^M A_M \right) + \alpha A_u\, ,\\
  \{\delta_\epsilon,\, \delta_{\tilde{\epsilon}}\} A_{\bar{u}} & = \xi^\nu \partial_\nu A_{\bar{u}} - D_{\bar{u}} \left(\xi^M A_M \right) - \alpha A_{\bar{u}}\, ,\\
  \{\delta_\epsilon,\, \delta_{\tilde{\epsilon}}\} A_z & = \xi^\nu \partial_\nu A_z + [\xi^M A_M,\, A_z] + \rho A_z\, ,\\
  \{\delta_\epsilon,\, \delta_{\tilde{\epsilon}}\} A_{\bar{z}} & = \xi^\nu \partial_\nu A_{\bar{z}} + [\xi^M A_M,\, A_{\bar{z}}] - \rho A_{\bar{z}}\, ,\\
  \{\delta_\epsilon,\, \delta_{\tilde{\epsilon}}\} P_L \lambda & = \xi^\nu \partial_\nu P_L \lambda + [\xi^M A_M,\, P_L \lambda] + \frac{1}{2} \alpha P_L \lambda + \frac{1}{2} \rho P_L \lambda\, ,\\
  \{\delta_\epsilon,\, \delta_{\tilde{\epsilon}}\} P_R \widetilde{\lambda} & = \xi^\nu \partial_\nu P_R \widetilde{\lambda} + [\xi^M A_M,\, P_R \widetilde{\lambda}] - \frac{1}{2} \alpha P_R \widetilde{\lambda} - \frac{1}{2} \rho P_R \widetilde{\lambda}\, ,\\
  \{\delta_\epsilon,\, \delta_{\tilde{\epsilon}}\} P_R \lambda & = \xi^\nu \partial_\nu P_R \lambda + [\xi^M A_M,\, P_R \lambda] - \frac{1}{2} \alpha P_R \lambda + \frac{1}{2} \rho P_R \lambda\, ,\\
  \{\delta_\epsilon,\, \delta_{\tilde{\epsilon}}\} P_L \widetilde{\lambda} & = \xi^\nu \partial_\nu P_L \widetilde{\lambda} + [\xi^M A_M,\, P_L \widetilde{\lambda}] + \frac{1}{2} \alpha P_L \widetilde{\lambda} - \frac{1}{2} \rho P_L \widetilde{\lambda}\, ,\\
  \{\delta_\epsilon,\, \delta_{\tilde{\epsilon}}\} D & = \xi^\nu \partial_\nu D + [\xi^M A_M,\, D]\, .
\end{split}
\end{align}

\subsection{Gauged (Anti-)Chiral Multiplet}\label{app:2dSUSYchiral}

Similar to the vector multiplet, we apply the following decomposition of the spinor to the chiral multiplet:
\be\label{eq:PsiDecomp}
  \Psi = \psi \otimes \zeta_+ + \widetilde{\psi} \otimes \zeta_-\, .
\ee
Consequently,
\be
  P_L \Psi = P_L \psi \otimes \zeta_+ + P_R \widetilde{\psi} \otimes \zeta_-\, ,
\ee
where $P_L$ on the left-hand side denotes the 4d projection operator, while $P_{L, R}$ on the right-hand side stand for the 2d projection operators.

Unlike in the main text, now we assume that the 2d fields are independent of the coordinates $(z,\, \bar{z})$ along the $\mathbb{R}^2$ directions, then we will obtain a 2d $\mathcal{N} = (2, 2)$ supersymmetry, similar to the one from dimensional reduction. Under these assumptions, we obtain the SUSY transformations for the chiral multiplet on $\mathbb{R}^4$ in terms of the 2d fields:
\begin{align}
\begin{split}\label{eq:chiralR22SUSYtrafo}
  \delta \phi & = \frac{1}{\sqrt{2}} (\tilde{\epsilon}^T C_2 P_L \psi + \epsilon^T C_2 P_R \widetilde{\psi})\, ,\\
  \delta P_L \psi & = \frac{1}{\sqrt{2}} (P_L \sigma^\mu \epsilon) D_\mu \phi + \sqrt{2} (P_L \tilde{\epsilon}) D_z \phi + \frac{1}{\sqrt{2}} (P_L \epsilon) F\, ,\\
  \delta P_R \widetilde{\psi} & = \frac{1}{\sqrt{2}} (P_R \sigma^\mu \tilde{\epsilon}) D_\mu \phi - \sqrt{2} (P_R \epsilon) D_{\bar{z}} \phi + \frac{1}{\sqrt{2}} (P_R \tilde{\epsilon}) F\, ,\\
  \delta F & = \frac{1}{\sqrt{2}} (\tilde{\epsilon}^T C_2 \sigma^\mu D_\mu P_L \psi + \epsilon^T C_2 \sigma^\mu D_\mu P_R \widetilde{\psi}) + \sqrt{2} (\tilde{\epsilon}^T C_2 \sigma_3 D_z P_R \widetilde{\psi} + \epsilon^T C_2 \sigma_3 D_{\bar{z}} P_L \psi)\\
  {} & \quad - (\tilde{\epsilon}^T C_2 P_R \lambda + \epsilon^T C_2 P_L \widetilde{\lambda}) \phi\, ,
\end{split}
\end{align}
where $\phi$ and $F$ denote the 2d scalar field and the auxiliary field respectively, and the 2d projection operators are in Eq.~\eqref{eq:2dProj}.

The transformations above satisfy the following commutation relations:
\be
  \{\delta_{\epsilon_1},\, \delta_{\epsilon_2}\} = 0 = \{\delta_{\tilde{\epsilon}_1},\, \delta_{\tilde{\epsilon}_2}\}\, ,
\ee
and
\begin{align}
\begin{split}
  \{\delta_\epsilon,\, \delta_{\tilde{\epsilon}}\} \phi & = \xi^\mu D_\mu \phi\, ,\\
  \{\delta_\epsilon,\, \delta_{\tilde{\epsilon}}\} \psi & = \xi^\mu D_\mu \psi\, ,\\
  \{\delta_\epsilon,\, \delta_{\tilde{\epsilon}}\} \widetilde{\psi} & = \xi^\mu D_\mu \widetilde{\psi}\, ,\\
  \{\delta_\epsilon,\, \delta_{\tilde{\epsilon}}\} F & = \xi^\mu D_\mu F\, .
\end{split}
\end{align}

In order to define the chiral multiplet consistently on $S^2 \times \mathbb{R}^2$, similar to the vector multiplet, we have to introduce some additional terms to the transformations $\delta \psi$, $\delta \widetilde{\psi}$ and $\delta F$:
\begin{align}
\begin{split}
  \delta' \psi & = - \frac{c}{\sqrt{2}} (P_L \sigma^\mu D_\mu \epsilon) \phi\, ,\\
  \delta' \widetilde{\psi} & = - \frac{\tilde{c}}{\sqrt{2}} (P_R \sigma^\mu D_\mu \tilde{\epsilon}) \phi\, ,\\
  \delta' F & = - \frac{d}{\sqrt{2}} (D_\mu \tilde{\epsilon})^T C_2 \sigma^\mu \psi - \frac{\tilde{d}}{\sqrt{2}} (D_\mu \epsilon)^T C_2 \sigma^\mu \widetilde{\psi}\, ,
\end{split}
\end{align}
while the transformation $\delta \phi$ remains the same as Eq.~\eqref{eq:chiralR22SUSYtrafo}. By requiring the closure of the SUSY algebra, one can fix the constants $c$, $\tilde{c}$, $d$ and $\tilde{d}$:
\be
  c = \tilde{c} = d = \tilde{d} = - q\, ,
\ee
where $q$ is a constant. Therefore, the SUSY transformations of the chiral multiplet on $S^2 \times \mathbb{R}^2$ become
\begin{align}
\begin{split}\label{eq:chiralS2R2SUSYtrafoNew}
  \delta \phi & = \frac{1}{\sqrt{2}} (\tilde{\epsilon}^T C_2 P_L \psi + \epsilon^T C_2 P_R \widetilde{\psi})\, ,\\
  \delta P_L \psi & = \frac{1}{\sqrt{2}} (P_L \sigma^\mu \epsilon) D_\mu \phi + \sqrt{2} (P_L \tilde{\epsilon}) D_z \phi + \frac{1}{\sqrt{2}} (P_L \epsilon) F + \frac{q}{\sqrt{2}} (P_L \sigma^\mu D_\mu \epsilon) \phi\, ,\\
  \delta P_R \widetilde{\psi} & = \frac{1}{\sqrt{2}} (P_R \sigma^\mu \tilde{\epsilon}) D_\mu \phi - \sqrt{2} (P_R \epsilon) D_{\bar{z}} \phi + \frac{1}{\sqrt{2}} (P_R \tilde{\epsilon}) F + \frac{q}{\sqrt{2}} (P_R \sigma^\mu D_\mu \tilde{\epsilon}) \phi\, ,\\
  \delta F & = \frac{1}{\sqrt{2}} (\tilde{\epsilon}^T C_2 \sigma^\mu D_\mu P_L \psi + \epsilon^T C_2 \sigma^\mu D_\mu P_R \widetilde{\psi}) + \sqrt{2} (\tilde{\epsilon}^T C_2 \sigma_3 D_z P_R \widetilde{\psi} + \epsilon^T C_2 \sigma_3 D_{\bar{z}} P_L \psi)\\
  {} & \quad - (\tilde{\epsilon}^T C_2 P_R \lambda + \epsilon^T C_2 P_L \widetilde{\lambda}) \phi + \frac{q}{\sqrt{2}} (D_\mu \tilde{\epsilon})^T C_2 \sigma^\mu P_L \psi + \frac{q}{\sqrt{2}} (D_\mu \epsilon)^T C_2 \sigma^\mu P_R \widetilde{\psi}\, .
\end{split}
\end{align}
The corresponding commutation relations are
\be
  \{\delta_{\epsilon_1},\, \delta_{\epsilon_2}\} = 0 = \{\delta_{\tilde{\epsilon}_1},\, \delta_{\tilde{\epsilon}_2}\}\, ,
\ee
and
\begin{align}
\begin{split}
  \{\delta_\epsilon,\, \delta_{\tilde{\epsilon}}\} \phi & = \xi^\nu \partial_\nu \phi + [\xi^M A_M,\, \phi]\, ,\\
  \{\delta_\epsilon,\, \delta_{\tilde{\epsilon}}\} P_L \psi & = \xi^\nu \partial_\nu P_L \psi + [\xi^M A_M,\, P_L \psi] + \frac{1}{2} \alpha P_L \psi + \frac{1}{2} \rho P_L \psi\, ,\\
  \{\delta_\epsilon,\, \delta_{\tilde{\epsilon}}\} P_R \widetilde{\psi} & = \xi^\nu \partial_\nu P_R \widetilde{\psi} + [\xi^M A_M,\, P_R \widetilde{\psi}] - \frac{1}{2} \alpha P_R \widetilde{\psi} - \frac{1}{2} \rho P_R \widetilde{\psi}\, ,\\
  \{\delta_\epsilon,\, \delta_{\tilde{\epsilon}}\} F & = \xi^\nu \partial_\nu F + [\xi^M A_M,\, F]\, ,
\end{split}
\end{align}
where $\xi^M$, $\rho$ and $\alpha$ are defined the same as before.

For the anti-chiral multiplet $(\bar{\Phi},\, P_R \Psi,\, \bar{F})$ on $\mathbb{R}^4$, the SUSY transformations are given by Eqs.~\eqref{eq:antichiralR4SUSYtrafo}. We can apply the same decomposition \eqref{eq:PsiDecomp} and consequently
\be
  P_R \Psi = P_R \psi \otimes \zeta_+ + P_L \widetilde{\psi} \otimes \zeta_-\, .
\ee
Different from the main text, if we assume that these 2d fields are independent of the coordinates $(z,\, \bar{z})$ along the $\mathbb{R}^2$ directions, we obtain the SUSY transformations for the anti-chiral multiplet on $S^2 \times \mathbb{R}^2$ as follows:
\begin{align}
\begin{split}
  \delta \bar{\phi} & = \frac{1}{\sqrt{2}} (\epsilon^T C_2 P_L \widetilde{\psi} + \tilde{\epsilon}^T C_2 P_R \psi)\, ,\\
  \delta P_R \psi & = \frac{1}{\sqrt{2}} (P_R \sigma^\mu \epsilon) D_\mu \bar{\phi} - \sqrt{2} (P_R \tilde{\epsilon}) D_z \bar{\phi} + \frac{1}{\sqrt{2}} (P_R \epsilon) \bar{F} + \frac{q}{\sqrt{2}} (P_R \sigma^\mu D_\mu \epsilon) \bar{\phi}\, ,\\
  \delta P_L \widetilde{\psi} & = \frac{1}{\sqrt{2}} (P_L \sigma^\mu \tilde{\epsilon}) D_\mu \bar{\phi} + \sqrt{2} (P_L \epsilon) D_{\bar{z}} \bar{\phi} + \frac{1}{\sqrt{2}} (P_L \tilde{\epsilon}) \bar{F} + \frac{q}{\sqrt{2}} (P_L \sigma^\mu D_\mu \tilde{\epsilon}) \bar{\phi}\, ,\\
  \delta \bar{F} & = \frac{1}{\sqrt{2}} (\epsilon^T C_2 \sigma^\mu D_\mu P_L \widetilde{\psi} + \tilde{\epsilon}^T C_2 \sigma^\mu D_\mu P_R \psi) + \sqrt{2} (\epsilon^T C_2 \sigma_3 D_{\bar{z}} P_R \psi + \tilde{\epsilon}^T C_2 \sigma_3 D_z P_L \widetilde{\psi})\\
  {} & \quad - (\epsilon^T C_2 P_R \widetilde{\lambda} + \tilde{\epsilon}^T C_2 P_L \lambda) \bar{\phi} + \frac{q}{\sqrt{2}} (D_\mu \epsilon)^T C_2 \sigma^\mu P_L \widetilde{\psi} + \frac{q}{\sqrt{2}} (D_\mu \tilde{\epsilon})^T C_2 \sigma^\mu P_R \psi\, .
\end{split}
\end{align}
The transformations satisfy the commutation relations:
\be
  \{\delta_{\epsilon_1},\, \delta_{\epsilon_2}\} = 0 = \{\delta_{\tilde{\epsilon}_1},\, \delta_{\tilde{\epsilon}_2}\}\, ,
\ee
and
\begin{align}
\begin{split}
  \{\delta_\epsilon,\, \delta_{\tilde{\epsilon}}\} \bar{\phi} & = \xi^\nu \partial_\nu \bar{\phi} + [\xi^M A_M,\, \bar{\phi}]\, ,\\
  \{\delta_\epsilon,\, \delta_{\tilde{\epsilon}}\} P_R \psi & = \xi^\nu \partial_\nu P_R \psi + [\xi^M A_M,\, P_R \psi] - \frac{1}{2} \alpha P_R \psi + \frac{1}{2} \rho P_R \psi\, ,\\
  \{\delta_\epsilon,\, \delta_{\tilde{\epsilon}}\} P_L \widetilde{\psi} & = \xi^\nu \partial_\nu P_L \widetilde{\psi} + [\xi^M A_M,\, P_L \widetilde{\psi}] + \frac{1}{2} \alpha P_L \widetilde{\psi} - \frac{1}{2} \rho P_L \widetilde{\psi}\, ,\\
  \{\delta_\epsilon,\, \delta_{\tilde{\epsilon}}\} \bar{F} & = \xi^\nu \partial_\nu \bar{F} + [\xi^M A_M,\, \bar{F}]\, .
\end{split}
\end{align}

%%%%%%%%%%%%%%%%%%%%%%%%%%%%%%%%%%%%%%%%%%%
%%%%%%%%%%%%%%%%%%%%%%%%%%%%%%%%%%%%%%%%%%%
\section{BPS Equations and Classical Solutions}\label{app:BPS}
%%%%%%%%%%%%%%%%%%%%%%%%%%%%%%%%%%%%%%%%%%%
%%%%%%%%%%%%%%%%%%%%%%%%%%%%%%%%%%%%%%%%%%%

Using the explicit expressions of the commuting Killing spinors \eqref{eq:ExplicitKS}, we obtain the following identities for the commuting Killing spinors $\Sigma$ and $\widetilde{\Sigma}$, which are useful in computing the $\delta$-exact part of the action in Subsection~\ref{sec:ExactAction}:
\be
  \Sigma^\dagger \Sigma = \widetilde{\Sigma}^\dagger \widetilde{\Sigma} = 2\, ,\quad \widetilde{\Sigma}^\dagger \Sigma = \Sigma^\dagger \widetilde{\Sigma} = 0\, ,\nonumber
\ee
\be
  \Sigma^\dagger \Gamma_5 \Sigma = \widetilde{\Sigma}^\dagger \Gamma_5 \widetilde{\Sigma} = - 2\, \textrm{cos}\, \theta\, ,\quad \widetilde{\Sigma}^\dagger \Gamma_5 \Sigma = \Sigma^\dagger \Gamma_5 \widetilde{\Sigma} = 0\, ,\nonumber
\ee
\be
  \Sigma^\dagger \Gamma_M \Sigma = \widetilde{\Sigma}^\dagger \Gamma_M \widetilde{\Sigma} =  -2\, \textrm{sin}\, \theta\, \delta_{M 1}\, ,\quad \widetilde{\Sigma}^\dagger \Gamma_M \Sigma = \Sigma^\dagger \Gamma_M \widetilde{\Sigma} = 0\, ,\nonumber
\ee
\be
  \Sigma^\dagger \Gamma_M \Gamma_5 \Sigma = -\widetilde{\Sigma}^\dagger \Gamma_M \Gamma_5 \widetilde{\Sigma} = -2 i\, \textrm{sin}\, \theta\, \delta_{M 2}\, ,\nonumber
\ee
\be
  \widetilde{\Sigma}^\dagger \Gamma_M \Gamma_5 \Sigma = 2\, \textrm{sin}\, \theta\, e^{-i \varphi}\, \delta_{M 3} + 2 i\, \textrm{sin}\, \theta\, e^{-i \varphi}\, \delta_{M 4}\, ,\nonumber
\ee
\be
  \Sigma^\dagger \Gamma_M \Gamma_5 \widetilde{\Sigma} = -2\, \textrm{sin}\, \theta\, e^{i \varphi}\, \delta_{M 3} + 2 i\, \textrm{sin}\, \theta\, e^{i \varphi}\, \delta_{M 4}\, ,\nonumber
\ee
\be
  \Sigma^\dagger \Gamma_{\mu\nu} \Sigma = - 2 i\, \textrm{cos}\, \theta\, \epsilon_{\mu\nu}\, ,\quad \Sigma^\dagger \Gamma_{\mu a} \Sigma = 0\, ,\quad \Sigma^\dagger \Sigma_{34} \Sigma = 2 i\, ,\nonumber
\ee
\be
  \widetilde{\Sigma}^\dagger \Gamma_{\mu\nu} \widetilde{\Sigma} = 2 i\, \textrm{cos}\, \theta\, \epsilon_{\mu\nu}\, ,\quad \widetilde{\Sigma}^\dagger \Gamma_{\mu a} \widetilde{\Sigma} = 0\, ,\quad \widetilde{\Sigma}^\dagger \Gamma_{34} \widetilde{\Sigma} = - 2 i\, ,\nonumber
\ee
\be
  \widetilde{\Sigma}^\dagger \Gamma_{\mu\nu} \Sigma = 0\, ,\quad \widetilde{\Sigma}^\dagger \Gamma_{34} \Sigma = 0\, ,
\ee
\be
  \widetilde{\Sigma}^\dagger \Gamma_{\mu a} \Sigma = \bigg\{
  \begin{array}{c}
    2\, \textrm{cos}\, \theta\, e^{-i \varphi} \delta_{\mu 1} - 2 i\, e^{- i \varphi} \delta_{\mu 2}\, ,\, \textrm{ for } a = 3\, ,\\
    2 i\, \textrm{cos}\, \theta\, e^{-i \varphi} \delta_{\mu 1} + 2\, e^{- i \varphi} \delta_{\mu 2}\, ,\, \textrm{ for } a = 4\, ,
  \end{array}\nonumber
\ee
\be
  \Sigma^\dagger \Gamma_{\mu\nu} \widetilde{\Sigma} = 0\, ,\quad \Sigma^\dagger \Gamma_{34} \widetilde{\Sigma} = 0\, ,\nonumber
\ee
\be
  \Sigma^\dagger \Gamma_{\mu a} \widetilde{\Sigma} = \bigg\{
  \begin{array}{c}
    - 2\, \textrm{cos}\, \theta\, e^{i \varphi} \delta_{\mu 1} - 2 i\, e^{i \varphi} \delta_{\mu 2}\, ,\, \textrm{ for } a = 3\, ,\\
    2 i\, \textrm{cos}\, \theta\, e^{i \varphi} \delta_{\mu 1} - 2\, e^{i \varphi} \delta_{\mu 2}\, ,\, \textrm{ for } a = 4\, ,
  \end{array}\nonumber
\ee
\be
  \Sigma^\dagger \Gamma_{\mu\nu} \Gamma_5 \Sigma = 2 i \epsilon_{\mu\nu}\, ,\quad \Sigma^\dagger \Gamma_{\mu a} \Gamma_5 \Sigma = 0\, ,\quad \Sigma^\dagger \Gamma_{34} \Gamma_5 \Sigma = - 2 i\, \textrm{cos}\, \theta\, ,\nonumber
\ee
\be
  \widetilde{\Sigma}^\dagger \Gamma_{\mu\nu} \Gamma_5 \widetilde{\Sigma} = - 2 i \epsilon_{\mu\nu}\, ,\quad \widetilde{\Sigma}^\dagger \Gamma_{\mu a} \Gamma_5 \widetilde{\Sigma} = 0\, ,\quad \widetilde{\Sigma}^\dagger \Gamma_{34} \Gamma_5 \widetilde{\Sigma} = 2 i\, \textrm{cos}\, \theta\, ,\nonumber
\ee
\be
  \widetilde{\Sigma}^\dagger \Gamma_{\mu\nu} \Gamma_5 \Sigma = 0\, ,\quad \widetilde{\Sigma}^\dagger \Gamma_{34} \Gamma_5 \Sigma = 0\, ,\nonumber
\ee
\be
  \widetilde{\Sigma}^\dagger \Gamma_{\mu a} \Gamma_5 \Sigma = \bigg\{
  \begin{array}{c}
    - 2\, e^{-i \varphi}\, \delta_{\mu 1} + 2 i\, \textrm{cos}\, \theta\, e^{-i \varphi}\, \delta_{\mu 2}\, ,\, \textrm{ for } a = 3\, ,\\
    - 2 i\, e^{-i \varphi}\, \delta_{\mu 1} - 2\, \textrm{cos}\, \theta\, e^{-i \varphi}\, \delta_{\mu 2}\, ,\, \textrm{ for } a =4\, ,
  \end{array}\nonumber
\ee
\be
  \Sigma^\dagger \Gamma_{\mu\nu} \Gamma_5 \widetilde{\Sigma} = 0\, ,\quad \Sigma^\dagger \Gamma_{34} \Gamma_5 \widetilde{\Sigma} = 0\, ,\nonumber
\ee
\be
  \Sigma^\dagger \Gamma_{\mu a} \Gamma_5 \widetilde{\Sigma} = \bigg\{
  \begin{array}{c}
    2\, e^{i \varphi}\, \delta_{\mu 1} + 2 i\, \textrm{cos}\, \theta\, e^{i \varphi}\, \delta_{\mu 2}\, ,\, \textrm{ for } a = 3\, ,\\
    - 2 i\, e^{i \varphi}\, \delta_{\mu 1} + 2\, \textrm{cos}\, \theta\, e^{i \varphi}\, \delta_{\mu 2}\, ,\, \textrm{ for } a = 4\, .
  \end{array}\nonumber
\ee

By applying the identities above, we obtain the $\delta$-exact part of the action \eqref{eq:LexactWithD} and consequently the BPS equations \eqref{eq:AbelianBPS-1} $\sim$ \eqref{eq:AbelianBPS-3}. In Subsection~\ref{sec:ClassicalSol}, we have discussed how to obtain the classical solutions to some of the BPS equations. In the following of this appendix, we 
focus on Eq.~\eqref{eq:AbelianBPS-4}:
\be\label{eq:BPS-1}
  F_{12} - g_{YM}^2 (| \Phi^I |^2 - \eta) = 0
\ee
and one of the equations in Eq.~\eqref{eq:AbelianBPS-3}:
\be\label{eq:BPS-2}
  D_{\bar{u}} \Phi^I = 0\, ,
\ee
and we discuss their (anti-)vortex solutions.

First, in the explicit coordinates Eqs.~\eqref{eq:BPS-1} and \eqref{eq:BPS-2} become
\be\label{eq:BPS-1-temp}
  \frac{1}{\ell^2\, \textrm{sin}\, \theta} F_{\theta \varphi} - g_{YM}^2 (| \Phi^I |^2 - \eta) = 0\, ,
\ee
\be\label{eq:BPS-2-temp}
  \left[\frac{1}{\ell} \partial_\theta + \frac{i}{\textrm{sin}\, \theta} (\partial_\varphi + i A_\varphi + i \widetilde{A}_\varphi^I) - \frac{\varepsilon}{\textrm{sin}\, \theta} w\, \partial_w + \frac{\varepsilon}{\textrm{sin}\, \theta} \bar{w}\, \partial_{\bar{w}} \right] \Phi^I = 0\, .
\ee

In Section~\ref{sec:ClassicalSol}, we have considered the vortex solution throughout the whole $S^2$, but we have also mentioned that there can be a configuration with a vortex solution at the north pole and an anti-vortex solution at the south pole. In this appendix let us consider this configuration in detail. To obtain the (anti-)vortex solutions \eqref{eq:VortexSol} \eqref{eq:AntiVortexSol}, we try to solve these two equations in different regions:
\begin{itemize}
\item $\theta \approx 0$ (near the core of the vortex at the north pole):

In this case, we consider a vortex solution located at the north pole of $S^2$ and at the same time at the origin of $\mathbb{R}^2_\varepsilon$. The vortex solution is given by
\be
  \Phi^I \simeq (\theta\, e^{i \varphi})^m w^k\, ,\quad A_\varphi \simeq \ell \widetilde{m}_I - k \ell \varepsilon \quad (k \geq 0)\, ,
\ee
where $k \geq 0$ is required by the regularity of the solution at $w = 0$. They solve Eq.~\eqref{eq:BPS-2-temp} exactly. For Eq.~\eqref{eq:BPS-1-temp}, to ensure a non-vanishing field strength at the north pole, we need to tune the FI paramter $\eta \to \infty$, which corresponds to a point-like vortex. The explicit form of the field strength $F_{\theta\varphi}$ is irrelevant.

\item $\theta \approx \pi$ (near the core of the anti-vortex at the south pole):

In this case, we consider an anti-vortex solution located at the south pole of $S^2$ and at the same time at the origin of $\mathbb{R}^2_\varepsilon$. The anti-vortex solution is given by
\be
  \Phi^I \simeq (\hat{\theta}\, e^{i \varphi})^n \bar{w}^{k'}\, ,\quad A_\varphi \simeq \ell \widetilde{m}_I + k' \ell \varepsilon \quad (k' \geq 0)\, ,
\ee
where $\hat{\theta} \equiv \pi - \theta$ , and $k' \geq 0$ is required by the regularity of the solution at $w = 0$. Similar to the previous case, they solve Eq.~\eqref{eq:BPS-2-temp} exactly, and for Eq.~\eqref{eq:BPS-1-temp} one has to tune the FI paramter $\eta \to \infty$, which corresponds to a point-like anti-vortex at the south pole.

\item $\theta \approx \frac{\pi}{2}$ (on the northern hemisphere far from the core of the vortex):

In this case, we consider the solution on the northern hemisphere far from the core of the vortex located at the north pole of $S^2$, which is given by
\be
  \Phi^I \simeq \sqrt{\eta}\, e^{i m \varphi} w^k\, ,\quad A_\varphi \simeq \ell \widetilde{m}_I - m - k \ell \varepsilon \quad (k \geq 0)\, .
\ee
They solve Eq.~\eqref{eq:BPS-2-temp} exactly, but cannot solve Eq.~\eqref{eq:BPS-1-temp} exactly unless $k = 0$.

\item $\theta \approx \frac{\pi}{2}$ (on the southern hemisphere far from the core of the anti-vortex):

In this case, we consider the solution on the southern hemisphere far from the core of the anti-vortex located at the south pole of $S^2$, which is given by
\be
  \Phi^I \simeq \sqrt{\eta}\, e^{i n \varphi} \bar{w}^{k'}\, ,\quad A_\varphi \simeq \ell \widetilde{m}_I - n + k' \ell \varepsilon \quad (k' \geq 0)\, .
\ee
They solve Eq.~\eqref{eq:BPS-2-temp} exactly, but cannot solve Eq.~\eqref{eq:BPS-1-temp} exactly unless $k' = 0$.
\end{itemize}

Comparing the solutions of the last two cases, we find that in order to glue the solutions from the two hemispheres together, regularity requires that
\be\label{eq:GluingCond}
  k = k' = 0\, ,
\ee
which is consistent with the requirement discussed before that the solutions solve Eq.~\eqref{eq:BPS-1-temp} exactly at $k = k' = 0$. Hence, a vortex solution located at the north pole of $S^2$ and an anti-vortex solution located at the south pole of $S^2$ can be glued together through a gauge transformation along the equator ($\theta = \frac{\pi}{2}$), but these configurations are trivial on the $\mathbb{R}^2_\varepsilon$-plane.

%%%%%%%%%%%%%%%%%%%%%%%%%%%%%%%%%%%%%%%%%%%
%%%%%%%%%%%%%%%%%%%%%%%%%%%%%%%%%%%%%%%%%%%
\section{1-Loop Determinants via Index Theorem}\label{app:IndexThm}
%%%%%%%%%%%%%%%%%%%%%%%%%%%%%%%%%%%%%%%%%%%
%%%%%%%%%%%%%%%%%%%%%%%%%%%%%%%%%%%%%%%%%%%

In this appendix, we apply the index theorem to compute the 1-loop determinant of the partition function around the classical solutions. The method is similar to the one used in Refs.~\cite{S2-1, Benini-3d, Peelaers}.

First, the complex coordinate $w$ on $\mathbb{R}^2_\varepsilon$ can be parametrized as
\be\label{eq:parametrization}
  w = r\, e^{i \psi}\, .
\ee
From Eqs.~\eqref{eq:KillingVecS2R2epsilon} and \eqref{eq:FullSUSYalg} we see that the square of the supersymmetry that we constructed on $S^2 \times \mathbb{R}^2_\varepsilon$ is
\begin{align}
  \delta^2 & = -\frac{1}{\ell} \partial_\varphi - \varepsilon \partial_\psi + i\, \textrm{cos} \varphi\, \textrm{sin} \theta (\partial_w + \partial_{\bar{w}}) - \textrm{sin} \varphi\, \textrm{sin} \theta (\partial_w - \partial_{\bar{w}}) \nonumber\\
  {} & \quad + \Lambda - \frac{i}{\ell}\, \textrm{cos} \theta\, J_s^u - \frac{i}{2} (\varepsilon - \frac{1}{\ell}) R_2 + i \mathcal{F}_I \widetilde{m}_I\, ,\label{eq:SUSYalgebraFinal}
\end{align}
which has fixed points $(\theta, w) = (0, 0)$ and $(\pi, 0)$, corresponding to the north and the south poles of $S^2$ together with the origin of $\mathbb{R}^2_\varepsilon$. Hence, we can consider the indices at these fixed points to obtain the 1-loop determinants.

First, the term $- \varepsilon \partial_\psi$ in the algebra \eqref{eq:SUSYalgebraFinal} generates a rotation on $\mathbb{R}^2_\varepsilon$ around the origin, which contributes a sum $\sum_{p \in \mathbb{Z}} e^{- i p \varepsilon}$ to the index.

For the chiral multiplet, at the north pole ($\theta = 0$) the SUSY transformation is of the form $D_\theta + \frac{i}{\theta} D_\varphi \sim D_{\bar{u}}$ with $u = \theta\, e^{i \varphi}$. The index is the one for the Dolbeault operator with inverted grading, i.e. $-1 / (1 - u)$. Hence, by expanding $-1 / (1 - u)$ in terms of $t = e^{i \varphi}$ and using the equivariant parameter $-1 / \ell$, we obtain the index for the chiral multiplet at the north pole:
\be\label{eq:AbelianIndexNP}
  \textrm{ind}_N^{\textrm{chiral}} = - \sum_{p \in \mathbb{Z}} e^{- i p \varepsilon} \sum_{q \geq 0} e^{-i q / \ell}\, e^{- i (\varepsilon - 1 / \ell) R_2 / 2}\, e^{i \mathcal{F}_J \widetilde{m}_J}\, e^{\Lambda_N}\, ,
\ee
where $J$ denotes an arbitrary flavor, and
\be
  \Lambda_N = - \frac{i}{\ell} A_\varphi = - i \mathcal{F}_I \widetilde{m}_I + \frac{i}{\ell} m + i k \varepsilon\, ,
\ee
with $m$ and $k$ denoting the numbers of vortices located at the north pole of $S^2$ and at the origin of $\mathbb{R}^2_\varepsilon$ respectively.

Similarly, at the south pole ($\theta = \pi$), the SUSY transformation is of the form $D_{\hat{\theta}} + \frac{i}{\hat{\theta}} D_\varphi \sim D_{\bar{u}}$ with $\hat{\theta} = \pi - \theta$ and $u = \hat{\theta}\, e^{i \varphi}$. The index is still the one for the Dolbeault operator with inverted grading, i.e. $-1 / (1 - u)$. Hence, by expanding $-1 / (1 - u)$ in terms of $t = e^{- i \varphi}$ and using the equivariant parameter $-1 / \ell$, we obtain the index for the chiral multiplet at the south pole:
\be\label{eq:AbelianIndexSP}
  \textrm{ind}_S^{\textrm{chiral}} = \sum_{p \in \mathbb{Z}} e^{i p \varepsilon} \sum_{q \geq 1} e^{i q / \ell}\, e^{- i (\varepsilon - 1 / \ell) R_2 / 2}\, e^{i \mathcal{F}_J \widetilde{m}_J}\, e^{\Lambda_S}\, ,
\ee
where $J$ denotes an arbitrary flavor, and
\be
  \Lambda_S = -\frac{i}{\ell} A_\varphi = - i \mathcal{F}_I \widetilde{m}_I + \frac{i}{\ell} n + i k \varepsilon\, ,
\ee
with $n$ and $k$ denoting the number of vortices located at the south pole of $S^2$ and at the origin of $\mathbb{R}^2_\varepsilon$ respectively. Both the index at the north pole \eqref{eq:AbelianIndexNP} and the index at the south pole \eqref{eq:AbelianIndexSP} are generalized to the non-Abelian case, which are Eqs.~\eqref{eq:indexChiralNP} and \eqref{eq:indexChiralSP} respectively.

For the vector multiplet, its index is given by the one of the de Rham operator. Hence, we obtain the indices at the north and the south poles as follows:
\be
  \textrm{ind}_N^{\textrm{vec}} = \frac{1}{2} \sum_{\vec{\alpha} \in \Delta_G} \sum_{p \in \mathbb{Z}} e^{- i p \varepsilon}\, e^{i \vec{\alpha} \cdot \vec{\sigma}^N}\, ,\quad \textrm{ind}_S^{\textrm{vec}} = \frac{1}{2} \sum_{\vec{\alpha} \in \Delta_G} \sum_{p \in \mathbb{Z}} e^{i p \varepsilon}\, e^{i \vec{\alpha} \cdot \vec{\sigma}^S}\, ,
\ee
where $\vec{\alpha}$ denotes the root vectors of the gauge group, and the constants $\vec{\sigma}^{N,\, S}$ are given by Eqs.~\eqref{eq:FixConstNP} and \eqref{eq:FixConstSP}.

%%%%%%%%%%%%%%%%%%%%%%%%%%%%%%%%%%%%%%%%%%%
%%%%%%%%%%%%%%%%%%%%%%%%%%%%%%%%%%%%%%%%%%%
\section{Special Functions}\label{app:SpFct}
%%%%%%%%%%%%%%%%%%%%%%%%%%%%%%%%%%%%%%%%%%%
%%%%%%%%%%%%%%%%%%%%%%%%%%%%%%%%%%%%%%%%%%%

In this appendix let us summarize some relevant facts about the double gamma function $\Gamma_2 (x | \varepsilon_1, \varepsilon_2)$ from the math literature.

First, the Barnes double zeta function is defined as
\be\label{eq:defineBarnesDoubleZeta}
  \zeta_2 (s;\, x | \varepsilon_1, \varepsilon_2) = \frac{1}{\Gamma (s)} \int_0^\infty \frac{dt}{t} t^s \frac{e^{-t x}}{(1 - e^{- \varepsilon_1 t}) (1 - e^{-\varepsilon_2 t})}\, .
\ee
The double gamma function is then defined as
\be
  \Gamma_2 (x | \varepsilon_1, \varepsilon_2) = \textrm{exp} \frac{d}{ds} \bigg|_0 \zeta_2 (s;\, x | \varepsilon_1, \varepsilon_2)\, .
\ee

The function $\zeta_2 (s;\, x | \varepsilon_1, \varepsilon_2)$ can be viewed as the regularization of the infinite sum:
\be
  \zeta_2 (s;\, x | \varepsilon_1, \varepsilon_2) = \sum_{m, n \geq 0} (x + m \varepsilon_1 + n \varepsilon_2)^{-s}\, .
\ee
The function $\Gamma_2 (x | \varepsilon_1, \varepsilon_2)$ is analytic in $x$ except at the poles
\be
  x = - (m \varepsilon_1 + n \varepsilon_2)\quad (m,\, n \in \mathbb{Z}_{\geq 0})\, .
\ee
Hence, $\Gamma_2 (x | \varepsilon_1, \varepsilon_2)$ can be viewed as a regularized infinite product depending on the signs of $\varepsilon_1$, $\varepsilon_2$:
\be
  \Gamma_2 (x | \varepsilon_1, \varepsilon_2) \propto \left\{
  \begin{aligned}
    & \prod_{m, n \geq 0} (x + m \varepsilon_1 + n \varepsilon_2)^{-1}\, , & \textrm{for } \varepsilon_1 > 0, \varepsilon_2 > 0\, ; \\
    & \prod_{m, n \geq 0} \left(x + m \varepsilon_1 - (n+1) \varepsilon_2 \right)\, , & \textrm{for } \varepsilon_1 > 0, \varepsilon_2 < 0\, ; \\
    & \prod_{m, n \geq 0} \left(x - (m+1) \varepsilon_1 + n \varepsilon_2 \right)\, , & \textrm{for } \varepsilon_1 < 0, \varepsilon_2 > 0\, ; \\
    & \prod_{m, n \geq 0} \left(x - (m+1) \varepsilon_1 - (n+1) \varepsilon_2 \right)^{-1}\, , & \textrm{for } \varepsilon_1 < 0, \varepsilon_2 < 0\, .
  \end{aligned} \right.
\ee

\bibliographystyle{utphys}
\bibliography{N=1Loc}

\end{document}